\documentclass[usenatbib]{mn2e}
\usepackage{graphicx,amsmath,multirow,amssymb}
\usepackage{subfig}
\usepackage{natbib}
\newcommand{\comment}[1]{}

\def\simgt{\lower.5ex\hbox{$\; \buildrel > \over \sim \;$}}
\def\simlt{\lower.5ex\hbox{$\; \buildrel < \over \sim \;$}}

\newcommand{\Msun}{\ensuremath{{\rm M}_{\sun}}}

\newcommand{\iso}[2]{\hbox{${}^{#1}{\rm #2}$}}

\title[Solar metallicity AGBs]{Gas and dust from solar metallicity AGB stars}
\author[Ventura et al.]{P. Ventura$^1$, A. Karakas$^2$, F. Dell'Agli$^{3,4}$, 
D. A. Garc\'{\i}a--Hern\'andez$^{3,4}$, 
\newauthor
L. Guzman-Ramirez$^{5}$   \\ 
$^{1}$INAF -- Osservatorio Astronomico di Roma, Via Frascati 33, 00040, Monte Porzio Catone (RM), Italy \\
$^{2}$Monash Centre for Astrophysics (MoCA), School of Physics and Astronomy, Monash University, Victoria 3800, Australia \\
$^{3}$Instituto de Astrof\'{\i}sica de Canarias, E-38205 La Laguna, Tenerife, Spain \\
$^{4}$Departamento de Astrof\'{\i}sica, Universidad de La Laguna (ULL), E-38206 La Laguna, Tenerife, Spain\\
$^{5}$Leiden Observatory, Leiden University, Niels Bohrweg 2, 2333 CA Leiden, The Netherlands \\
}

\begin{document}

\date{Accepted, Received; in original form }

\pagerange{\pageref{firstpage}--\pageref{lastpage}} \pubyear{2012}

\maketitle

\label{firstpage}

\begin{abstract}
We study the asymptotic giant branch (AGB) evolution of stars with masses between
$1~M_{\odot} - 8.5~M_{\odot}$. We focus on stars with a solar chemical
composition, which allows us to interpret evolved stars in the
Galaxy.  We present a detailed comparison with models of the same 
chemistry, calculated with a different evolution code and based 
on a different set of physical assumptions.
We find that stars of mass $\ge 3.5~M_{\odot}$ experience hot bottom burning at the base
of the envelope. They have AGB lifetimes shorter than $\sim 3\times 10^5$ yr and
eject into their surroundings gas contaminated by proton-capture nucleosynthesis, at an
extent sensitive to the treatment of convection.
Low mass stars with $1.5~M_{\odot} \le M \le 3~M_{\odot}$ become carbon stars. 
During the final phases the C/O ratio grows to $\sim 3$. We find a remarkable agreement 
between the two codes for the low-mass models and conclude that predictions for
the physical and chemical properties of these stars, and the AGB lifetime,
are not that sensitive to the modelling of the AGB phase.
The dust produced is also dependent on the mass: low-mass stars produce mainly
solid carbon and silicon carbide dust, whereas higher mass stars produce silicates and 
alumina dust.
Possible future observations potentially able to add more robustness to the present
results are also discussed.
\end{abstract}

140.252.118.146

\section{Introduction}
The recent years have witnessed a growing interest in the evolution of
asymptotic giant branch (AGB) stars. This is because AGB stars play an important
role in various contexts of interest for the astrophysical community. In studies focused 
on Galaxy evolution, AGB yields are crucial for the interpretation
of the chemical trends traced by stars in different parts of the Milky Way \citep{romano10, 
kobayashi11}. Still in the context of the Galaxy, massive AGB stars have been proposed 
as the main actors in the formation of multiple populations in Globular Clusters 
\citep{ventura01, ventura16b}. Moving out of the Galaxy, it is generally believed
that AGB stars provide an important contribution to dust production at
high redshift \citep{valiante09, valiante11}. 

The research focused on AGB evolution has made significant progress in
recent years. This is partly due to the improvement in computer performance, which allows
faster and more exhaustive explorations of the parameter
space. However, stellar evolutionary modelling is still plagued by
major uncertainties in the input physics. It is now generally accepted that
the treatment of convection and the description of mass loss are the two
most relevant phenomena on the determination of the physical evolution of this class of
objects and on the modality with which they contaminate their
surroundings \citep[for recent reviews][]{herwig05,karakas14b}.
 
Some research groups have recently completed models of the AGB
phase with the inclusion of dust formation processes in the wind 
\citep{nanni13a, nanni13b, nanni14, paperI, paperII}. This is a welcome result, given
that the circumstellar envelopes of AGB stars are a favourable environment for the condensation 
of gas molecules into solid particles \citep{gail99}. This approach is
crucial for determining the type and amount of dust produced by AGB stars, and in a broader context,
how they participate in the lifecycle of the Universe.  This research is also
necessary for interpreting the results from infrared (IR) space missions, 
considering that some of the brightest nearby objects in the IR are
mass-losing dusty AGB stars.

To assess the reliability of results from the current generation of AGB
models, we have recently started a research project aimed at understanding how the
interpretation of the observations is sensitive to the numerical and physical
input adopted to compute the evolutionary sequences. This approach, based on a
comparison between results from two codes that are well known to model
AGB stars and their yields, was applied to interpret and characterise
the most obscured stars in the Magellanic Clouds (MC) \citep{ventura15a, ventura16a}.
This choice was motivated by the fact that the research on the AGB phase has been traditionally
focused on the MC instead of the Milky Way, given the largely unknown 
distances to the Galactic sources. The comparison was based on the metallicities
$Z=4,8\times 10^{-3}$, typical of MC stars.

We intend to apply this approach to study other environments, external to 
the MC. This step is of extreme importance if we consider that Gaia and the 
incoming launch of the James Webb Space Telescope (JWST) will definitively provide 
a boost in the research on AGB stars. Gaia will provide the distances to several classes 
of AGB stars in the Milky Way with great precision; this will allow us
to overcome the major difficulty in the study of AGB stars in the
Galaxy, which is their unknown distances. A lack of accurate distances 
prevents an exhaustive and reliable interpretation
of the data. Furthermore, thanks to the JWST, we will soon have a 
considerable amount of IR data on resolved AGB populations in nearby galaxies, 
spanning a range of mean metallicities \citep{jones17}.

In order to be prepared for these upcoming observational challenges, it is
important to fix the critical and most uncertain points in the description of the
AGB evolution and to select the results for which different studies reach
similar conclusions. To this
aim, here we provide a step forward by extending the analysis done by
\citet{ventura15a} and \citet{ventura16a} to stars of solar metallicity.  We
compare the results published by \citet{karakas14a} and \citet{karakas16},
calculated with the MONASH code, with new, updated models of solar metallicity,
calculated with the ATON code. These ATON models have been calculated with the
same metallicity ($Z=0.014$) and the same mixture \citep{asplund09} adopted by
\citet{karakas14a}, to allow a straightforward comparison. The analysis will
be focused on the physical properties of stars of different progenitor mass, on
the chemistry of the gas expelled into the circumstellar environment, and on the
dust produced. The comparison with the recent explorations by \citet{marcella16}
and Dell'Agli et al. (2017, hereinafter D17), based on the solar mixture by
\citet{gs98}, will be used to assess the differences due to the particular solar
mixture adopted.

The paper is structured as follows. The description of the input physics used to
build the evolutionary sequences and to model dust formation are given in
Sections 2 and 3, respectively. In Section 4 we present an overview of the
main physical phenomena affecting the description of the AGB phase, while the
physical and chemical properties of the AGB stars presented here are discussed
in Sections 5 and 6, respectively. Section 7 presents the final chemical
composition, a summary of the observational limitations regarding present
chemical abundances determinations in Galactic AGB stars and their descendants
such as post-AGB stars and planetary nebulae (PNe) as well as some future
observational directions that would be useful to test the theoretical AGB
models. The gas and dust yields are presented in Section 8 and 9,
respectively, while in Section 10 we discuss the metallicity effects on the 
evolutionary properties of AGB stars. Finally, the conclusions are given in Section 11.

\section{Stellar modelling}
The models presented in this work were calculated with the ATON code
\citep{ventura98}. Each model was evolved from the main sequence until the almost total
consumption of the external mantle. The results will be compared with models by
\citet{karakas14a}, calculated with the MONASH version of the Mount Stromlo Stellar 
Structure Program \citep{frost96}. In the following we will refer to the two sets of models
as ATON and MONASH models, respectively. An exhaustive description of
the numerical details of the codes (along with the most recent
updates) can be found in \citet{ventura13} and in
\citet{karakas14a}. Here we proviede a brief summary of the physical input most relevant to 
this work and outline the differences between ATON and MONASH.

\subsection{Initial chemistry}
The models calculated span the mass interval $1M_{\odot} \leq M \leq 8.5M_{\odot}$. The 
metallicity used is $Z=0.014$ and the mixture adopted is taken from \citet{asplund09}. 
The initial helium is $Y=0.265$ in the ATON case, whereas the MONASH models are computed
with $Y=0.28$. This difference in the initial helium has some effects on the extent
of the third dgredge-up (hereinafter TDU), which is more efficient the lower is the
helium in the star.

\subsection{Convection}
In the ATON case the temperature gradient within regions unstable to convection is
found via the Full Spectrum of Turbulence (FST) model \citep{cm91}. The MONASH sequences 
are based on the Mixing Length Theory (MLT), with the mixing length parameter 
$\alpha=1.86$. For the determination of the extension of the mixing region, 
in the ATON case it is assumed that the velocity of convective eddies decay exponentially 
beyond the neutrality point, fixed via the Schwartzschild criterion: the e-folding 
distance of the velocity decays during the core (hydrogen and helium) burning phases and 
during the AGB phase is taken as $0.02H_P$ and $0.002H_P$,
respectively. 

In the MONASH model we apply the algorithm described by
\citet{lattanzio86} in order to search for convective neutrality at
the border between all radiative and convective regions. This method has
been shown to increase the depth of TDU relative to
schemes that apply the Schwartzschild criterion
\citep[e.g.,][]{frost96}. However Kamath et al. (2012) showed that
this algorithm does not provide enough TDU at an small
enough core mass to match the observations of AGB stars in Magellanic
Cloud Clusters. Some overshoot is needed, especially at the lowest
masses to experience it. For this reason, a simple overshoot scheme
is applied to the base of the convective envelope
during the AGB in models of mass $M=1.5, 1.75\Msun$ to allow these
cases to become carbon rich \citep[we refer to][for details]{karakas16}.
No overshoot is however used in models with $M \ge 2\Msun$.

\subsection{Mass loss}
The mass loss rate for oxygen-rich ATON models is determined via the 
\citet{blocker95} treatment; the parameter entering the \citet{blocker95}'s recipe is
set to $\eta=0.02$, following the calibration given in \citet{ventura00}. For carbon stars 
the ATON calculations are based on the description of mass loss from the Berlin group 
\citep{wachter02, wachter08}. In the MONASH case the mass-loss formulation by \citet{vw93} 
is adopted.

\subsection{Opacities} 
In both codes the radiative opacities are calculated according to the OPAL
release, in the version documented by \citet{opal}. The molecular opacities in the 
low-temperature regime ($T < 10^4$ K) are calculated by means of the AESOPUS tool \citep{marigo09}. 
The opacities are suitably constructed to follow the changes in the chemical composition 
of the envelope, particularly of the individual abundances of carbon, nitrogen, and oxygen.

\section{The description of dust formation}
\label{dustin}
The description of dust formation is based on the pioneering formalism
introduced  by the Heidelberg group \citep{fg01, fg02, fg06}. The full set of equations, with an
exhaustive discussion of the role played by the different physical quantities, can be
found in previous papers \citep{paperI, paperII, paperIII}.

The model is based on the assumption that the outflow expands symmetrically from the surface
of the star and that dust formation occurs within the condensation zone, where the
temperatures are sufficiently low that the rate of growth of dust grains overcomes
the rate of vaporisation.

On the mathematical side we consider two independent variables, namely the velocity of
the gas and the optical depth, whose behaviour is described by two differential equations.

The first equation is the expression of momentum conservation: the acceleration of gas
particles is given by the balance between gravity and radiation pressure.

\begin{equation}
v{dv\over{dr}}=-{GM_*\over r^2}(1-\Gamma),
\label{eqvel}
\end{equation}

where $\Gamma$ represents the effects of radiation pressure on dust particles. When $\Gamma$ is
above unity the wind is accelerated. The expression for $\Gamma$ is the following:

\begin{equation}
\Gamma={kL_*\over{4\pi cGM_*}},
\label{eqgamma}
\end{equation}

where $k$, $M_*$ and $L_*$ indicate, respectively, the extinction coefficient, the mass
and the luminosity of the star.

The equation for the optical depth is the following:

\begin{equation}
{d\tau\over{dr}}=-k\rho\left({R_*\over r}\right)^2,
\end{equation}

where $\rho$ is the density of the gas.

The two above equations are completed by the mass conservation equation, for density, and the
relationship giving the radial variation of temperature as a function of the effective 
temperature of the star:

\begin{equation}
\rho={\dot M\over{4\pi r^2 v}},
\label{eqrho}
\end{equation}

\begin{equation}
T^4={1\over 2}T_{eff}^4\left[1-\sqrt{1-\left({R_*\over r}\right)^2}+{3\over 2}\tau \right].
\label{eqt}
\end{equation}

The growth of dust grains is given by the difference between the rate
of the addition of gas molecules on pre-existing solid particles and
the vaporisation rate.  This requires the
introduction of additional differential equations, one for each dust species
considered.

The choice of the dust species is based on the argument of molecular stability.
The most relevant assumption is the stability of the CO
molecule, which absorbs entirely into CO molecules the least abundant between C and O.

In oxygen--rich environments we consider the formation of alumina dust 
($Al_2O_3$), silicates and solid iron. The key elements for the formation of these 
dust species are aluminium, silicon and iron. For carbon stars we follow the formation of 
solid carbon grains, silicon carbide and solid iron; in this case the
key elements are carbon, silicon and iron.

\begin{table*}
\caption{Physical properties of solar metallicity AGB models.}                                       
\begin{tabular}{c c c c c c c c c c}        
\hline\hline                        
$M/ \Msun$  &  $\tau_{MS}$  &  $\tau_{AGB}$  &  $\tau_{TP-AGB}$  & $M_{c}/\Msun^a$ &
$L_{max}/L_{\odot}$  &  $T_{bce}^{max} (MK)$ & $N_{TP}$ & $\lambda$ & $M_f/\Msun$  \\
\hline       
1.00 & 1.0E+10 & 2.92E+07 & 9.12E+05 & 0.515 & 3.80E+03 &   1 & 5  & 0.04 & 0.534 \\
1.25 & 4.4E+09 & 2.73E+07 & 1.63E+06 & 0.523 & 6.53E+03 &   3 & 11 & 0.05 & 0.589 \\
1.50 & 2.6E+09 & 2.67E+07 & 1.85E+06 & 0.523 & 8.41E+03 &   4 & 16 & 0.25 & 0.618 \\
1.75 & 1.7E+09 & 2.79E+07 & 2.22E+06 & 0.527 & 1.00E+04 &   6 & 19 & 0.34 & 0.636 \\
2.00 & 1.2E+09 & 3.27E+07 & 2.74E+06 & 0.516 & 1.05E+04 &   7 & 23 & 0.31 & 0.646 \\
2.25 & 8.4E+08 & 3.15E+07 & 4.25E+06 & 0.488 & 1.23E+04 &  10 & 34 & 0.37 & 0.673 \\
2.50 & 6.3E+08 & 2.42E+07 & 3.52E+06 & 0.500 & 1.32E+04 &  13 & 35 & 0.40 & 0.669 \\
2.75 & 4.8E+08 & 1.76E+07 & 2.75E+06 & 0.526 & 1.43E+04 &  18 & 35 & 0.40 & 0.699 \\
3.00 & 3.9E+08 & 1.25E+07 & 1.91E+06 & 0.565 & 1.53E+04 &  22 & 31 & 0.43 & 0.709 \\
3.50 & 2.6E+08 & 6.00E+06 & 9.16E+05 & 0.670 & 2.34E+04 &  67 & 41 & 0.28 & 0.822 \\
4.00 & 1.7E+08 & 3.10E+06 & 3.15E+05 & 0.793 & 3.16E+04 &  84 & 36 & 0.27 & 0.875 \\
4.50 & 1.4E+08 & 1.98E+06 & 2.26E+05 & 0.834 & 3.89E+04 &  86 & 35 & 0.13 & 0.903 \\
5.00 & 1.1E+08 & 1.35E+06 & 2.11E+05 & 0.864 & 4.68E+04 &  90 & 33 & 0.12 & 0.935 \\
5.50 & 8.5E+07 & 9.31E+05 & 1.31E+05 & 0.899 & 5.50E+04 &  91 & 31 & 0.10 & 0.955 \\
6.00 & 7.0E+07 & 6.64E+05 & 7.78E+04 & 0.938 & 6.46E+04 &  93 & 28 & 0.10 & 0.980 \\
6.50 & 5.9E+07 & 4.76E+05 & 4.86E+04 & 0.986 & 7.76E+04 &  96 & 26 & 0.10 & 1.020 \\
7.00 & 5.0E+07 & 3.47E+05 & 4.32E+04 & 1.045 & 8.91E+04 &  99 & 26 & 0.10 & 1.084 \\
7.50 & 4.4E+07 & 2.88E+05 & 5.30E+04 & 1.110 & 9.33E+04 & 100 & 26 & 0.08 & 1.141 \\
8.00 & 3.8E+07 & 2.26E+05 & 4.19E+04 & 1.230 & 1.07E+05 & 104 & 24 & 0.05 & 1.248 \\
8.50 & 3.4E+07 & 1.78E+05 & 2.80E+04 & 1.310 & 1.29E+05 & 118 & 21 & 0.03 & 1.315 \\
\hline     
\label{tabfis}
\end{tabular}

$^a$ Core mass at the beginning of the TP-AGB phase.

\end{table*}

\section{The key factors affecting AGB evolution modelling}

\begin{figure}
\resizebox{1.\hsize}{!}{\includegraphics{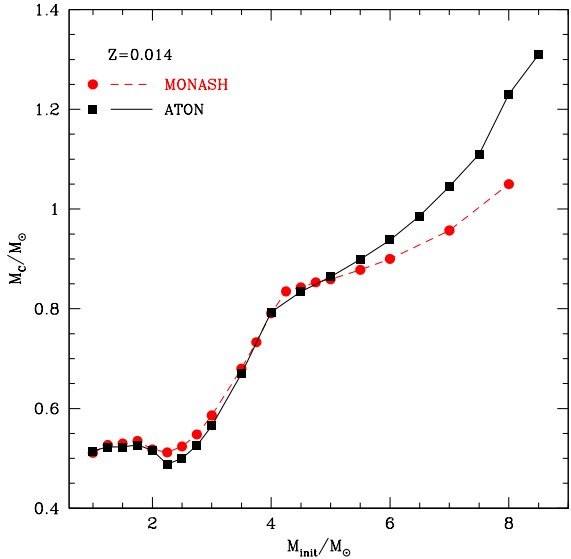}}
\vskip-60pt
\caption{The core mass at the beginning of the TP-AGB phase for the
solar metallicity models presented here.  Full, black squares connected with a solid line,
indicate the results obtained with the ATON code, whereas the red points, connected
by a dashed line, indicate the results published in \citet{karakas14a} and 
\citet{karakas16}, obtained with the MONASH code.}
\label{fmcore}
\end{figure}

Fig.~\ref{fmcore} shows the core mass $M_{\rm c}$ at the beginning of the TP-AGB phase for the models
discussed in the present work. This quantity is reported into col.~5 of Table 1.
We show for comparison the results from \citet{karakas14a} and \citet{karakas16}. 
In the mass domain $M \leq 5~M_{\odot}$ the results are very similar. Conversely,
for $M > 5~M_{\odot}$ the results diverge, with the ATON models reaching the
TP-AGB phase with a more massive core. The largest difference of $\sim 0.2~M_{\odot}$
is reached for $M = 8~M_{\odot}$.

The core-mass threshold for hot bottom burning (HBB) is  $\sim 0.7~M_{\odot}$ in
the ATON code \citep{ventura13}, which is lower than in the MONASH code, where
the threshold is $\gtrsim 0.85\Msun$. The ignition of HBB has an important
effect on the luminosity evolution of the star \citep{renzini81, blocker91}, 
and on the surface chemical composition.

Before entering the general discussion of the properties of AGB stars of solar
metallicity, we present the main features of the evolution of stars undergoing
HBB and their lower mass counterparts. We select the 5$\Msun$ and the $3\Msun$
models from the ATON and MONASH codes as being representative of stars with HBB
and stars that become carbon rich.  As shown in Fig.~\ref{fmcore} the core
masses at the beginning of the TP-AGB phase are very similar in the  ATON and
MONASH models: this will allow us to disentangle the effects of the various
physics input adopted, without taking care of possibile differences arising
from the pre-TP-AGB phase.

\subsection{The evolution of massive AGB stars: the role of HBB}
\label{hmagb}

\begin{figure*}
\begin{minipage}{0.48\textwidth}
\resizebox{1.\hsize}{!}{\includegraphics{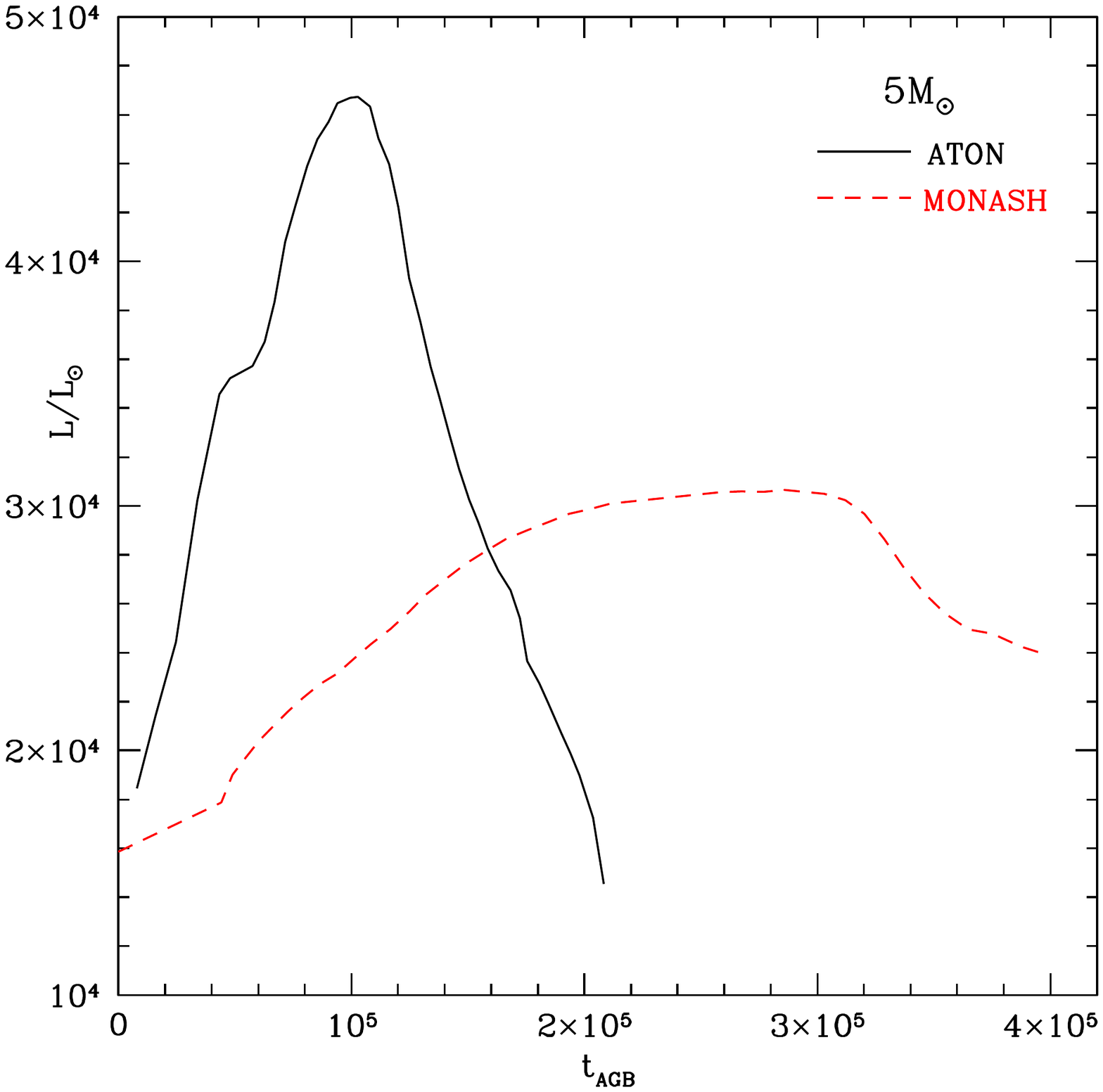}}
\end{minipage}
\begin{minipage}{0.48\textwidth}
\resizebox{1.\hsize}{!}{\includegraphics{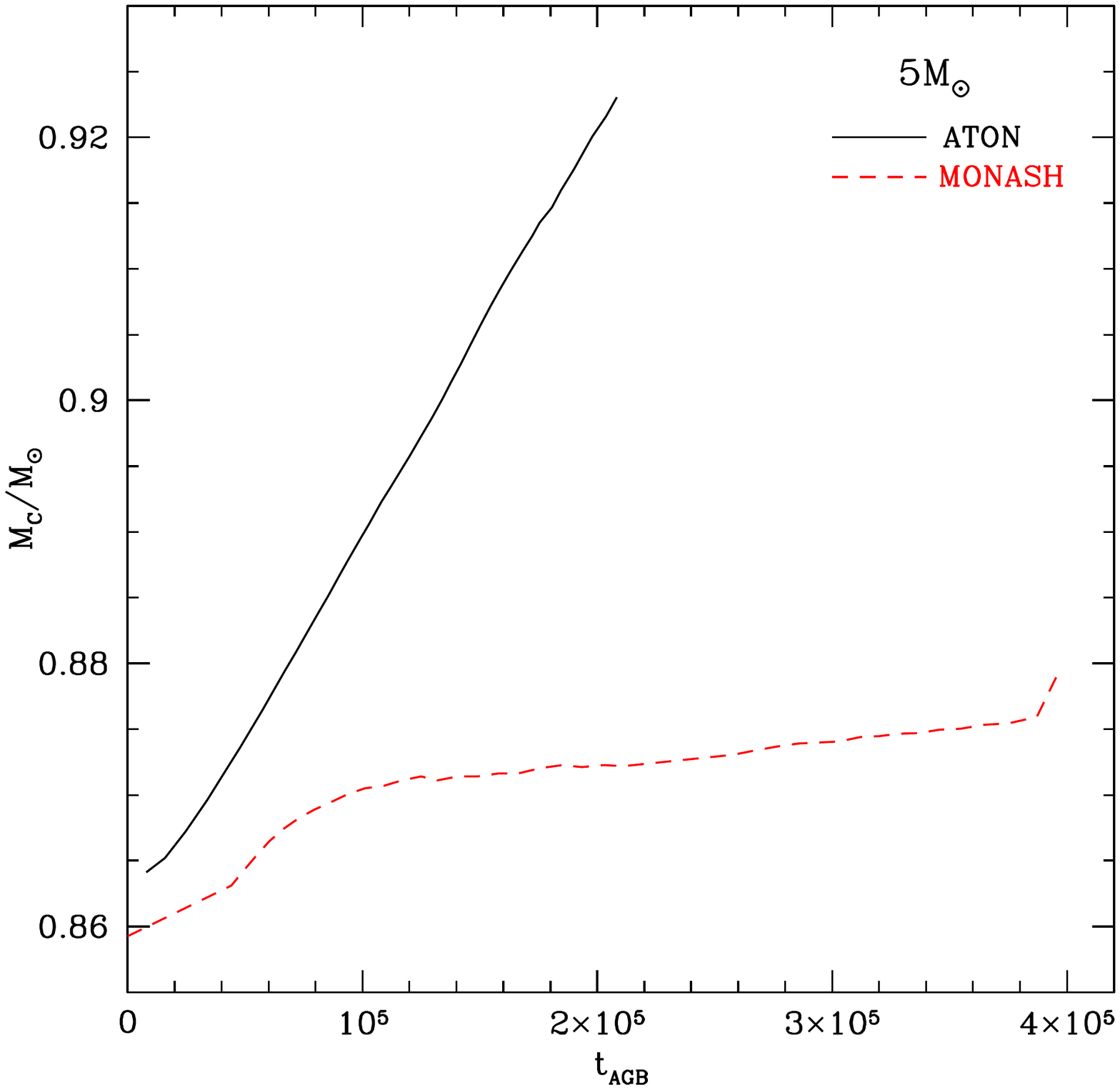}}
\end{minipage}
\vskip-60pt
\caption{The variation with time (counted from the beginning of the TP-AGB phase) of 
luminosity (left panel) and core mass (right) for 5$\Msun$ stars.
The results from ATON are shown with a black, solid line, whereas the
red, dashed line indicates the results from MONASH code.}
\label{f5fis}
\end{figure*}

To understand the main features of the evolution of stars experiencing HBB we show
in Fig.~\ref{f5fis} the variation of the luminosity and of the core
mass for a 5$\Msun$ model, compared with the corresponding model by \citet{karakas14a}. 

We see that 
the maximum luminosity reached and the overall duration of the AGB phase differ. 
The ATON model reaches a luminosity $L_{\rm max} \sim 4.5\times 10^4 L_{\odot}$ significantly 
higher than MONASH ($L_{\rm max} \sim 3\times 10^4 L_{\odot}$). This is
a direct result of the FST model of convection in the ATON case: as shown by \citet{vd05}, FST models 
experience a stronger HBB and evolve at larger luminosities in comparison with 
models calculated with the MLT. 

The difference in the luminosity in turn affects the overall duration of the AGB phase.
Owing to the larger luminosities, the ATON model is exposed to larger rates of mass
loss, thus the envelope is lost faster and the duration of this phase is shorter.
This is clearly shown in both panels of Fig.~\ref{f5fis}, where we see that in the
MONASH model the envelope is lost in $\sim 0.4$Myr, approximately double the evolution
time of the FST model. 

The present results confirm the analysis by \citet{ventura15a}, which outlined the effects
of convection modelling on the luminosity and the duration of AGB models experiencing HBB,
with metallicities typical of LMC stars.

The description of convective zones affects the luminosity and consequently the growth 
rate of the core, $\dot M_{\rm C}$, because the luminosity determines the rapidity with 
which the CNO-burning shell moves outwards (in mass). This is confirmed by the results 
shown in the right panel of Fig.~\ref{f5fis}, indicating that $\dot
M_{\rm C}$ is higher in  the ATON case. A direct consequence of the
higher core-mass growth is the final mass of the star, which is larger
in the ATON case.  We will go back to this point in section 
\ref{agbfis}.

\begin{figure*}
\begin{minipage}{0.48\textwidth}
\resizebox{1.\hsize}{!}{\includegraphics{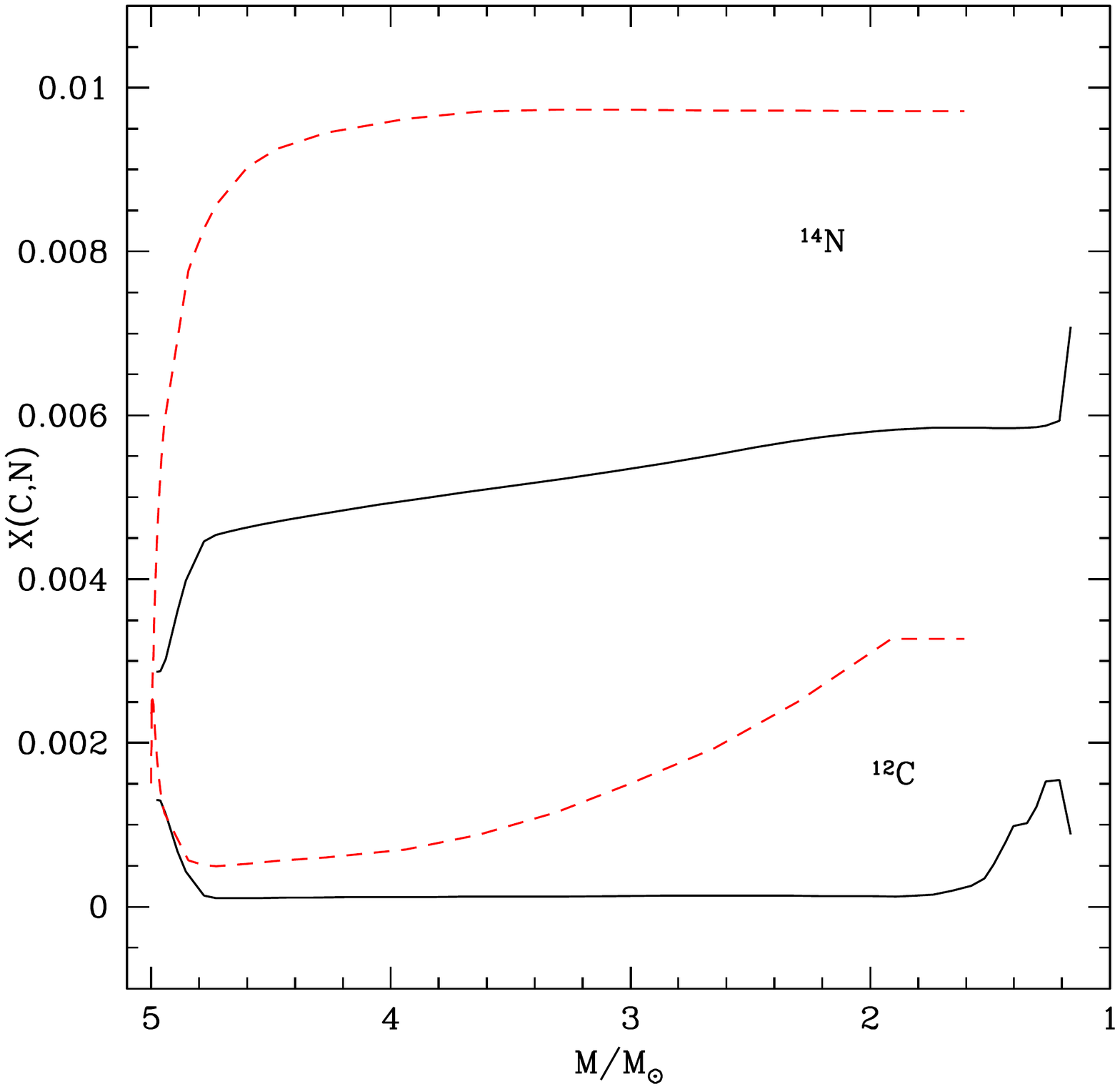}}
\end{minipage}
\begin{minipage}{0.48\textwidth}
\resizebox{1.\hsize}{!}{\includegraphics{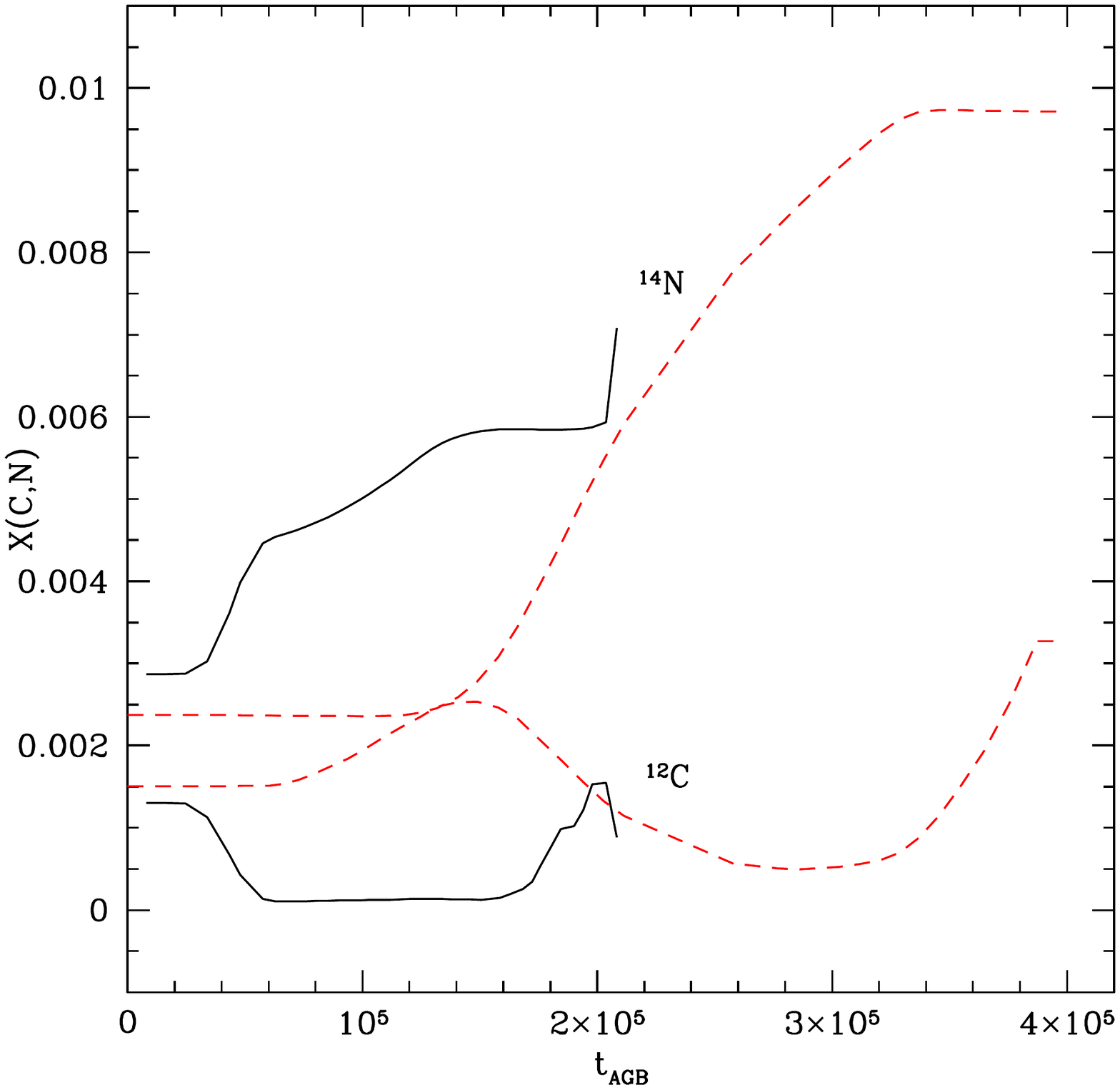}}
\end{minipage}
\vskip-60pt
\caption{The variation of the surface abundances of carbon and nitrogen during the
AGB phase for the same models shown in Fig.~\ref{f5fis} calculated with the ATON 
(black, solid line) and
MONASH  (red, dashed track) codes. In the left panel in the abscissa
we show the current mass of the star, whereas in the left panel we show the time,
counted from the beginning of the TP-AGB phase.}
\label{f5cn}
\end{figure*}

The differences in the evolution of the main physical properties of massive AGB stars
affect the variation of the surface chemical composition. 
Fig.~\ref{f5cn} shows the evolution of the surface mass fraction of carbon and nitrogen.
In the left panel we show the current mass of the star on the abscissa, to have an idea 
of the contamination of the interstellar medium from these objects. Generally speaking, we find
nitrogen production, a clear signature of the activation of HBB. The
main difference we observe is in the behaviour of carbon. This is
because in the ATON models carbon is destroyed from the very first
thermal pusles (hereinafter TP) and is found to be a factor $\sim 20$
lower than the initial value for most of the AGB  phase. Only in the
very final evolutionary stages do we see some carbon transported to
the surface by TDU. It is clear from the left panel of
Fig.~\ref{f5cn} that the ejecta from this star is carbon poor.

When the MLT model for convection is used, the situation is considerably different.
We note that the surface carbon first increases owing to the
action of TDU during the first few TPs before HBB is activated. 
Furthermore, the destruction of carbon is milder during the total AGB
phase, and in the final evolutionary stages the surface carbon
abundance grows to be larger than at the beginning of the AGB phase. 
The net yield is negative where the average
C mass fraction in the wind is $\approx 60$~\% lower than the initial.

\begin{figure}
\resizebox{1.\hsize}{!}{\includegraphics{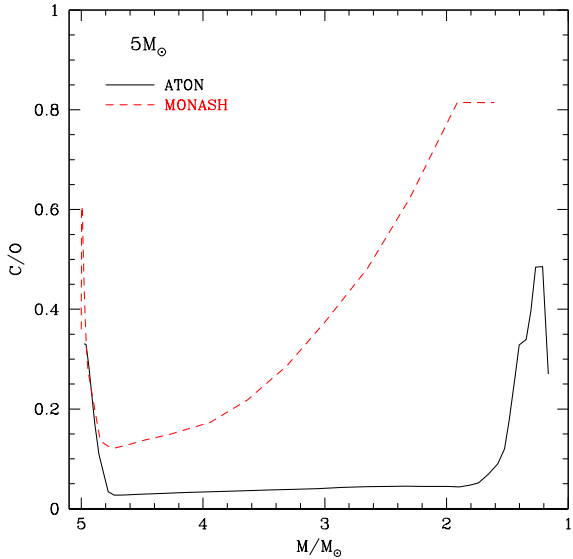}}
\vskip-60pt
\caption{The variation of the surface C/O ratio in the same models shown in Fig.~\ref{f5fis}. 
}
\label{f5co}
\end{figure}

In Fig~\ref{f5co} we show the evolution of the surface C/O ratio,
which we will see is important for a number of issues and is also
deeply affected by convection modelling. In the
ATON case, the strong HBB conditions ensures that model evolves with
C/O $< 0.05$ for most of the AGB phase. In the MONASH model, after an initial phase of 
decrease, the surface C/O gradually increases until reaching C/O $\sim 0.8$ towards the 
end of the evolution.

A general result outlined by these models is the synthesis of nitrogen. In the MONASH
case the quantity of nitrogen synthesized is higher, because of a dominant primary component,
produced by proton captures on carbon nuclei synthesized in the convective shell formed during
each TP and convected to the envelope via TDU. In the ATON case, because
TDU has only modest effects, the secondary component is dominant in this range of mass.

\begin{figure*}
\begin{minipage}{0.48\textwidth}
\resizebox{1.\hsize}{!}{\includegraphics{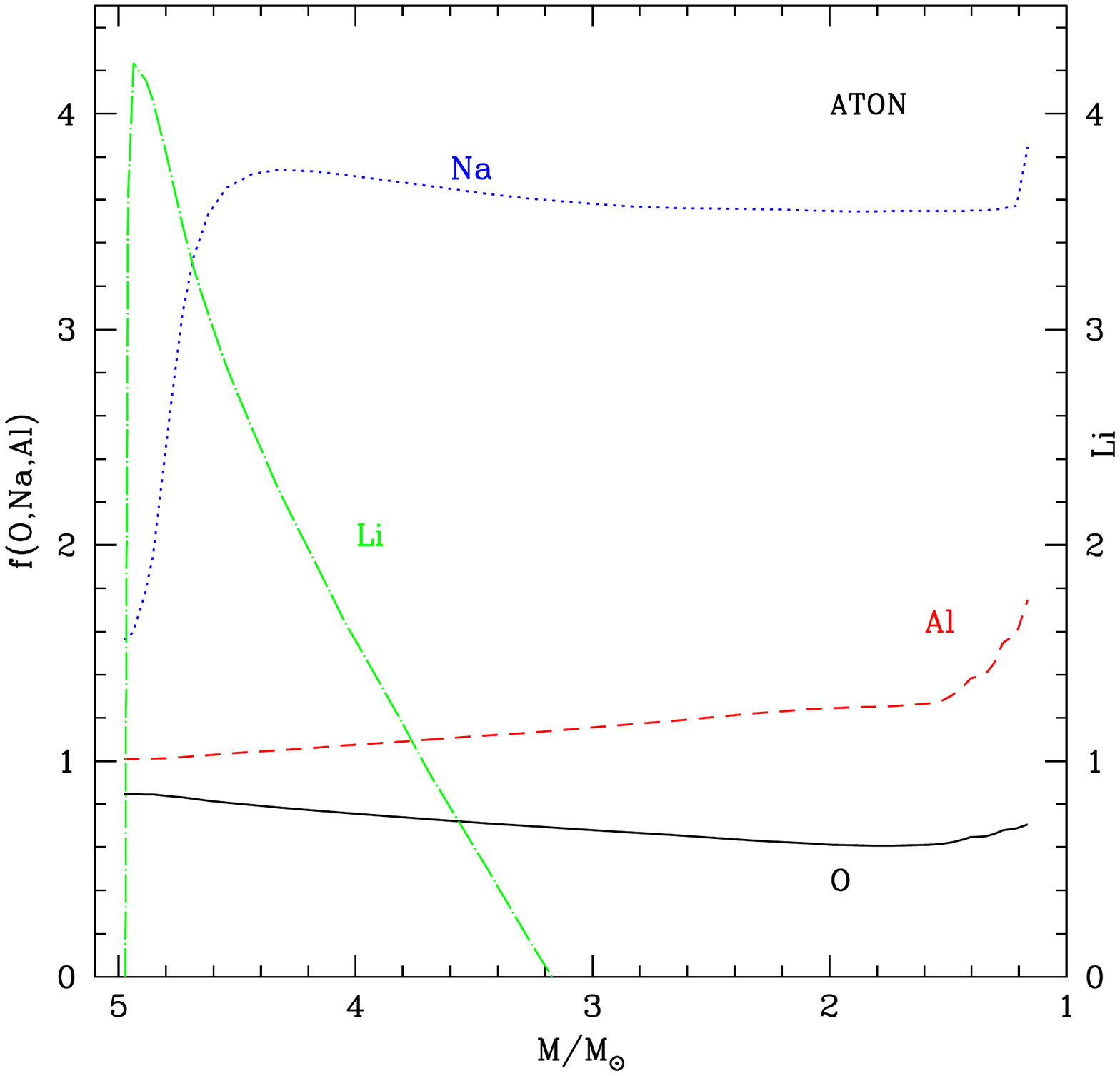}}
\end{minipage}
\begin{minipage}{0.48\textwidth}
\resizebox{1.\hsize}{!}{\includegraphics{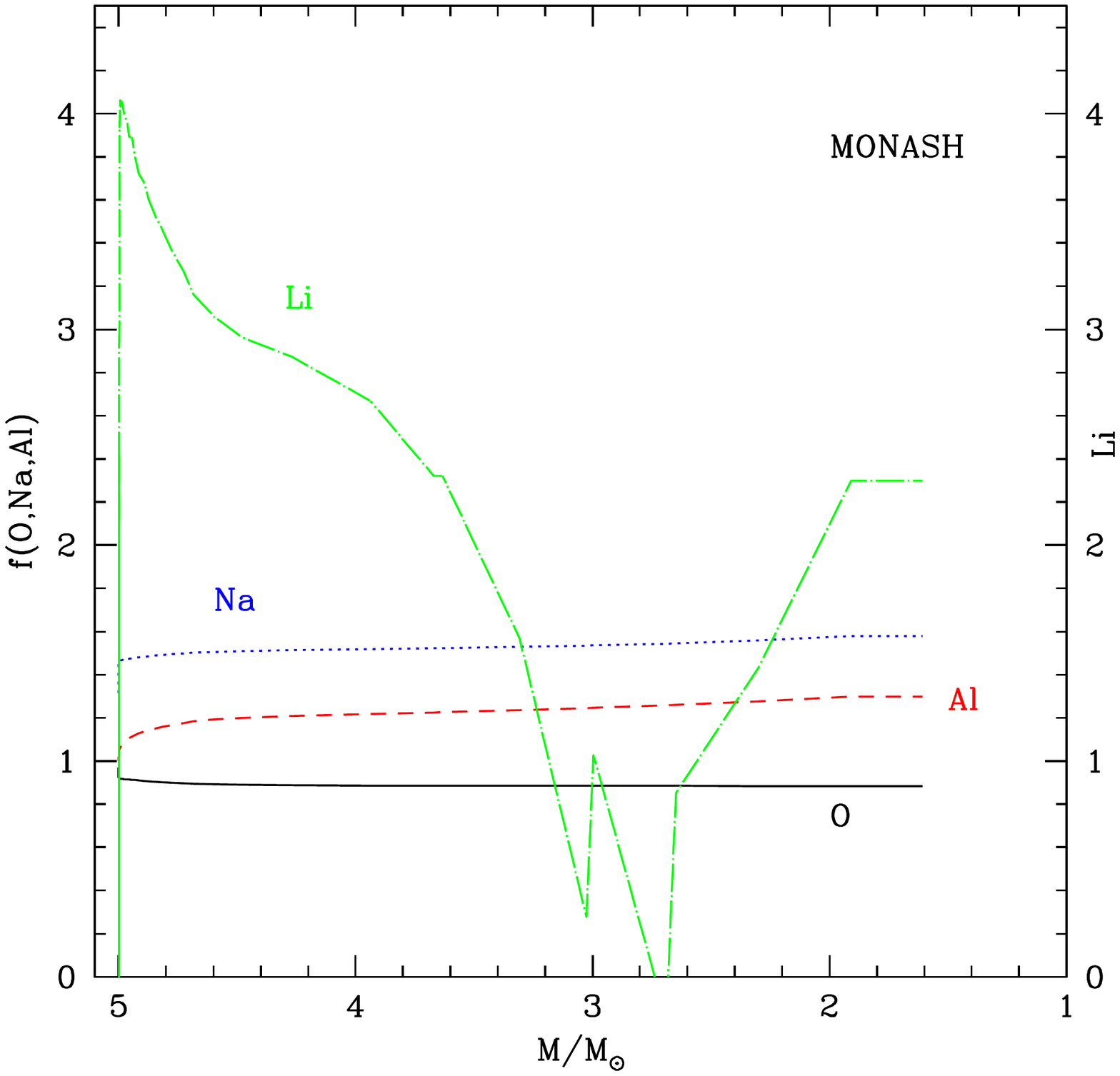}}
\end{minipage}
\vskip-60pt
\caption{The evolution of the production factors of oxygen (black, solid 
line), sodium (blue, dotted), aluminium (red, dashed) during the AGB phase of a
$5~M_{\odot}$  model calculated with the ATON (left panel) and MONASH (right) codes. 
The green, dotted-dashed lines indicates the surface abundance of lithium, in the standard 
units, $\log \epsilon(^7Li)=12+\log (n(^7Li)/n(H))$ (scale on the right).
}
\label{f5hbb}
\end{figure*}

If the temperatures exceed $\sim 80$MK oxygen can be destroyed via
proton captures, while the sodium and aluminium may be produced
\citep{izzard07}. It is generally recognized that this occurs in Pop II, 
massive AGB stars, given the large HBB temperatures experienced
\citep{ventura13, fishlock14}. To check whether this advanced nucleosynthesis occurs 
at solar metallicites, we show in Fig.~\ref{f5hbb} the production 
factors of oxygen, sodium and aluminium\footnote{We define the production factor of a 
given element as the ratio between the surface mass fraction of
that element at a given time and the initial abundance, with which the star formed.}
for the $5~M_{\odot}$ models presented in Fig.~\ref{f5fis}, \ref{f5cn} and \ref{f5co},
calculated with ATON (left) and MONASH (right). We also show, for completeness, the 
evolution of the surface lithium, which will be discussed in more details in section 
\ref{lithium}. 

The depletion of the surface oxygen is higher in the ATON case, owing to the stronger
HBB conditions; however, the overall oxygen destruction is below $\sim 0.2$ dex: the gas 
ejected by these stars, independently of the description of convection used, is characterized 
by only a modest depletion in the oxygen content. For what attains sodium, in the ATON model 
we find a significant production, almost by a factor $\sim 4$, a signature of the activation 
of $^{22}Ne + p$ at the base of the envelope; accordingly, the gas ejected by these stars 
is expected to be sodium-rich; in the MONASH case a much smaller increase in the sodium
content is found. The difference in the behaviour of sodium is due to the combined
effects of convection modelling and of the cross-section adopted; the ATON models have
been calculated by assuming the upper limits given by \citet{hale02} for the $^{22}Ne + p$
reaction rates, wheres the MONASH results are based on the recommended values.
Finally, we see in Fig.~\ref{f5hbb} that only a modest production of aluminium is
expected, consistently with the low efficiency of HBB at solar metallicities.

\begin{figure*}
\begin{minipage}{0.48\textwidth}
\resizebox{1.\hsize}{!}{\includegraphics{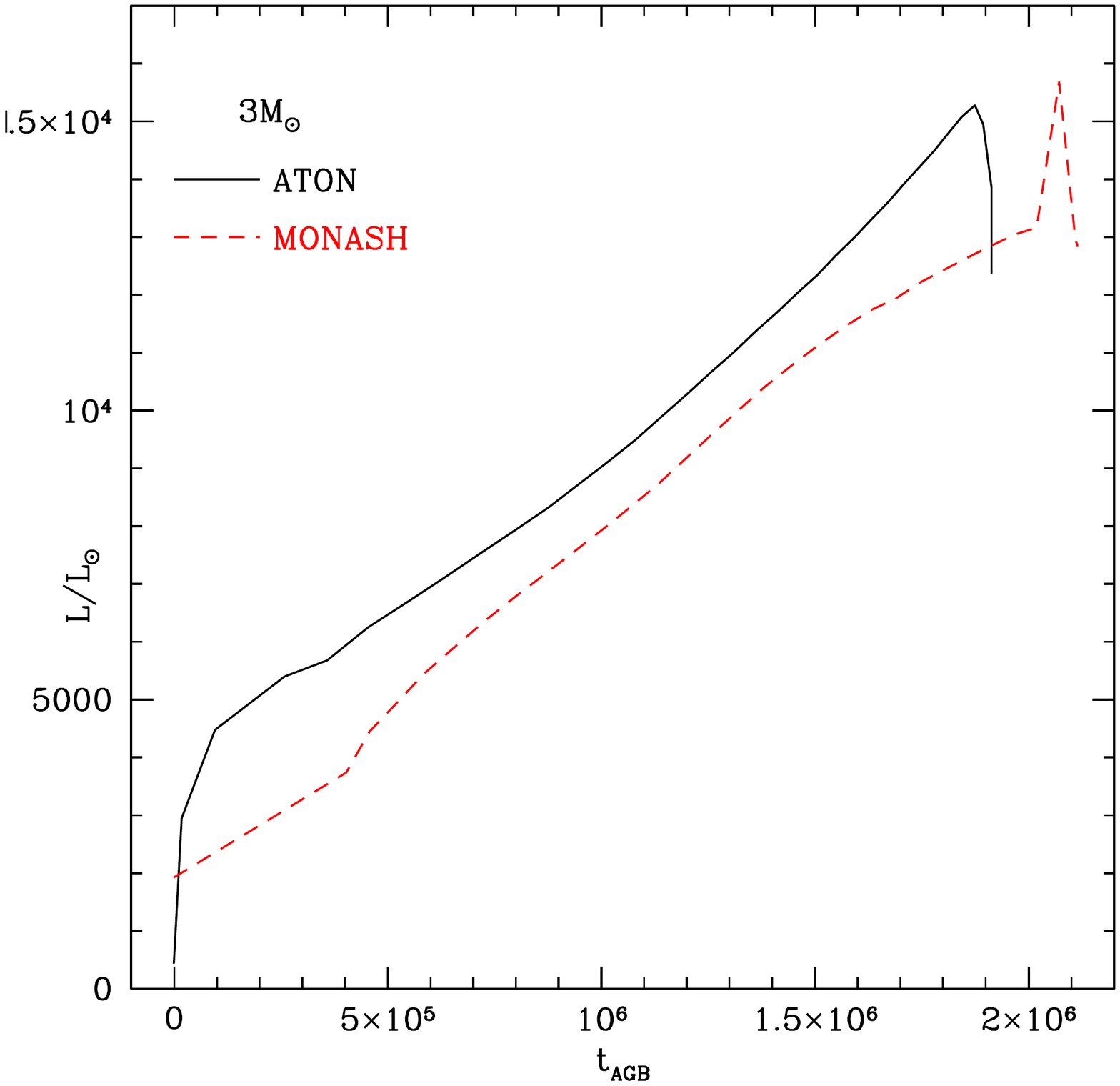}}
\end{minipage}
\begin{minipage}{0.48\textwidth}
\resizebox{1.\hsize}{!}{\includegraphics{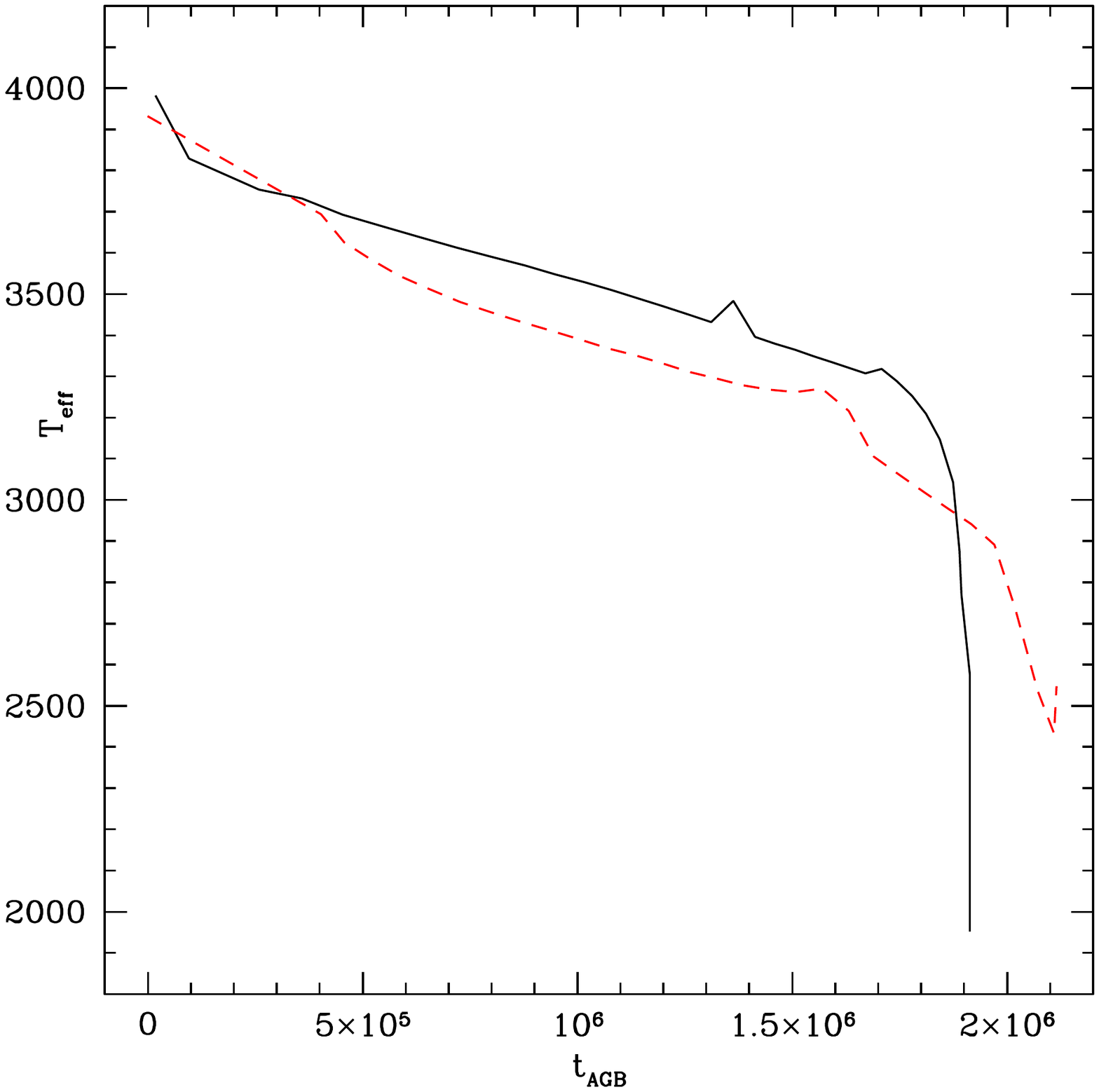}}
\end{minipage}
\vskip-80pt
\begin{minipage}{0.48\textwidth}
\resizebox{1.\hsize}{!}{\includegraphics{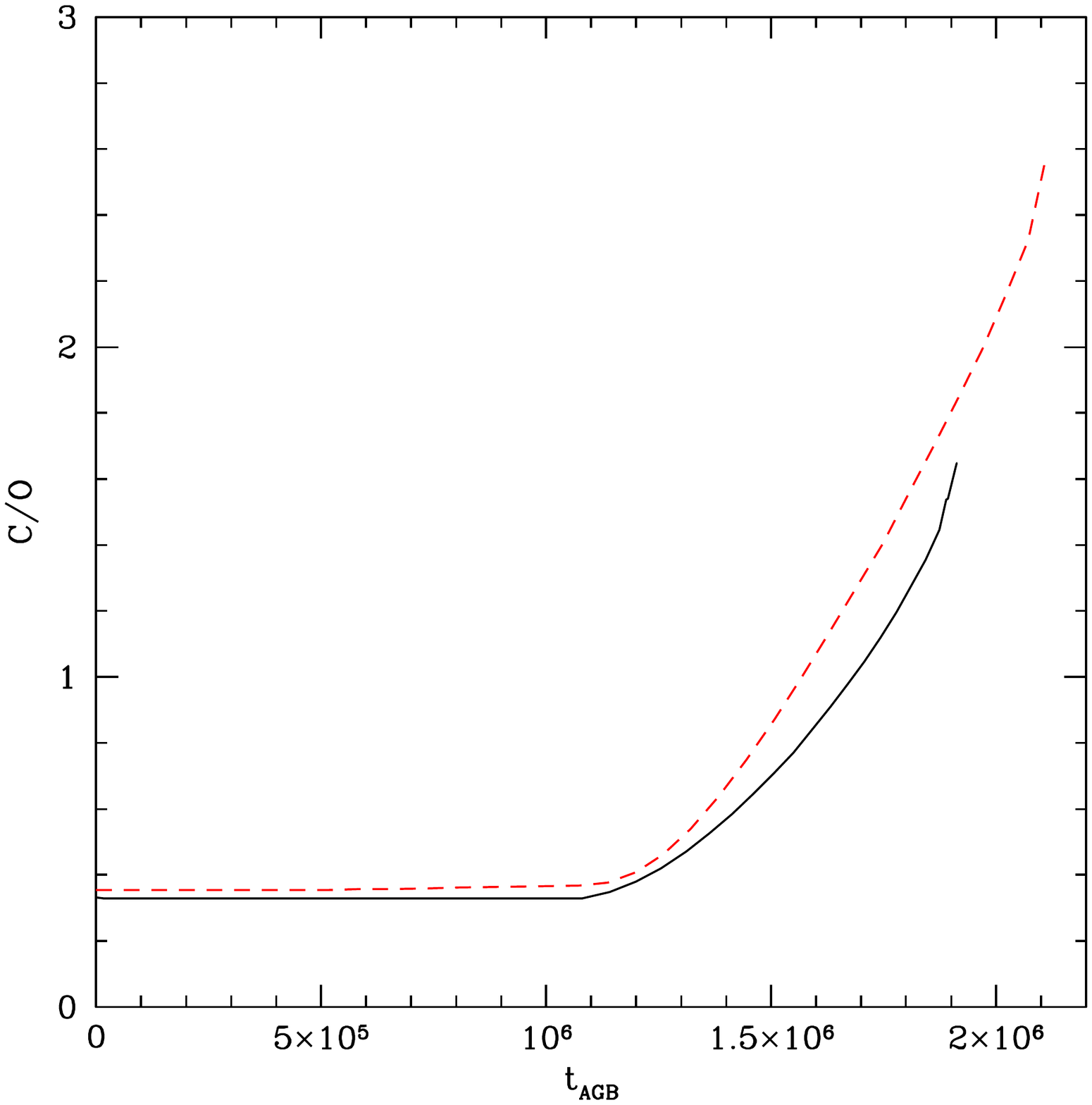}}
\end{minipage}
\begin{minipage}{0.48\textwidth}
\resizebox{1.\hsize}{!}{\includegraphics{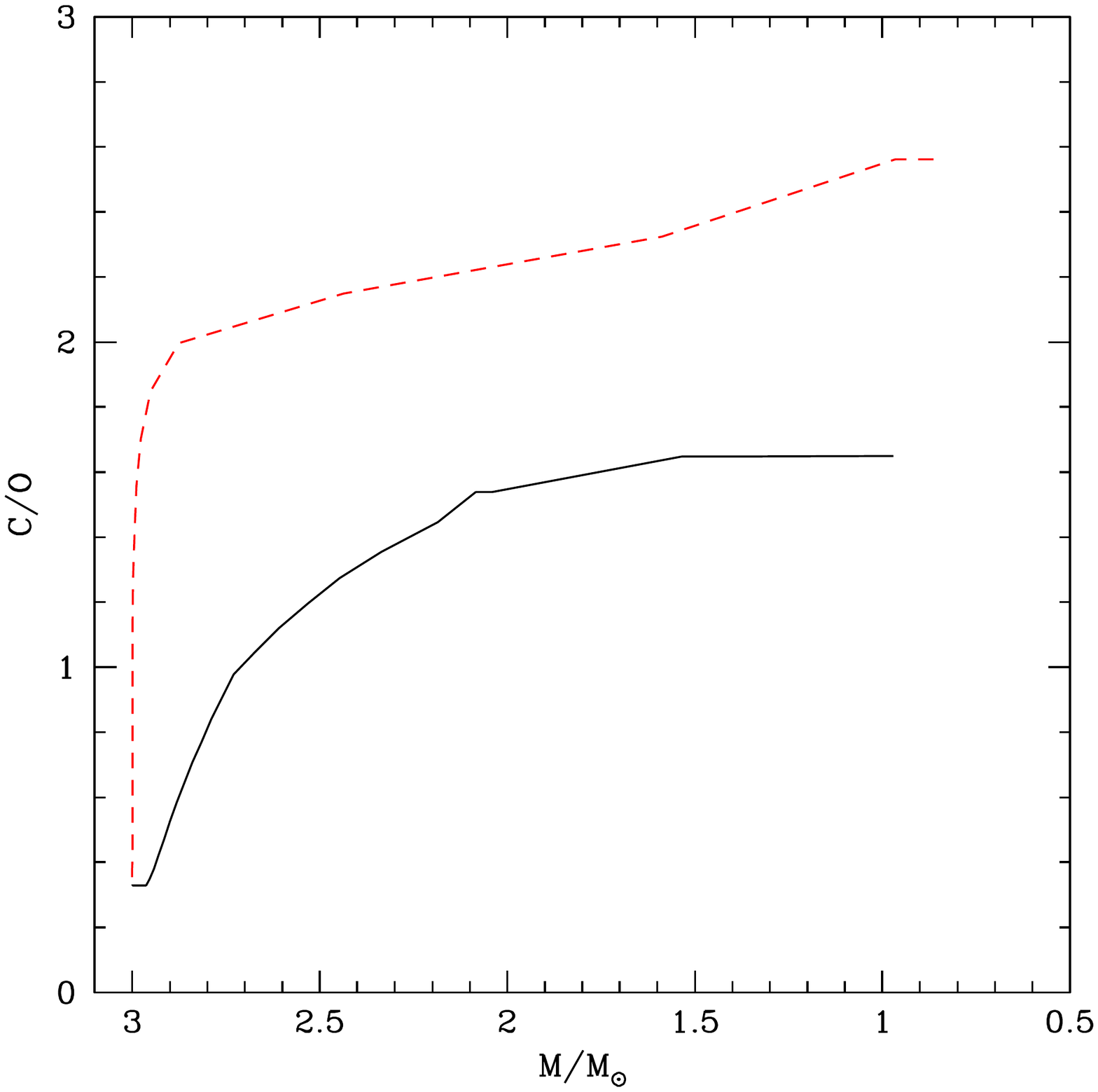}}
\end{minipage}
\vskip-50pt
\caption{The variation with time of the luminosity (left, top panel), effective temperature
(right, top), surface C/O (left, bottom) for two models of initial mass $3~M_{\odot}$ 
calculated with the ATON (black, solid track) and MONASH (red, dotted) codes.
The evolution of the surface C/O is also shown 
as a function of the current mass of the star in
the bottom-right panel.
}
\label{f3msun}
\end{figure*}

\subsection{Low mass stars and the C-star phase}
\label{lmagb}
Stars of mass $\lesssim 4~M_{\odot}$ do not experience HBB and their surface chemistry
is affected only by TDU episodes, which may eventually turn the star into a carbon star.

The top panels of Fig.~\ref{f3msun} shows the evolution of the luminosity and effective 
temperature of the $3~M_{\odot}$ models. The luminosity increases steadily during the AGB
phase, from $\sim 5 \times 10^3 L_{\odot}$ to $\sim 1.5 \times 10^4 L_{\odot}$. At the 
same time the effective temperature decreases as the star expands,
starting from $T_{eff} \sim 4000$ at the beginning of the TP-AGB evolution. 
The cooling of the external regions of the star are particularly important
after the C-star stage is reached: this is caused by
the significant increase in the molecular opacities in carbon-rich gas 
\citep{marigo02, vm09, vm10}.

When comparing the present results with \citet{karakas14a} we note that, unlike more 
massive stars, the luminosities are independent of convective
modelling. This is because no HBB is experienced, which means no contribution from
the internal regions of the envelope to the overall energy release.

The evolution of the effective temperature is illustrated in the
right, top panel of Fig.~\ref{f3msun}. Here we see some similarities
but also important differences between the ATON and MONASH results,
suggesting that the treatment of convection may have some effect here. For $\sim 90\%$ of 
the AGB phase the effective temperatures are rather similar, with $T_{eff}$ decreasing 
from $\sim 4000$K to $T_{eff} \sim 3000$K. 

When the surface C/O overcomes $\sim 1.4$ the ATON 
temperatures become extremely cool, until reaching $T_{eff} \sim 2000$K in the very final 
evolutionary stages. In the model by \citet{karakas14a} the effective temperature is
above $2500$K for the whole AGB evolution.

This dissimilarity is due to the development of a region within the envelope where the
convective efficiency, $\Sigma$ is extremely small\footnote{In the present work we use the 
same definition of the convective efficiency adopted by \citet{cm91}, given in their Eq.~5.}, 
of the order of $\Sigma \sim 0.05$. In these conditions, the ratio between the convective
flux ($\Phi$) found via the FST model and the MLT flux is $\Phi_{FST}/\Phi_{MLT} \sim 0.1$
(see Fig.~5 in Canuto \& Mazzitelli 1991),
which implies that the FST description requires an overadiabaticity peak narrower and
higher than MLT.  Indeed we find in the FST case $\nabla - \nabla_{ad} \sim 10$, whereas 
in the MLT case we would find $\nabla - \nabla_{ad} \sim 1$ for the 
same physical conditions. This dissimilarity in the overadiabaticity peak is
the reason for the difference in the effective temperatures.
This is the first time within the context of AGB modelling that we encounter a situation 
where the treatment of convection has an impact on the temperature gradient within the
outermost regions of the star.

The smaller effective temperatures favour larger rates of mass loss, thus shorter time 
scales, independently of the mass loss description. 
While these differences are within $\sim 10\%$, we will see that this
will have an important impact on the production of dust by these stars.

A general result found here is that the carbon-star phase is shorter than the oxygen-rich
phase, accounting for only $\sim 15 \%$ of the total AGB evolution. For this
reason the chance of detecting these stars during the initial O-rich
phase is higher. On the other hand, the gas ejected by these stars is carbon rich.
This can be understood by looking at the right, bottom panel of Fig.~\ref{f3msun},
which shows the evolution of the surface C/O as a function of the (current) mass of the
star. We see that most of the mass expelled is carbon-rich, which
therefore means that the yields will also be similarly carbon rich
\citep[e.g.,][]{cristallo15, karakas16}.  This is due to the fact that most of mass loss occurs after the
carbon-star stage is reached. In the ATON case the C/O reached is smaller compared
to MONASH, because the fast mass loss occurring in the final AGB phase prevents 
additional TDU events.
 
\begin{figure}
\resizebox{1.\hsize}{!}{\includegraphics{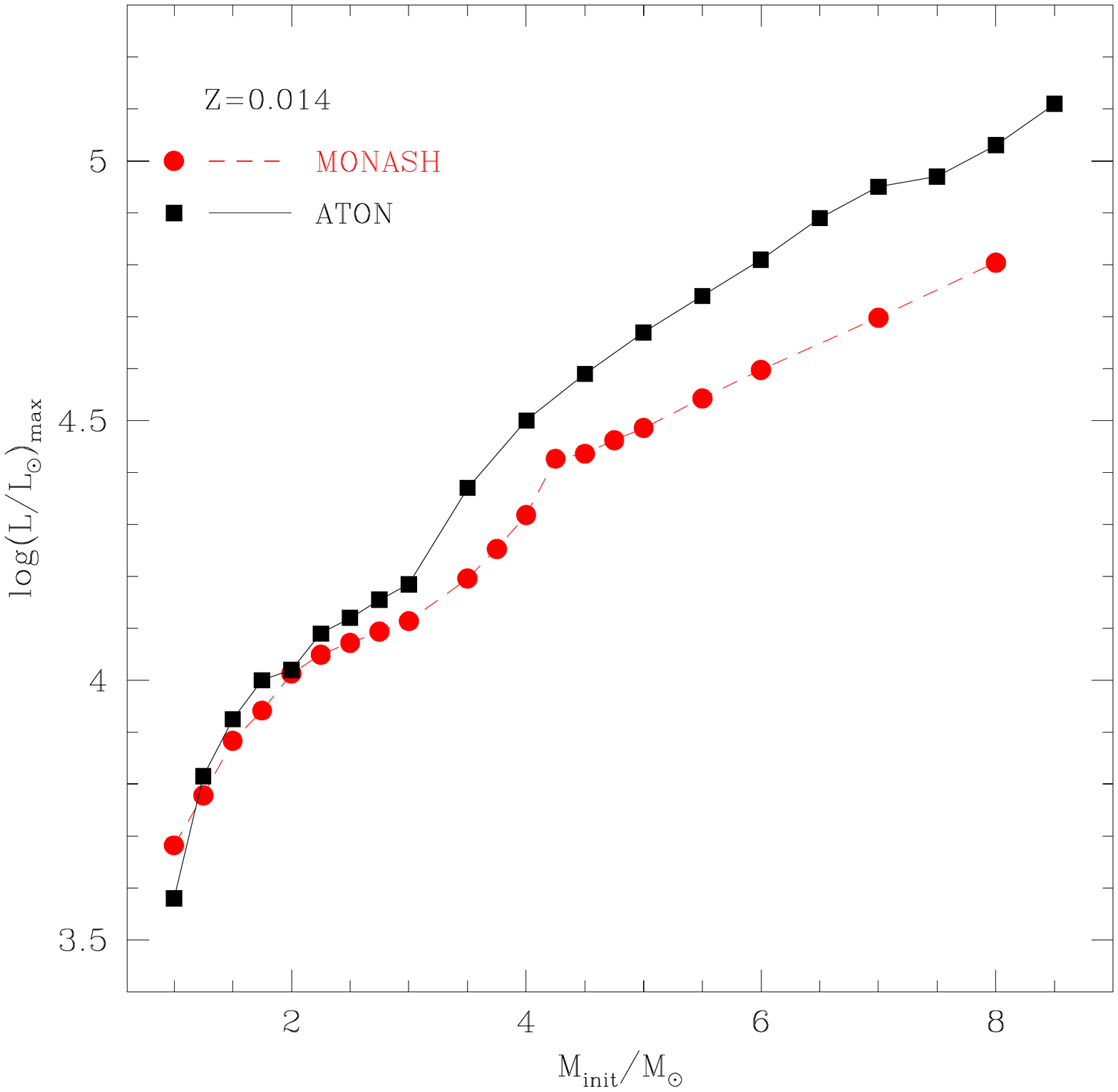}}
\vskip-60pt
\caption{The maximum luminosity reached during the TP-AGB phase by stars of 
different mass is shown with black squares, connected by a solid line. For comparison
we also show the results by \citet{karakas14a}, indicated with red circles, connected
with a dashed line.
}
\label{flum}
\end{figure}

\section{Physical properties of AGB stars}
\label{agbfis}
Table~\ref{tabfis} reports important physical quantities of the AGB
models presented here,  which includes the duration of the main sequence, AGB and TP-AGB phases, the core mass at the
beginning of the AGB (we have discussed this quantity when analyzing Fig.~\ref{fmcore}), the
maximum luminosity, the maximum temperature at the base of the envelope, the number of
thermal pulses experienced, the maximum TDU parameter, $\lambda$, and the final
mass of the star.

\subsection{The brightness of AGB stars}
Fig.~\ref{flum} shows the maximum luminosity ($L_{max}$) reached during the TP-AGB evolution 
as a function of the initial mass ($M_{init}$). Stars not experiencing  
HBB evolve at luminosities below $1.5\times 10^4 L_{\odot}$. The sudden change in the
slope of the $L_{max}$ vs $M_{init}$ relationship occurring at $\sim 3.5~M_{\odot}$ is
because stars experiencing HBB deviate from 
\citet{paczynski}'s core mass - luminosity law \citep{blocker91}: in this mass domain  
we find $2\times 10^4L_{\odot} < L < 1.2\times 10^5 L_{\odot}$.

\begin{figure}
\resizebox{1.\hsize}{!}{\includegraphics{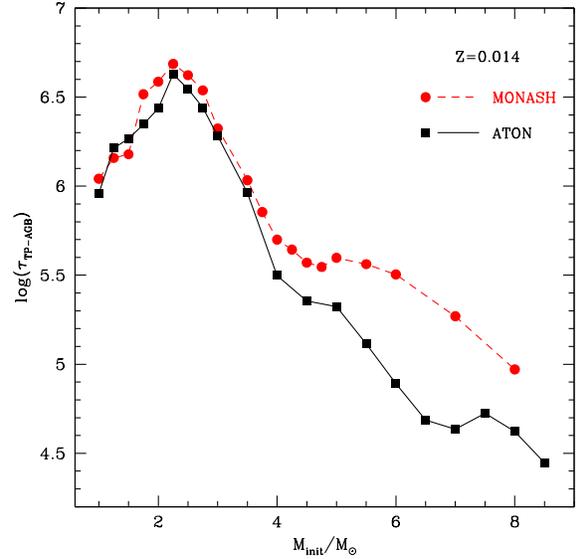}}
\vskip-60pt
\caption{The duration of the TP-AGB phase of the AGB models presented here. The
meaning of the symbols is the same as Fig.~\ref{flum}.
}
\label{ftau}
\end{figure}

\subsection{The evolutionary time scale}
The duration of the TP-AGB phase, $\tau_{TP-AGB}$, is shown in Fig.~\ref{ftau}.
Low-mass AGB stars evolve on time scales above $\sim 1$ Myr. In this mass range the 
evolutionary time scale is determined by two factors, which have opposite effects
on $\tau_{TP-AGB}$. The mass of the envelope (higher masses
require longer times to be lost) and the luminosity (which, as shown in Fig,~\ref{flum}, 
increases with the mass of the star). This is the reason why the trend 
with mass is not monotonic. The stars with the longest TP-AGB evolution, of the order
of $\sim 5$ Myr, are those with $M_{init} \sim 2~M_{\odot}$.

For stars experiencing HBB the time scale of the TP-AGB evolution is determined 
essentially by the luminosity. $\tau_{TP-AGB}$ decreases with $M_{init}$,
because higher mass models have larger luminosities.
The $8.5~M_{\odot}$ is the fastest evolving model, with a TP-AGB
duration of only $\sim 3\times 10^4$ yr.

\begin{figure}
\resizebox{1.\hsize}{!}{\includegraphics{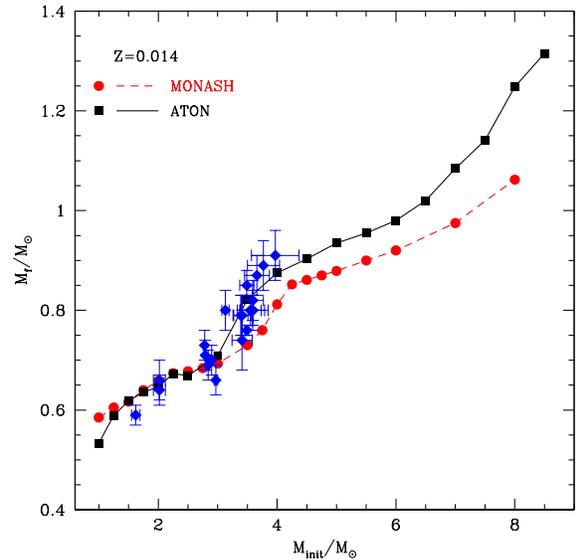}}
\vskip-60pt
\caption{The initial - final mass relationship for the AGB models presented 
in Fig. \ref{flum} and \ref{ftau}. Blue diamonds indicate data from open clusters
White Dwarfs by \citet{kalirai14}.
}
\label{fmfin}
\end{figure}

\subsection{The initial-final mass relationship}
Fig.~\ref{fmfin} shows the initial-final mass relationship.
The mass of the envelope and the luminosity affect the final mass of the star, hence
the mass of the remnant. The pre-AGB evolution is also important for the final mass of the 
star, because the initial mass of the core at the beginning of the TP-AGB phase depends 
on $M_{init}$, as shown in Fig.~\ref{fmcore}.

Stars of initial mass $M_{init} \leq 3~M_{\odot}$ develop core 
masses in the range $0.55~M_{\odot} - 0.7~M_{\odot}$; the final mass increases with
$M_{init}$, for almost the whole range of masses involved. 
Turning to the stars experiencing HBB the results shown in Fig.~\ref{fmfin} outline
a sudden rise in the final core mass, which
increases from $0.7~M_{\odot}$ (for $3~M_{\odot}$ stars) to $0.85~M_{\odot}$ 
($3.5~M_{\odot}$ stars). For the stars in the range $3.5~M_{\odot} < M_{init} < 8.5~M_{\odot}$,
the final core mass increases monotonically, from $0.85~M_{\odot}$ to 
$\sim 1.3~M_{\odot}$.

In Fig.~\ref{fmfin} we show the results from \citet{kalirai14},
where the authors report the analysis of White Dwarfs in the clusters Hyades,
Praesepe, NGC 6819 and NGC 7789. From their analysis, an initial-final
mass relation was determined for  low and intermediate mass stars in
the initial mass range $M_{init} \leq 4~M_{\odot}$. The comparison
with the results from the current investigation shows
a satisfactory agreement in the range of initial masses covered by the
observations.

\subsection{Common findings and differences in AGB modelling}
To assess how the results presented here depend on the numerical
details with which the AGB phase is modelled, in Fig.~\ref{flum},
\ref{ftau} and \ref{fmfin} we compare the present findings with those 
published in \citet{karakas14a}.

In the large mass domain the luminosities reached by ATON models are generally higher
than MONASH (see discussion in Section~\ref{hmagb} and
Fig.~\ref{f5fis}).  The differences, as shown in Fig.~\ref{flum}, increase with the mass
of the star. For an $8~M_{\odot}$ model the luminosity is $\sim 50\%$ larger than in \citet{karakas14a}. 
For stars of mass $3.5~M_{\odot} \leq M_{init} \leq 5~M_{\odot}$ the main actor is convection 
modelling, which affects the strength of the HBB experienced and thus
the overall luminosity. Note that we do not need to consider the
pre-AGB here because the core masses at the first TP are very similar  
(see Fig.~\ref{fmcore}). For higher mass stars the gap between the ATON and 
the MONASH luminosities is
determined by the combined effects of convection modelling and the treatment of core
overshoot during the MS phase. Overshoot during the main sequence means
that the ATON models start the TP-AGB phase with
larger core masses, which can be seen in Fig.~\ref{fmcore}. 

The differences in the luminosity reflect into the duration of the whole evolutionary
phase. As shown in Fig.~\ref{ftau}, the AGB evolutionary times of $M_{init} \geq 4~M_{\odot}$
stars are shorter in the ATON case. For the same reasons given above, the difference
increases with the mass of the star, reaching a factor $\sim 3$ for the
most massive AGB stars. 

In the low mass domain the luminosities are very similar between the
ATON and MONASH models because these stars do not experience HBB. The same holds
for the duration of the entire AGB phase, reported in Fig.~\ref{ftau}. 

For the initial-final mass relationship (see Fig.~\ref{fmfin}), we
find once again similar results for models that do not have HBB. For
the stars experiencing HBB we find that the ATON models 
develop more massive remnants compared to the MONASH case. As for the luminosity, we may
associate these differences due the larger growth rate of the core mass of the ATON models 
(see the right panel of Fig.~\ref{f5fis}) and, for stars of mass $ M_{init} > 6~M_{\odot}$, 
to the difference in the core mass between ATON and MONASH models, present at the 
beginning of the TP-AGB phase.

\begin{figure*}
\begin{minipage}{0.48\textwidth}
\resizebox{1.\hsize}{!}{\includegraphics{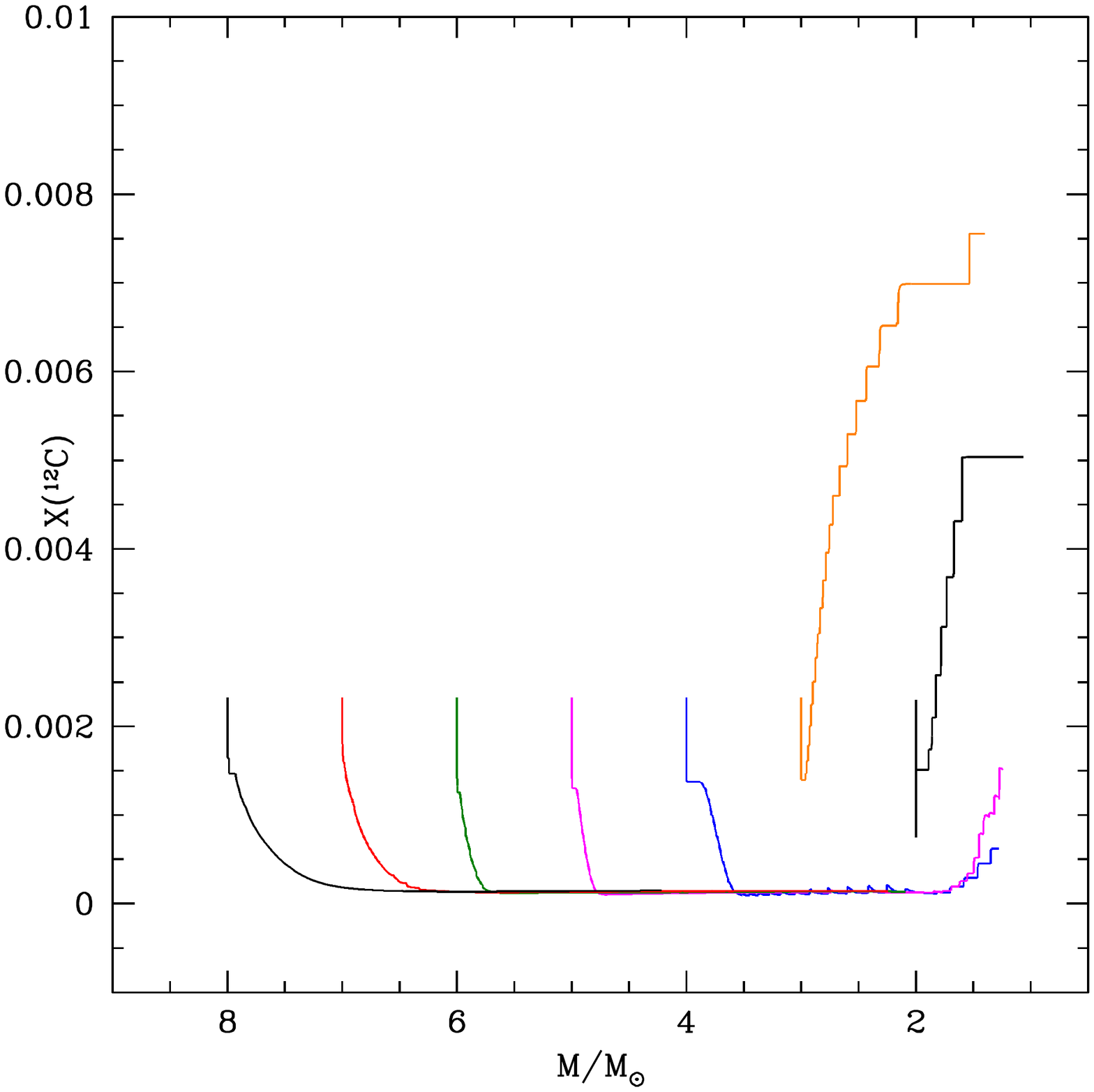}}
\end{minipage}
\begin{minipage}{0.48\textwidth}
\resizebox{1.\hsize}{!}{\includegraphics{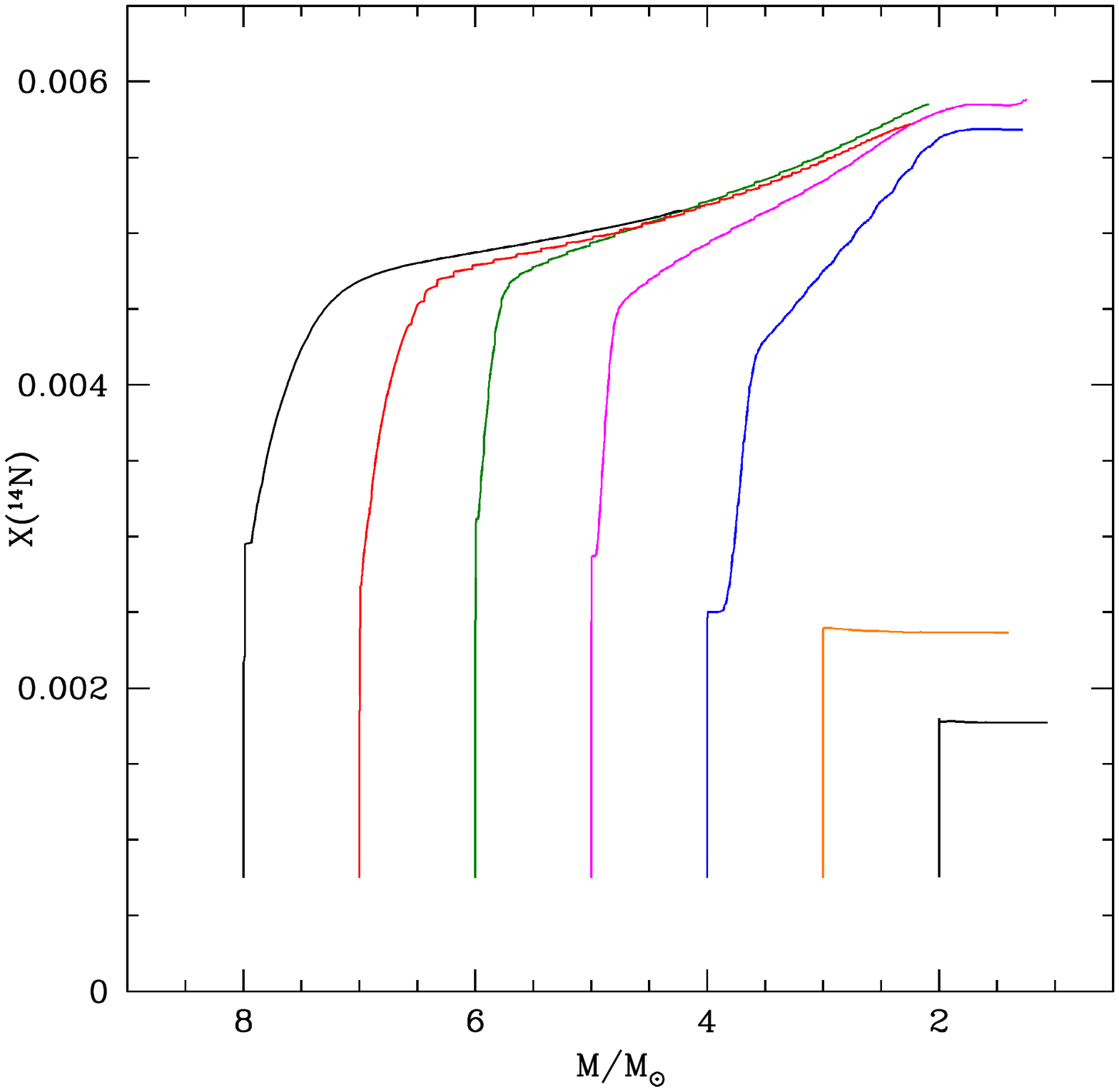}}
\end{minipage}
\vskip-80pt
\begin{minipage}{0.48\textwidth}
\resizebox{1.\hsize}{!}{\includegraphics{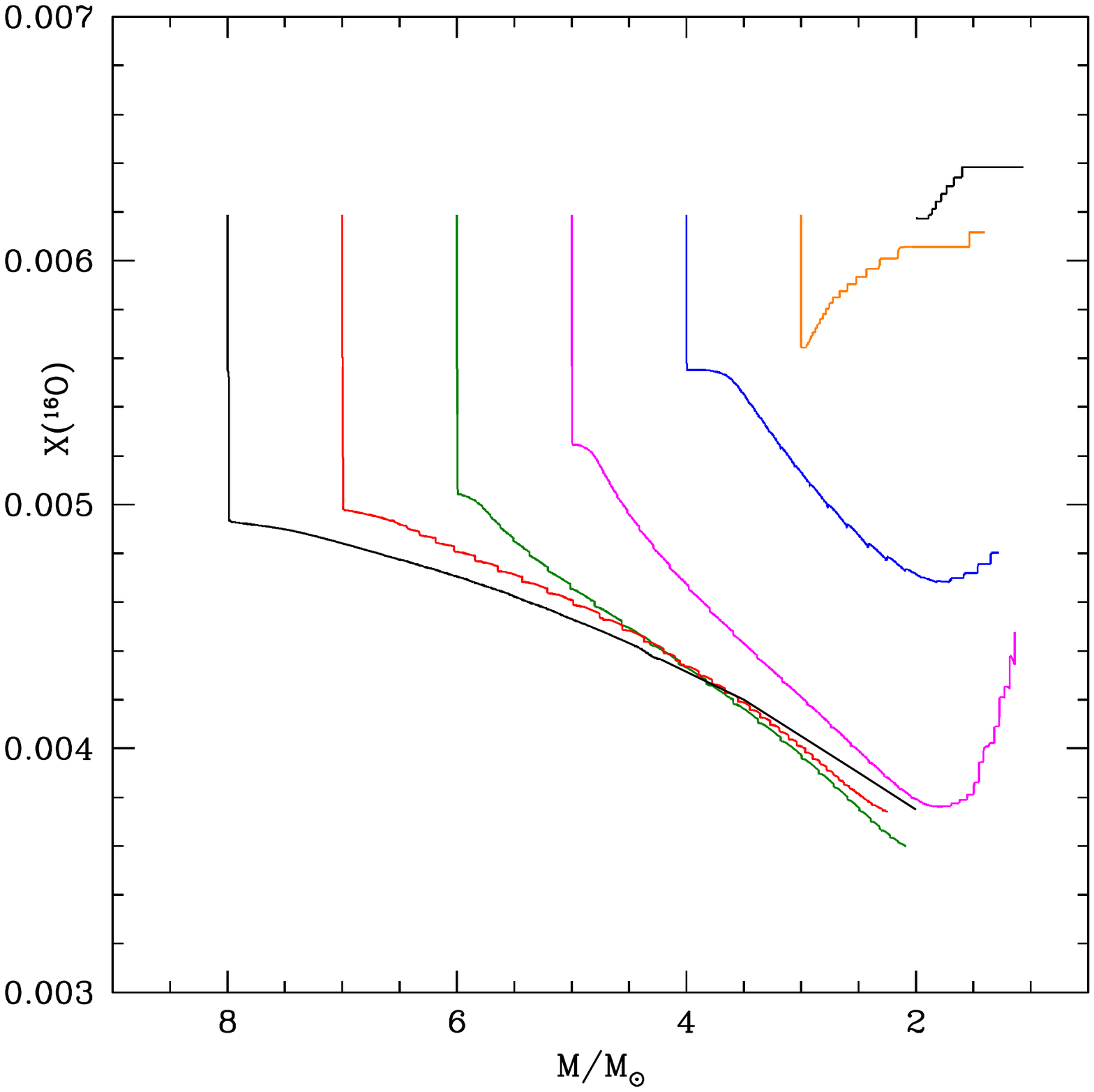}}
\end{minipage}
\begin{minipage}{0.48\textwidth}
\resizebox{1.\hsize}{!}{\includegraphics{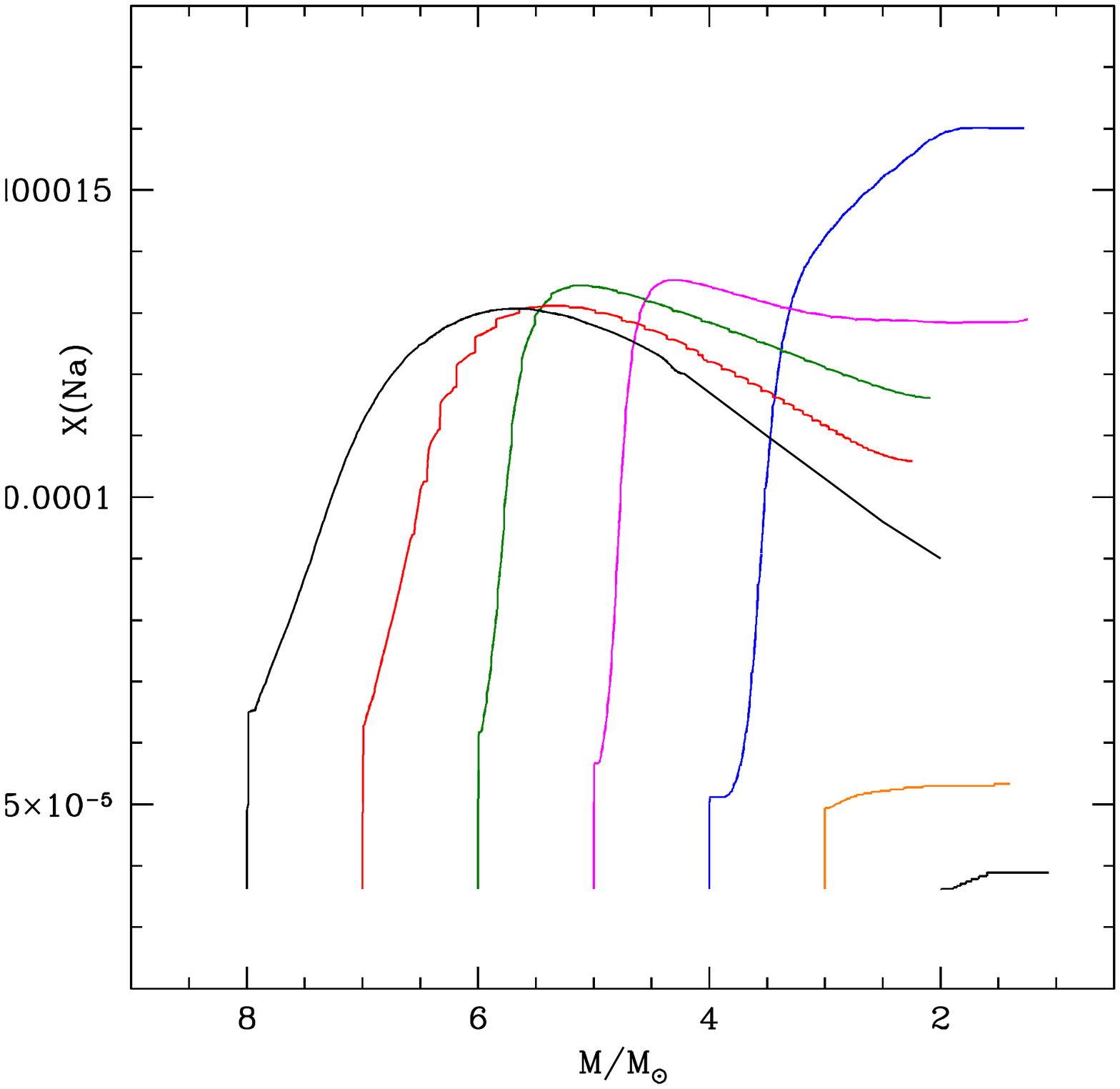}}
\end{minipage}
\vskip-50pt
\caption{The variation of the surface mass fraction of the CNO elements and of
sodium in AGB models of different initial mass. We report the current mass of the
star on the abscissa to deduce the chemical composition of the ejecta. The starting
point of each line marks the initial mass of the star.
}
\label{fchim}
\end{figure*}

\section{The chemical composition of AGB stars}
\label{chimagb}
We focus now on the surface chemical composition, which is crucial to assess the
role played by this class of objects in the pollution of the interstellar 
medium. Understanding how the surface chemistry changes as these stars evolve is
also important to establish which kind of dust particles form in their wind
\citep{fg06}. The latter will be discussed in more detail in Section
\ref{agbdust}, while some observational facts are discussed in Section \ref{pne}
(with the exception of Li, which is already discussed at the end of Section
\ref{lithium}).

Fig.~\ref{fchim} shows the evolution of the surface abundances of carbon, nitrogen,
oxygen, and sodium. For the CNO elements we refer to the most abundant isotopes, namely
$^{12}C$, $^{14}N$ and $^{16}O$. The behaviour of carbon highlights the different 
evolution of $M_{init} \leq 3~M_{\odot}$ models from their higher-mass counterparts.

\subsection{Third dredge-up events: the formation of carbon stars}
Low-mass stars may undergo several TDU episodes, which increases the
surface abundance of carbon. The maximum carbon abundance increases
with increasing mass, up to 3$\Msun$.   The overall increase in the 
surface carbon is a factor $\sim 2$, for $M=1.5~M_{\odot}$ models and up to a factor $\sim 4$, 
for models of $M=3~M_{\odot}$. The gas ejected by these stars is also enriched in nitrogen
because of the first dredge-up (FDU); this can be seen in the steep rise of the surface 
nitrogen in the lines corresponding to
$2~M_{\odot}$ and $3~M_{\odot}$ stars, in the right, top panel of Fig.~\ref{fchim}.

\begin{figure}
\resizebox{1.\hsize}{!}{\includegraphics{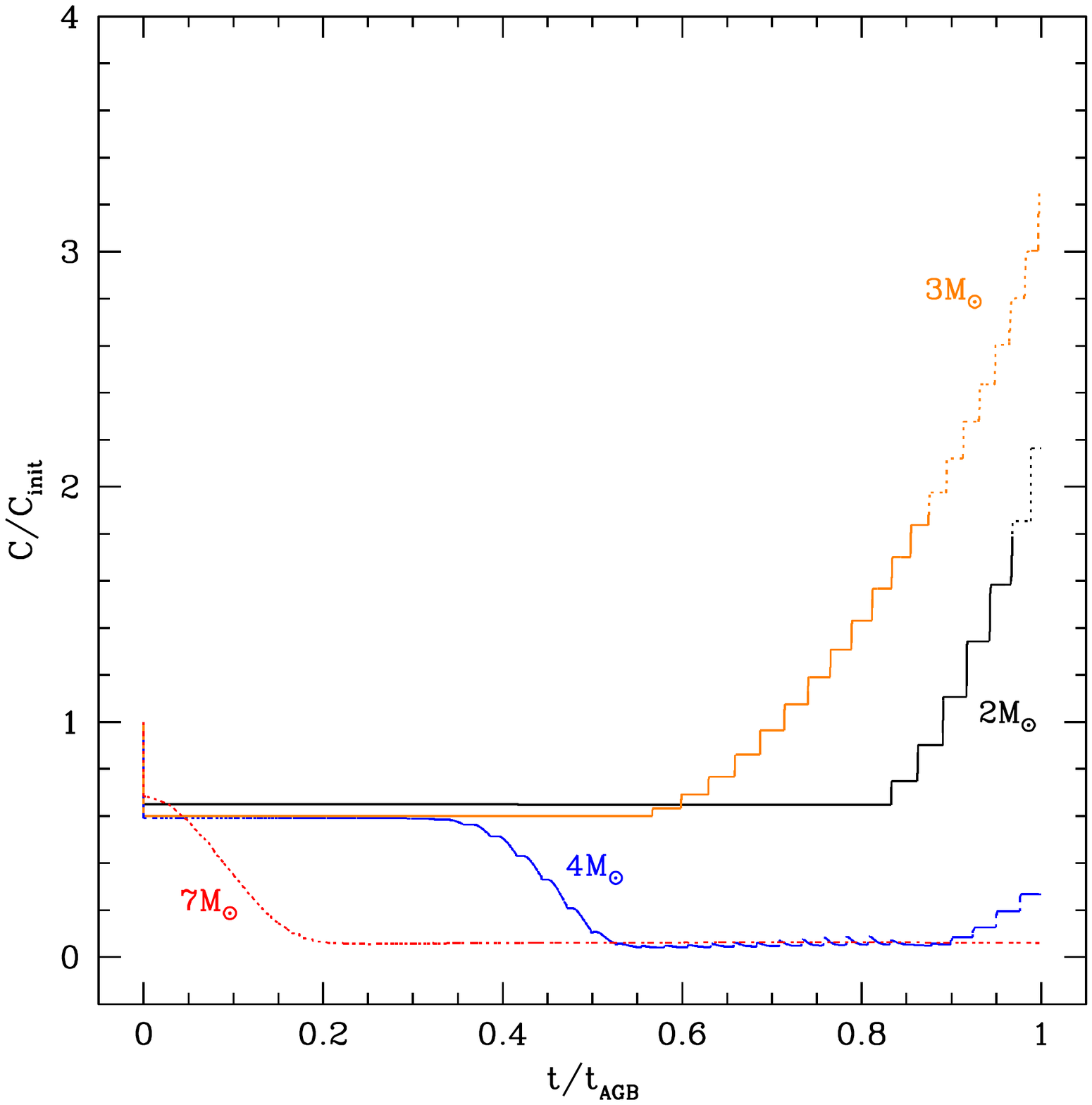}}
\vskip-60pt
\caption{The variation of the surface carbon (normalized to the initial mass
fraction) in AGB models of initial
mass $2~M_{\odot}$ (black line), $3~M_{\odot}$ (orange), $4~M_{\odot}$ (blue),
$7~M_{\odot}$ (red). Times on the abscissa are normalized to the duration of the
AGB phase. The C-star phases are shown with a dotted line.
}
\label{fctime}
\end{figure}

The enrichment in carbon favours the formation of carbon stars, when the surface
C/O $\ge 1$. We find that at solar metallicities the minimum mass required 
to reach the carbon star stage is $1.5~M_{\odot}$. Stars below this limit loose their 
mantle before the C/O $>1$ condition is reached.  
For the majority of the AGB phase low-mass AGB stars are observed as
oxygen-rich, as illustrated in Fig.~\ref{fctime}. The duration of the
C-rich phase is below $10\%$ for the  $2~M_{\odot}$ stars, whereas 
it is $\sim 15\%$ for the $3~M_{\odot}$ star \citep[see also][]{karakas14b}.

\subsection{The imprinting of HBB on the surface chemistry of massive AGB stars}
The HBB operating in massive AGB stars prevents the formation of a
C-rich atmosphere and sets an upper limit for C-star formation. The
upper mass limit is model dependent and is $3~M_{\odot}$ in the ATON
models and $4.5\Msun$ in the MONASH models. 

The left, top panel  of Fig.~\ref{fchim} shows that the mass expelled
by these stars is carbon-poor, with a 
carbon content $\sim 20$ times smaller than the initial mass fraction. In 
the left, bottom panel of Fig.~\ref{fchim} we notice that the ejecta
of massive AGB stars
present traces of oxygen destruction: the most massive stars exhibit the largest depletion 
of oxygen, $\sim 30\%$ lower than the initial abundance. The
activation of the CNO cycles also results in a significant rise in the
nitrogen abundance (see right, top  panel of Fig.~\ref{fchim}), 
which increases by a factor $\sim 20$ during the AGB evolution. 
The surface sodium abundance, shown in the right, bottom panel of 
Fig.~\ref{fchim}, is seen to increase during the AGB phase, with production factors
of the order of $\sim 4$. As discussed in section \ref{hmagb}, the ejecta of these
stars are sodium rich, owing to extremely favourable conditions to the synthesis of
sodium.

While the gas expelled by massive AGB stars is expected to show the signature of
proton-capture processing, the percentage of the AGB phase during which the surface
chemical composition of the star is substantially altered by HBB is sensitive to the
mass of the star. This can be deduced by focusing on the lines corresponding to
the $4~M_{\odot}$ and $7~M_{\odot}$ stars in Fig.~\ref{fchim}. In the former case 
the surface chemistry is practically unchanged for the first half of the evolution,
whereas in the $7~M_{\odot}$ star, owing to an early activation of HBB, the surface
chemical composition show traces of HBB from the first TPs \citep[see
also][]{karakas16}.  
We conclude that in the massive AGB domain we shift gradually from the stars with mass just above
the threshold to activate HBB, which spend about half of their AGB evolution with the
original chemical composition, to the most massive AGB stars, which show the imprinting
of HBB for most of the TP-AGB phase.

\begin{figure*}
\begin{minipage}{0.32\textwidth}
\resizebox{1.\hsize}{!}{\includegraphics{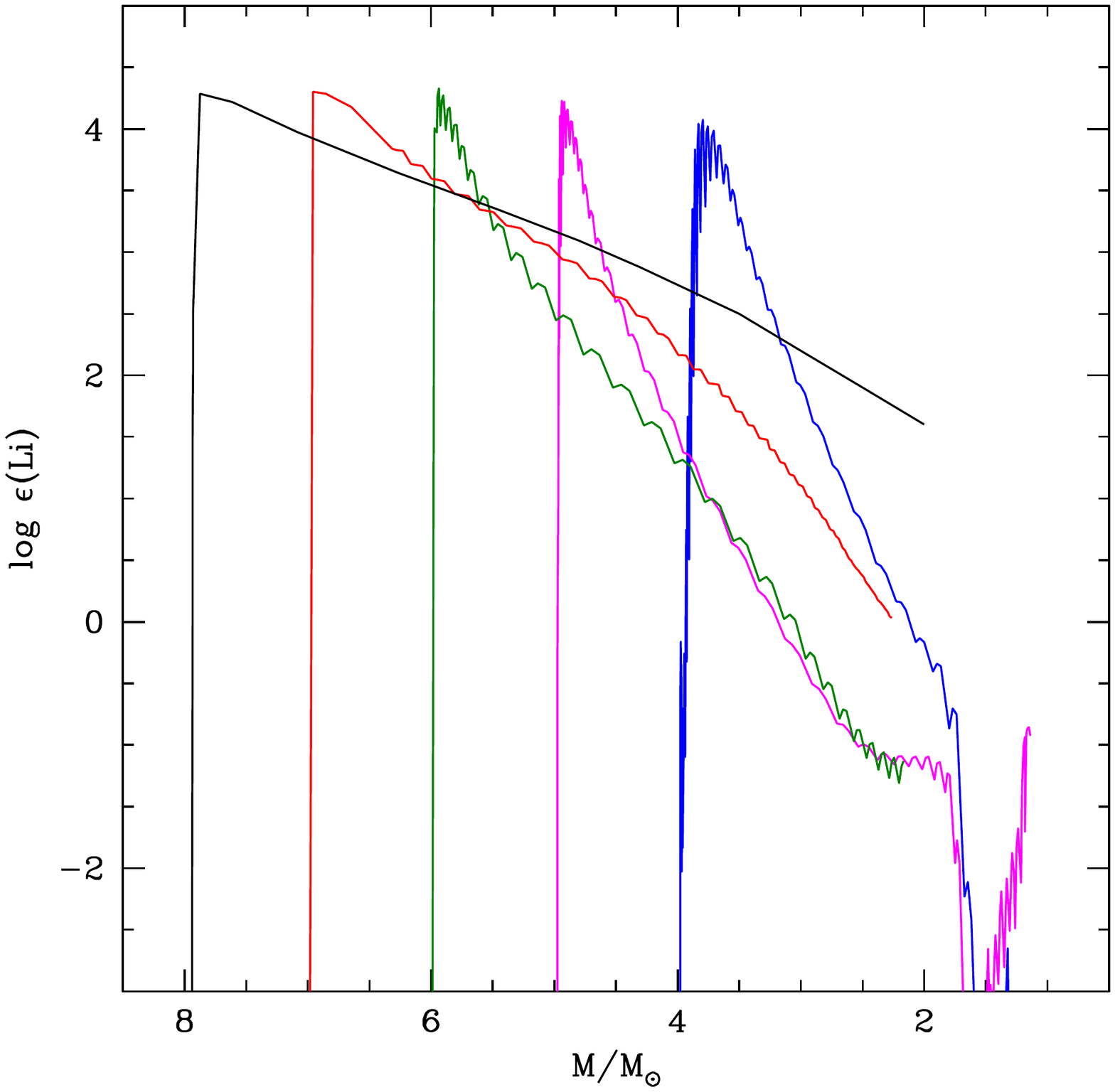}}
\end{minipage}
\begin{minipage}{0.32\textwidth}
\resizebox{1.\hsize}{!}{\includegraphics{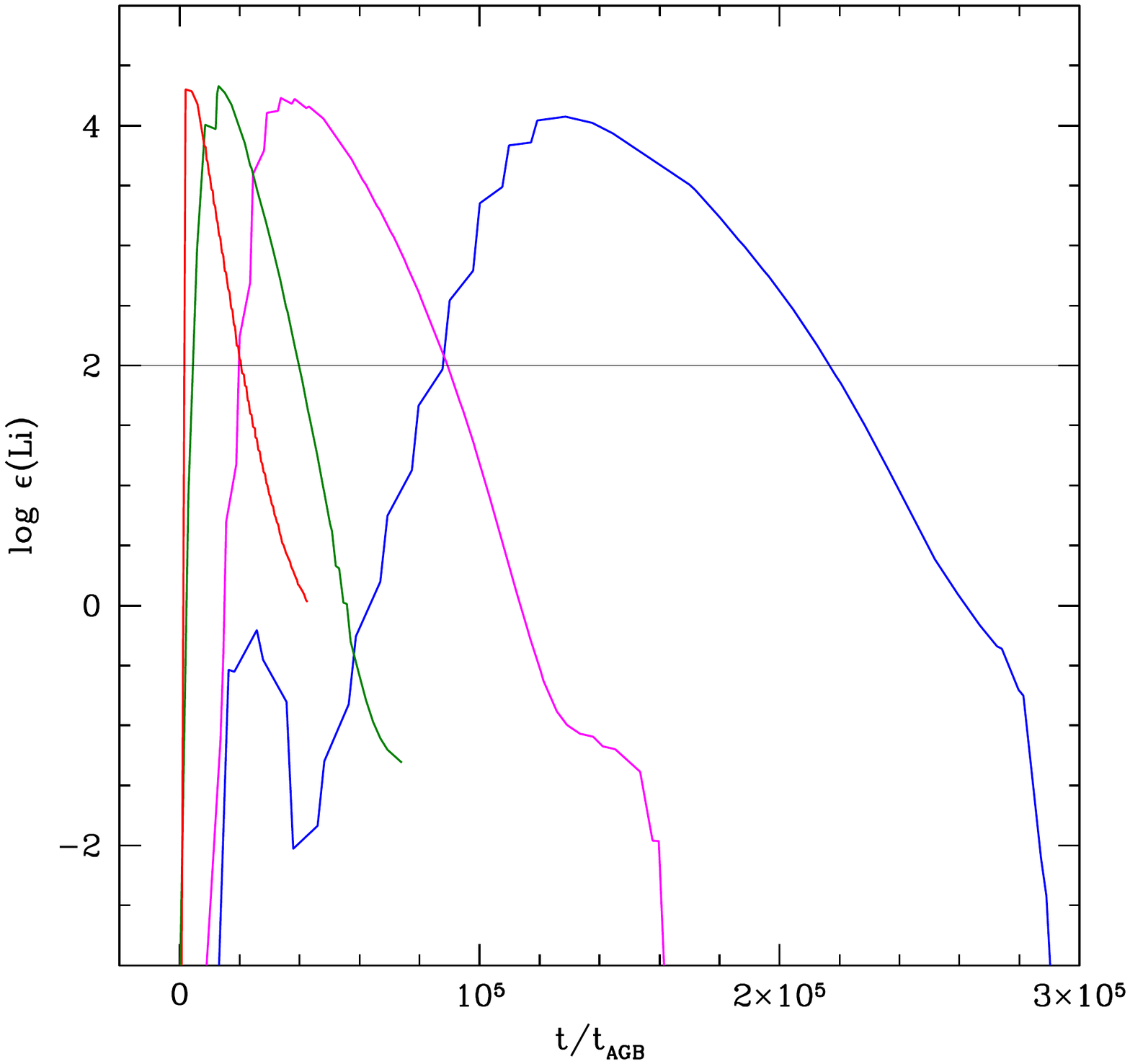}}
\end{minipage}
\begin{minipage}{0.32\textwidth}
\resizebox{1.\hsize}{!}{\includegraphics{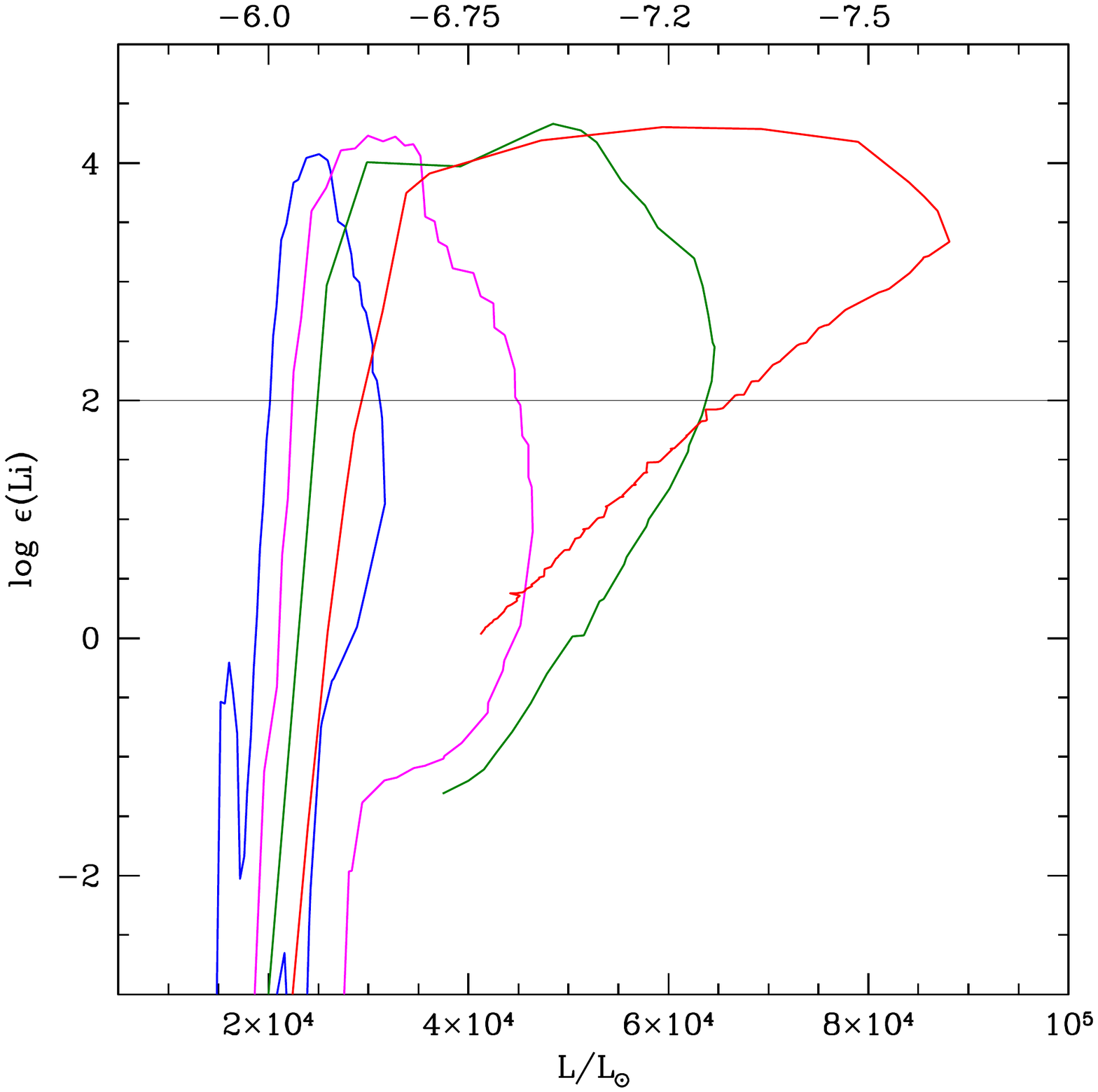}}
\end{minipage}
\vskip-30pt
\caption{The variation of the surface lithium in AGB stars experiencing HBB.
The various colours, the same used in Fig.~\ref{fchim}, correspond to different initial 
masses. The surface lithium is shown as a function of
the current mass of the star (left panel), of the time counted since the beginning of the
AGB evolution, normalized to the total duration of the AGB phase (middle panel) and
of the luminosity of the star. On top of the right panel we show the bolometric magnitudes
corresponding to the luminosities reported on the abscissa. In the middle and right
panels the track of the $8~M_{\odot}$ star was omitted for clarity reasons, as it would
largely overlay with the $7~M_{\odot}$ line.
}
\label{flitio}
\end{figure*}

\subsection{Lithium}
\label{lithium}
The discovery of bright red giants stars enriched in lithium in the
Magellanic Clouds \citep{smith89, smith90} and our Milky Way Galaxy
\citep{garcia07, garcia13} showed that AGB stars with HBB may be important
factories for the production of Li, at least for part of the AGB phase.

The mechanism upon which lithium production is based was first identified by 
\citet{cameron71} and confirmed by AGB modelling by
\citet{sackmann92}.  
When the temperature at the base of the envelope, $T_{\rm bce} \gtrsim 30$MK, the production
of beryllium via the $^3He + ^4He \rightarrow ^7Be$ reaction is activated. Owing to the
rapidity of convective motions, part of the beryllium is transported
to cooler regions in the envelope where it can capture an electron to
form lithium.  The newly formed lithium will survive in the outer most
layers although eventually convection will mix it down to hotter
regions, where it will be destroyed. Lithium production will continue
until the star runs out of \iso{3}He.

The temperatures given above require the ignition of HBB.  Therefore lithium synthesis
is limited to intermediate-mass AGB stars, which is consistent with the existence of a lower
limit in the luminosity of the lithium-rich sources in the Magellanic Clouds 
discovered by \citet{smith89, smith90}. The luminosity function of lithium-rich stars
in the MCs was used by \citet{ventura00} to calibrate the rate of mass loss of oxygen-rich AGB
stars.

Fig.~\ref{flitio} shows the variation of the surface lithium in models experiencing
HBB. In the $y$-axis of the three panels we show the standard quantity used to quantify the 
lithium content, namely $\log (\epsilon (^7Li)) = 12 + \log (n(^7Li) / n(H))$. We note 
three points in common to all the stars considered: a) the stars enters the AGB phase
with practically no lithium, as it is destroyed prior to the AGB; b) 
lithium production begins after the beginning of the AGB phase and the amount of 
lithium at the surface reaches a maximum abundance of the order of 
$\log (\epsilon (^7Li)) \sim 4.3$; c) the surface lithium declines in the final part 
of the AGB phase, when there is no $^3He$ available.

The left panel of Fig.~\ref{flitio} shows that the matter expelled by these stars is
lithium-rich. The amount of Li enrichment increases with increasing
stellar mass, because in massive AGB stars the rate at which mass loss occurs exceeds the 
rate in which $^3He$ is destroyed. 

If we fix a threshold of $\log (\epsilon (^7Li)) = 2$, above which we consider the
star as being lithium-rich, we see in the middle panel of Fig.~\ref{flitio} that the
lithium-rich phase is about half of the total AGB evolution. The most massive stars 
start to produce lithium during the very first interpulse phases, whereas in stars of lower
mass the synthesis of lithium begins after $\sim 30\%$ of the TP-AGB
time has been completed, which is
the time required to reach HBB conditions.

The right panel of Fig.~\ref{flitio} shows the lithium versus luminosity trend. A clear
indication we get from this plot is that lithium-rich abundances are expected when
the stars reach a luminosity of $20000~L_{\odot}$, i.e. $M_{bol}=-6$, almost
independently of the initial mass. This is the threshold above which we expect to find
lithium-rich AGB stars. The upper limit in luminosity where we expect to observe
lithium-rich sources is sensitive to the mass of the stars, and is
higher in stars of higher initial mass.

The Lithium predictions, both for ATON and MONASH models \citep[see e.g.][]{garcia13}, qualitatively
agree with existing spectroscopic observations of massive Galactic HBB-AGB stars
\citep{garcia07, garcia13}, which show that the most luminous and O-rich AGB
stars (obscured OH/IR stars) in our Galaxy are Li-rich \citep{garcia07} and that
these stars can reach $\log (\epsilon (Li)) \sim 4$ at the beginning of the
TP-AGB phase \citep{garcia13}. The s-process element Rb, being a good indicator
of the progenitor mass in AGB stars \citep[see e.g.][]{garcia06, perezmesa17},
has been also measured in these stars. Contrary to the synthesis of Li, strong
Rb production is expected towards the end of the AGB phase, when a significant
number of TPs have been experienced \citep[see e.g.][]{garcia13}. The
observations show that the presence of Li is not always correlated with Rb,
indicating that the observed Galactic samples contain massive AGB stars with
different progenitor masses and/or at several AGB evolutionary stages. A more
detailed comparison with the observations is hampered by the uncertain distances
(and so the their luminosities) to these Galactic massive AGB stars. Precise
Gaia distances (and luminosities) to these Galactic massive AGB stars would
permit to disentangle the evolutionary stage and progenitor mass of these
Galactic Li-rich AGB stars.

\section{The final chemical composition}
\label{pne}

\subsection{Model predictions}
\label{predic}
The final chemical composition is a key indicator of the relative efficiency of
HBB and TDU in altering the surface chemistry of these stars. 

\begin{figure}
\resizebox{1.\hsize}{!}{\includegraphics{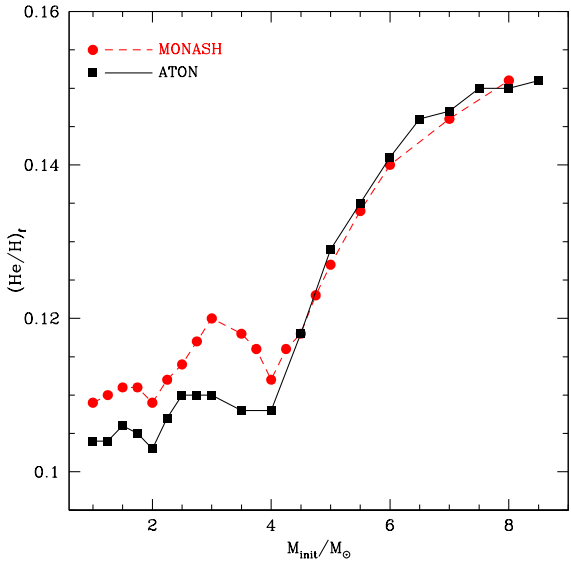}}
\vskip-60pt
\caption{The final $He/H$ fraction of AGB models of different mass.  
}
\label{fhef}
\end{figure}

Helium is a peculiar element among the various chemical species, because the surface
abundance is not strongly sensitive to the details of AGB modelling. 
The modification of the surface helium content is mainly determined by the
efficiency of the FDU and of the second dredge-up (SDU) episode.

Fig.~\ref{fhef} shows the final surface $He/H$ of the models discussed here.
Stars of mass below $\sim 4~M_{\odot}$ do not experience any SDU. In this case the
final $He/H$ ranges from $\sim 0.1$ to $\sim 0.11$, with little dependence on the
mass of the star. In more massive stars the SDU, taking place shortly
after the end of core He-burning \citep{karakas14b}, favours the increase in the surface helium. 
The strength of the SDU depends on the initial mass of the star and is
more efficient for higher mass objects \citep{ventura10}.
As shown in Fig.~\ref{fhef} the final $He/H$ increases monotonically from
$He/H \sim 0.11$, for a $4~M_{\odot}$ star, to $He/H \sim 0.15$ for
the most massive stars. 

The comparison with the MONASH results is shown in Fig.~\ref{fhef} and outlines the
following: a) in the low-mass domain the final $He/H$ is $\sim 0.05$ higher than
the present models, owing to the higher helium assumed in the MONASH computations;
b) for massive AGB stars we find a remarkable agreement between the ATON and the
MONASH results. The helium enrichment of the surface regions of these stars
turn out to be substantially independent of AGB modelling.

\begin{figure*}
\begin{minipage}{0.48\textwidth}
\resizebox{1.\hsize}{!}{\includegraphics{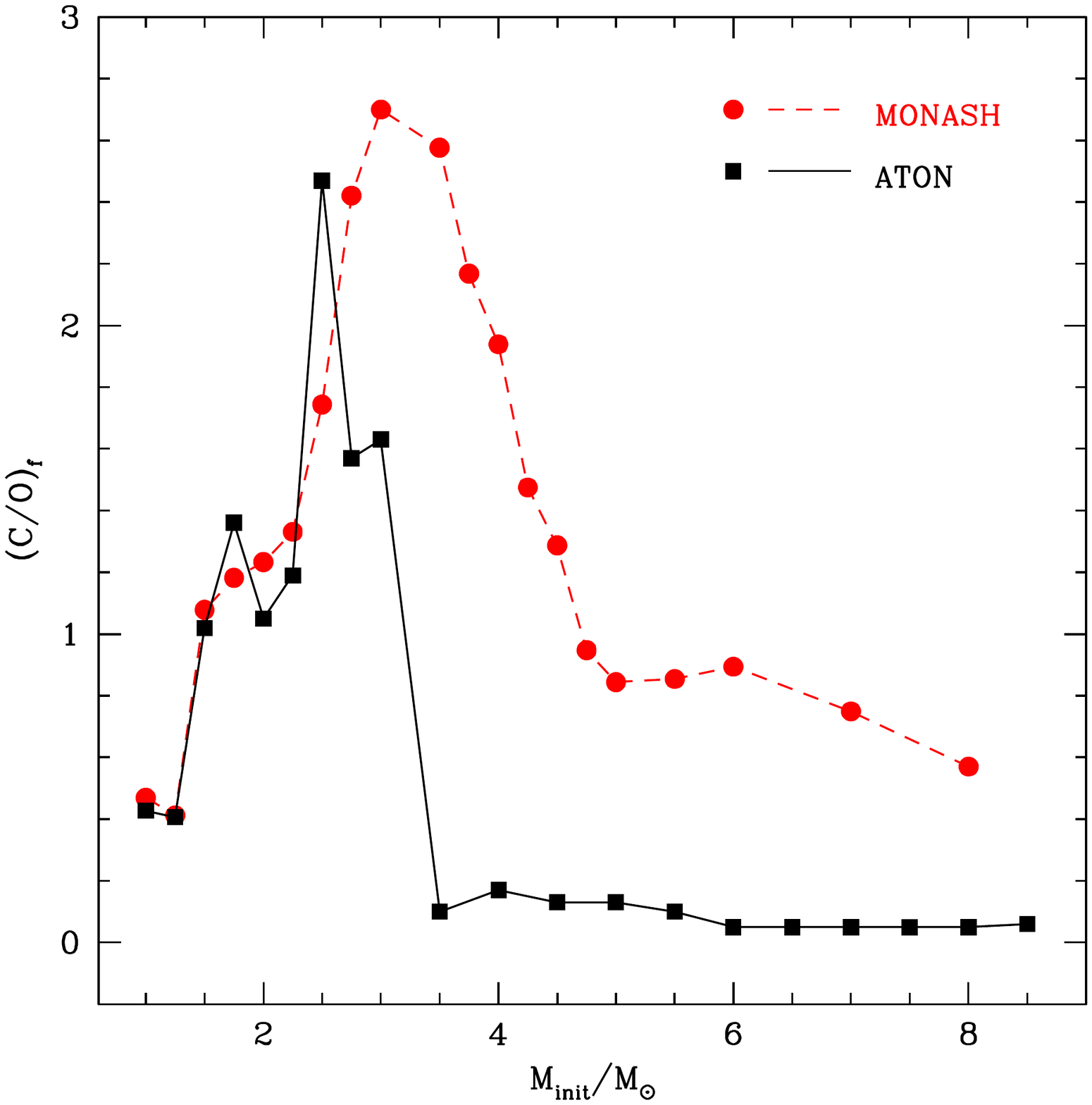}}
\end{minipage}
\begin{minipage}{0.48\textwidth}
\resizebox{1.\hsize}{!}{\includegraphics{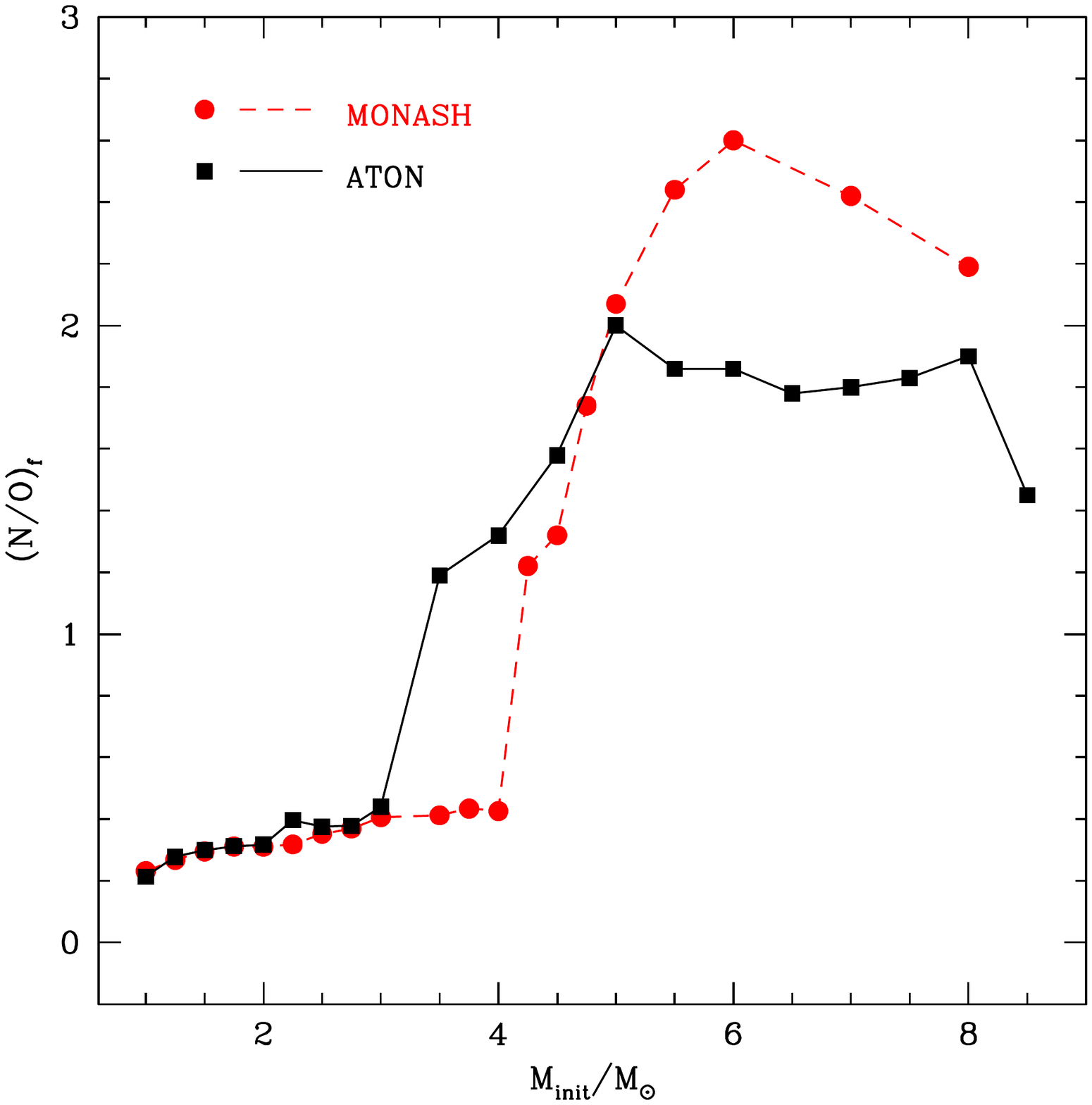}}
\end{minipage}
\vskip-60pt
\caption{The final C/O (left panel) and N/O (right) ratios for AGB stars
of different mass}
\label{fcnof}
\end{figure*}

For the elements involved in CNO cycling Fig.~\ref{fcnof} shows the
final surface C/O and N/O ratios. These results can be easily interpreted based on 
the discussion in Section~\ref{chimagb}.

For stars of mass below $3.5~M_{\odot}$ the final C/O increases with the mass of the star,
ranging from 0.4 ($1~M_{\odot}$ star) to 2.5 ($3~M_{\odot}$). This is because the number 
of TDU episodes increases with stellar mass, which in turn increases
the final C/O ratio.  We have seen that stars in the mass range 
$1.5~M_{\odot} \leq M_{init} \leq 3~M_{\odot}$ 
become carbon stars; this is consistent with their final C/O ratios above unity. The final 
N/O shows up only a mild dependance on the stellar mass, which is
caused by the efficiency of the FDU \citep[e.g.,][]{boothroyd99}. 
The final N/O for these stars spans the range $0.2 < N/O < 0.4$.

The stars experiencing HBB follow a completely different behaviour. As shown in the left 
panel of Fig.~\ref{fcnof} we find C/O ratios below $0.1$, independently of the stellar 
mass. These stars also show a significant increase in the final N/O, 
with values in the range $1.5 < N/O < 2$.

The comparison with the results from \citet{karakas14a} outlines strong similarities
in the low-mass domain, whereas the ATON findings for massive AGB stars reveal significant 
differences compared to MONASH models.

First, we find that in the range of mass $3~M_{\odot} \leq M_{init} \leq 4~M_{\odot}$ the
MONASH C/O ratios are $\lesssim 2$, whereas the corresponding ATON values are 
C/O $< 0.2$. This is due to a shift in the threshold mass required to ignite HBB,
which is $\sim 1~M_{\odot}$ higher in the MONASH models. 

For stars of mass above $4~M_{\odot}$, while the present models are characterized
by final C/O ratios below $0.1$, in the MONASH models we find $0.5 <
C/O < 1.5$ \citep[see Fig 2 in][]{karakas16}.
This is partly due to the stronger HBB found in the present models, owing to the use
of the FST model for convection. An additional explanation is that the TDU efficiency
is extremely poor in this mass domain, whereas in the MONASH models some carbon is
transported to the surface via TDU, despite the fact that some of the
carbon is subsequently destroyed by HBB during the
following interpulse phase. This explanation finds additional confirmation in the 
comparison of the final N/O ratios, 
which are higher in the MONASH models.  This is because 
of the additional contribution of primary nitrogen, which is synthesized by fresh carbon
dredged-up from the He-shell. 

\subsection{Observational facts and future directions}
\label{obs}

A detailed comparison with the composition of solar metallicity AGB,
post-AGB stars and PNe, although out of the scope of the present paper, would
offer, in principle, the opportunity to test the theoretical models of this
still rather uncertain evolutionary phase. Ideally, the predicted abundances of
He, C, N, O, Ne, Na, Mg, and Al (as well as some key abundance ratios like C/O,
N/O, and C/N), from the two AGB models (ATON vs. MONASH) discussed here, could
be compared with the abundances of these elements as observed in solar
metallicitiy Galactic AGB, post-AGB stars and PNe and available in the
literature. However, this is not an easy task and there are several
observational limitations that depend on the source type and enumerated below.
Another serious observational problem is that Galactic AGB and post-AGB stars,
and PNe are plagued by distance uncertainties, which avoid detailed studies of
AGB nucleosynthesis at solar metallicity, depending on progenitor mass and
luminosity; the Gaia mission is thus expected to overcome the latter severe
observational problem.  

i) AGB stars: both C-rich and O-rich AGB stars may not display the final
chemical composition and their chemical abundance analysis (especially towards
the end of the AGB, where they are usually dust enshrouded; i.e., optically
invisible) is very complicated due to their complex dynamical atmospheres, which
can dramatically affect the derived abundances \citep[see e.g.][]{zamora14,
perezmesa17}. The CNO elemental and isotopic abundances, as obtained from
high-resolution optical and/or near-IR spectroscopy, are only available in some
Galactic C-rich AGB stars \citep[e.g.][and references therein]{hedrosa13,
abia17}. In the more massive O-rich HBB AGB stars the CNO
elemental/isotopic ratios can be derived in the near-IR wavelength region only
and such near-IR measurements have not been reported yet. Other
elements such as He
and Ne cannot be measured in AGB stars, while to the best of our knowledge, the
abundances of Na, Mg, and Al (although measurable from near-IR spectra) in
Galactic AGB stars have still to be reported. On-going massive high-resolution
near-IR spectroscopic surveys such as the second generation of The Apache Point
Observatory Galactic Evolution Experiment \citep[APOGEE-2; see e.g.][]{blanton17}
are expected to represent a major step forward in our understanding of AGB
nucleosynthesis, offering a invaluable test of the theoretical models presented
here. APOGEE-2 will provide homogeneous CNO elemental and isotopic abundances
(at least for the $^{12}$C/$^{13}$C ratios\footnote{\citet{marcella16} has recently
compared the observed C and O isotopic ratios (i.e., $^{12}$C/$^{13}$C and
$^{16}$O/$^{17}$O/$^{18}$O)  available in the literature for several types of
AGB stars with the AGB ATON predictions. The available C and O isotopic ratios,
however, are not homogeneous and they come from different observational data;
from optical/near-IR spectra in C-rich AGB stars to the far-IR (in a few massive
HBB O-rich stars) and to the radio domain.}) as well as Na, Mg, and Al
abundances for complete (flux-limited) samples of Galactic AGB stars (bulge,
disk, and halo), covering all progenitor masses. The possible circumstellar
effects (if any) on the near-IR molecular (CO, OH, CN) and atomic lines (Al, Mg,
Na) remained to be explored. Finally, observations of heavy neutron-rich
elements in AGB stars may provide clues to test these theoretical models but
their uncertainties are very large, ranging from 0.3$-$0.4 dex to as high as 0.7
dex for Rb \citep[e.g.,][]{abia01,garcia06, perezmesa17}, highlighting the need
for independent complementary observations (e.g., in post-AGB stars and PNe; see
below). Also, the simulations of the nucleosynthesis due to slow-neutron
captures (the s-process) in the ATON AGB models are still under construction
\citep[][]{yague16}.

ii) Post-AGB stars: The atmospheres of these stars \citep[stars in the fast transition 
phase between AGB stars and PNe; see e.g.,][for a review]{vanwinckel03}    
display the final chemical composition (i.e., the final result of chemical
enrichment from internal nucleosynthesis and dredge-up processes during the
entire stellar evolution), being, in principle, ideal probes to study AGB
stellar nucleosynthesis. Their photospheres \citep[spectral types from K to A; see e.g.,][]{suarez06}
are hotter than those in AGB stars, dominated by atomic spectral lines that
allow for more accurate abundance determinations of a larger number of elements,
including C, N, O, Na, Mg, and Al, among others, but also many neutron-rich
s-process elements \citep[see e.g.][]{desmedt16} (as mentioned above, the
s-process ATON simulations are still under construction). The use of the
chemical composition observed in post-AGB stars as valuable tests for the
theoretical AGB models, however, is also hampered by the non-homogeneous
chemical analysis reported in the literature and because the number of elements
that can be measured in post-AGB stars depend on the stellar effective
temperatures. Because of the fast AGB-PNe transition times
\citep[$\sim$10$^{2}$ $-$ 10$^{4}$ years, depending on the initial mass; see e.g.][]{vw94} 
only about hundred confirmed post-AGB stars are known in the Galaxy
\citep[e.g.,][]{szczerba07,szczerba12}. In addition, only a handful of post-AGB
stars have been observed at high-resolution in the optical range. Finally,
present spectroscopic optical observations of post-AGB stars are strongly biased
towards the lower mass progenitors (say $\sim$1$-$2 M$_{\odot}$); e.g., usually
high Galactic latitude (i.e., metal-poor) and optically bright s-process
enriched C-rich post-AGB stars  \citep[see e.g.][]{vanwinckel00,reyniers07,desmedt16}.
In contrast, higher mass post-AGB stars (above 2 M$_{\odot}$), evolving much
faster, may have systematically escaped detection in past high-resolution
optical surveys because they may remain hidden (dust enshrouded) during the
whole AGB-PN transition. Both optically bright and obscured post-AGB stars could
be studied in the near-IR, which provides an unexplored spectral window that
should be exploited in order to get homogeneous chemical analysis of a complete
sample of Galactic post-AGB stars. As in the case of Galactic AGB stars, the
SDSS-IV/APOGEE-2 survey could provide such an ambitious goal; e.g.,
SDSS-IV/APOGEE-2 may discover the coolest post-AGB stars (K and M spectral
types) in our Galaxy with access to the dust enshrouded ones (in principle the
more massive ones), no accesible in the optical.

iii) PNe: The comparison of the theoretical predictions with the chemical
composition observed in PNe (via their nebular emission lines) offers another
opportunity to test theoretical models of the still rather uncertain AGB phase.

A recent step in this direction was done by \citet{ventura17}, who used the observed
nebular chemical composition to estimate the mass and formation epoch of the progenitors 
of 142 Galactic PNe. This analysis was based on the comparison of the 
abundances data with the ATON AGB model predictions presented here, specifically on the 
final abundances of the various chemical species, discussed in this section.
On general grounds, the chemical abundances in 
PNe, typically more accurate than those in AGB stars,
have also their own problems/limitations; e.g., the chemical abundances
available in the literature, again, are not completely homogeneous and
ionization correction factors (ICFs), sometimes very uncertain, are needed to
estimate the contribution of unobserved ions to the total abundances 
\citep[see e.g.,][and references therein]{inglada14}.  
However, the main advantages of PNe (with respect to AGB and post-AGB stars, see
above) are that PNe can be easily observed at very large distances (because of
their emission-line nature) and that known PNe samples are more complete (e.g.,
they cover the full range of initial masses, despite the masses estimated are
more uncertain compared to post-AGB stars). Also, the abundances of key
elements such as He, C, N, O, and Ne are accesible for all types of PNe; recent 
studies outlined the possibility of measuring the surface Zn \citep{smith17}. 
On the other hand, the abundances of Na, Mg, and Al cannot be measured in PNe. The abundances
of He, N, O, and Ne (among others like Ar, Cl, and S) are easily extracted from
low-resolution optical spectroscopy \citep[see e.g.,][and references therein]{garcia14} 
and available in the literature. However, the derivation of C abundances needs
deep high-resolution optical spectra\footnote{Some heavy s-process elements like
Se, Kr, Xe, Rb, Cd, and Ge can be also obtained from deep high-resolution
optical and/or near-IR spectroscopy \citep[e.g.,][]{sharpee07,sterling08,sterling16}.}
and/or UV spectra (e.g., by using the Hubble Space Telescope, HST), which are
not easily obtained \citep[see e.g.,][]{ventura17}. For example, the
availability of accurate C abundances from HST-UV spectra in PNe of the
Magellanic Clouds, together with other observational data such as optical and
mid-IR spectra, have permitted detailed comparisons of their CNO elemental
abundances with the predictions from the ATON AGB models \citep[][]{ventura15b,ventura16c}.
Similar studies in complete samples of Galactic PNe are not still possible,
mainly due to the lack of UV spectra available for only a few Galactic
sources \citep[see e.g.,][]{ventura17}. Thus, the collection of deep
high-resolution optical/near-IR spectra and/or UV spectra in a complete sample
of Galactic PNe would permit to construct a unique homogeneous database of PNe
nebular abundances to test the AGB theoretical models. Unfortunately, deep
high-resolution optical/near-IR nebular spectroscopy is very time
consuming (even with 8$-$10 m class telescopes), while UV spectroscopy
requires the use of precious HST time.

\begin{table*}
\setlength{\tabcolsep}{5pt}
\caption{Chemical yields (see text for definition) for solar metallicity, AGB models}                                       
\begin{tabular}{c c c c c c c c c c c c c c c c c c}        
\hline                      
$M$ & $H$ & $He$ & $^{12}C$ & $^{13}C$ & $^{14}N$ & $^{16}O$ & $^{17}O$ & $^{18}O$ &
$Ne$ & $^{23}Na$ & $^{24}Mg$ & $^{25}Mg$ & $^{26}Mg$ & $^{27}Al$ \\
\hline       
1.00 & -1.3E-2 &  1.3E-2 & -1.8E-4 & -5.6E-6 &  2.1E-4 &     -   &     -   &     -   &      -   &     -   &     -   &     -   &     -   &     -   &    \\
1.25 & -1.7E-2 &  1.4E-2 &  1.1E-3 &  2.6E-5 &  3.2E-4 &  7.4E-5 &     -   &     -   &   4.5E-5 &  7.9E-7 &     -   &  3.4E-7 &  3.4E-7 &     -   &    \\
1.50 & -1.9E-2 &  1.8E-2 &  2.5E-4 & -4.2E-5 &  7.0E-4 &  1.1E-4 &  2.7E-6 & -2.1E-6 &   3.2E-5 &  7.1E-7 & -1.2E-8 &  2.2E-7 &  4.0E-7 &  2.5E-7 &    \\
1.75 & -2.3E-2 &  2.0E-2 &  1.9E-3 & -7.1E-5 &  1.1E-3 &  1.5E-4 &  1.2E-5 & -3.5E-6 &   1.3E-4 &  2.2E-6 & -2.3E-7 &  2.1E-6 &  2.1E-6 &  9.5E-7 &    \\
2.00 & -3.0E-2 &  2.6E-2 &  2.7E-3 & -9.2E-5 &  1.4E-3 &  2.1E-4 &  1.6E-5 & -4.6E-6 &   1.8E-4 &  2.9E-6 & -3.9E-7 &  3.3E-6 &  3.1E-6 &  1.4E-6 &    \\
2.25 & -4.5E-2 &  4.0E-2 &  3.6E-3 & -1.2E-4 &  2.2E-3 &  1.8E-4 &  2.4E-5 & -6.0E-6 &   2.3E-4 &  1.9E-5 & -8.4E-7 &  6.3E-6 &  6.3E-6 &  2.5E-6 &    \\
2.50 & -6.7E-2 &  5.4E-2 &  9.7E-3 & -1.5E-4 &  2.7E-3 &  1.1E-4 &  2.2E-5 & -7.3E-6 &   5.4E-4 &  2.8E-5 & -4.2E-6 &  3.1E-5 &  2.0E-5 &  9.9E-6 &    \\
2.75 & -7.4E-2 &  6.4E-2 &  7.4E-3 & -1.6E-4 &  3.2E-3 & -3.6E-4 &  2.1E-5 & -8.2E-6 &   4.7E-4 &  3.2E-5 & -3.9E-6 &  2.5E-5 &  1.8E-5 &  1.1E-5 &    \\
3.00 & -8.7E-2 &  7.4E-2 &  9.3E-3 & -1.9E-4 &  3.7E-3 & -4.1E-4 &  2.1E-5 & -9.4E-6 &   6.0E-4 &  3.8E-5 & -6.3E-6 &  3.8E-5 &  2.6E-5 &  1.6E-5 &    \\
3.50 & -8.2E-2 &  7.7E-2 & -3.7E-3 & -2.7E-4 &  1.0E-2 & -1.3E-3 &  1.8E-5 & -2.9E-5 &   8.2E-5 &  1.1E-4 & -6.3E-6 &  9.6E-6 &  1.5E-5 &  1.5E-5 &    \\
4.00 & -9.0E-2 &  8.8E-2 & -6.3E-3 & -7.3E-4 &  1.3E-2 & -3.7E-3 &  2.9E-5 & -4.0E-5 &  -2.3E-4 &  3.2E-4 & -2.0E-5 &  8.9E-6 &  1.6E-5 &  9.9E-6 &    \\
4.50 & -1.7E-1 &  1.7E-1 & -7.6E-3 & -8.8E-4 &  1.6E-2 & -6.0E-3 &  3.6E-5 & -4.7E-5 &  -3.0E-4 &  3.4E-4 & -6.8E-5 &  3.7E-5 &  2.2E-5 &  9.2E-6 &    \\
5.00 & -2.6E-1 &  2.6E-1 & -8.3E-3 & -9.5E-4 &  1.9E-2 & -7.5E-3 &  4.5E-5 & -5.3E-5 &  -2.8E-4 &  3.7E-4 & -1.5E-4 &  1.2E-4 &  2.0E-5 &  5.2E-5 &    \\
5.50 & -3.5E-1 &  3.5E-1 & -9.8E-3 & -1.1E-3 &  2.1E-2 & -9.2E-3 &  5.5E-5 & -6.0E-5 &  -3.8E-4 &  4.0E-4 & -2.5E-4 &  1.9E-4 &  4.1E-5 &  1.3E-5 &    \\
6.00 & -4.4E-1 &  4.4E-1 & -1.1E-2 & -1.2E-3 &  2.3E-2 & -1.0E-2 &  7.0E-5 & -6.6E-5 &  -4.1E-4 &  4.3E-4 & -3.8E-4 &  3.1E-4 &  5.1E-5 &  1.3E-5 &    \\
6.50 & -5.2E-1 &  5.2E-1 & -1.2E-2 & -1.4E-3 &  2.5E-2 & -1.1E-2 &  9.0E-5 & -7.2E-5 &  -4.4E-4 &  4.6E-4 & -5.3E-4 &  4.6E-4 &  3.7E-5 &  5.6E-5 &    \\
7.00 & -5.7E-1 &  5.7E-1 & -1.3E-2 & -1.4E-3 &  2.6E-2 & -1.1E-2 &  1.2E-4 & -7.8E-5 &  -4.6E-4 &  4.8E-4 & -6.9E-4 &  6.2E-4 &  4.0E-5 &  5.3E-5 &    \\
7.50 & -6.5E-1 &  6.5E-1 & -1.4E-2 & -1.5E-3 &  2.9E-2 & -1.3E-2 &  1.2E-4 & -8.2E-5 &  -5.0E-4 &  5.2E-4 & -7.4E-4 &  6.8E-4 &  3.9E-5 &  6.8E-5 &    \\
8.00 & -6.5E-1 &  6.5E-1 & -1.4E-2 & -1.6E-3 &  2.9E-2 & -1.2E-2 &  1.6E-4 & -8.8E-5 &  -5.2E-4 &  5.2E-4 & -8.7E-4 &  8.1E-4 &  3.7E-5 &  6.8E-5 &    \\
8.50 & -7.5E-1 &  7.5E-1 & -1.5E-2 & -1.5E-3 &  3.0E-2 & -1.2E-2 &  1.6E-4 & -9.3E-5 &  -3.4E-4 &  6.2E-4 & -1.2E-3 &  1.1E-3 &  3.6E-5 &  7.5E-5 &    \\
\hline     
\label{tabyield}
\end{tabular}
\end{table*}

\section{Yields from AGB stars}
The yields of the various chemical species are key quantities to understand the pollution
expected from a class of stars and the way they participate in the gas cycle
of the interstellar medium. 

In the following we will use the classic definition, according to which we indicate 
the yield $Y_i$ of the $i$-th element as

$$
Y_i=\int{[X_i-X_i^{init}] \dot{M} dt}.
$$

The integral is calculated over entire stellar lifetime and $X_i^{init}$ is the mass fraction
of species $i$ at the beginning of the evolution. Based on this
definition, the yield is negative if an element is destroyed and
positive if it is produced over the life of the star.

\begin{figure}
\resizebox{1.\hsize}{!}{\includegraphics{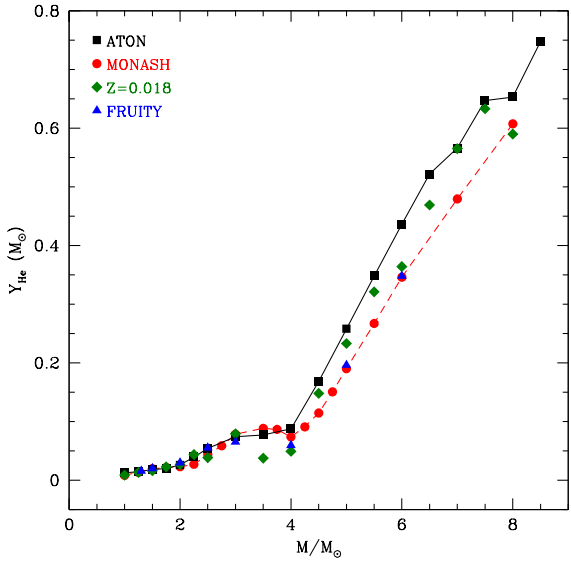}}
\vskip-60pt
\caption{The helium yields (see text for definition) of solar metallicity stars of
different mass are indicated with black squares. The results by \citet{karakas16}
(red points), \citet{cristallo15} (blue triangles) and \citet{marcella16} (green 
diamonds) are also shown for comparison.
}
\label{fyieldhe}
\end{figure}

Fig.~\ref{fyieldhe} shows the yields of helium, $Y_{He}$. It is evident the sudden 
increase in $Y_{He}$ occurring at $\sim 4~M_{\odot}$, representing the
lower limit for solar metallicity stars to experience SDU. The trend of $Y_{He}$ with the 
mass of the star is positive, ranging from $Y_{He} = 0.1~M_{\odot}$ to 
$Y_{He} = 0.75~M_{\odot}$. This is in agreement with the results shown in Fig.~\ref{fhef}. 
In the low-mass domain the helium yield is determined primarily by FDU and we find 
$Y_{He} < 0.1~M_{\odot}$ in all cases.

\begin{figure*}
\begin{minipage}{0.48\textwidth}
\resizebox{1.\hsize}{!}{\includegraphics{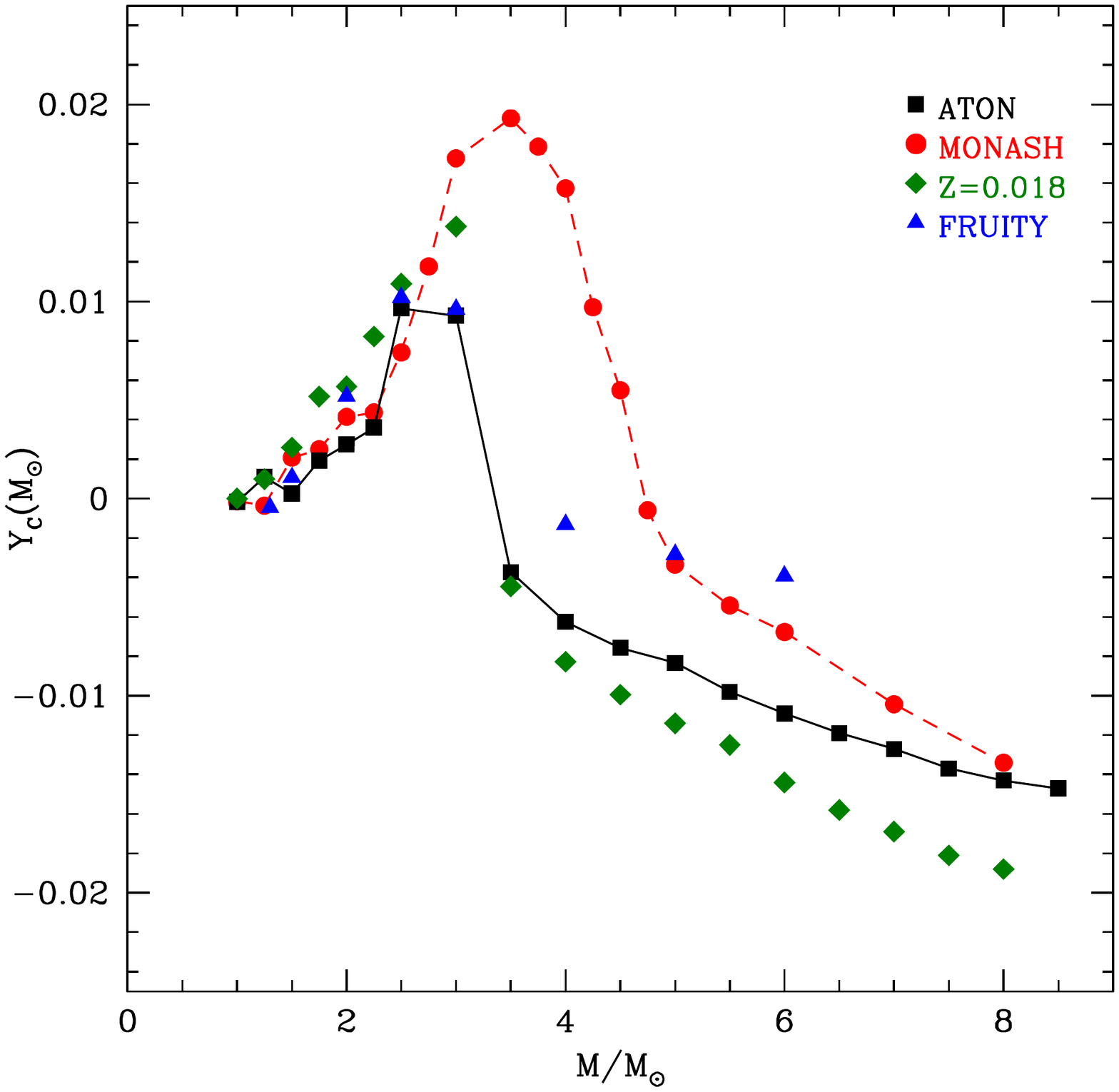}}
\end{minipage}
\begin{minipage}{0.48\textwidth}
\resizebox{1.\hsize}{!}{\includegraphics{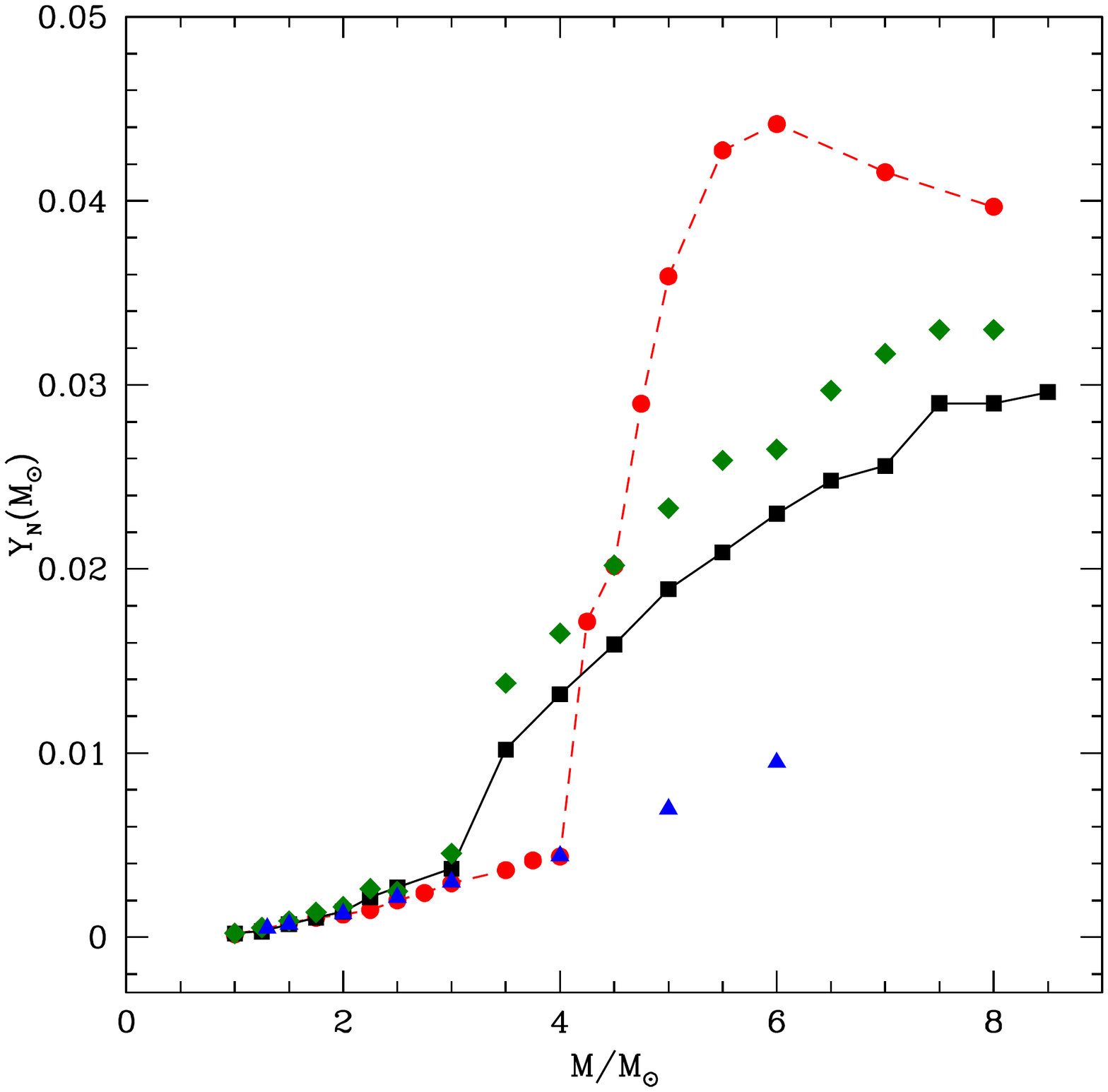}}
\end{minipage}
\vskip-80pt
\begin{minipage}{0.48\textwidth}
\resizebox{1.\hsize}{!}{\includegraphics{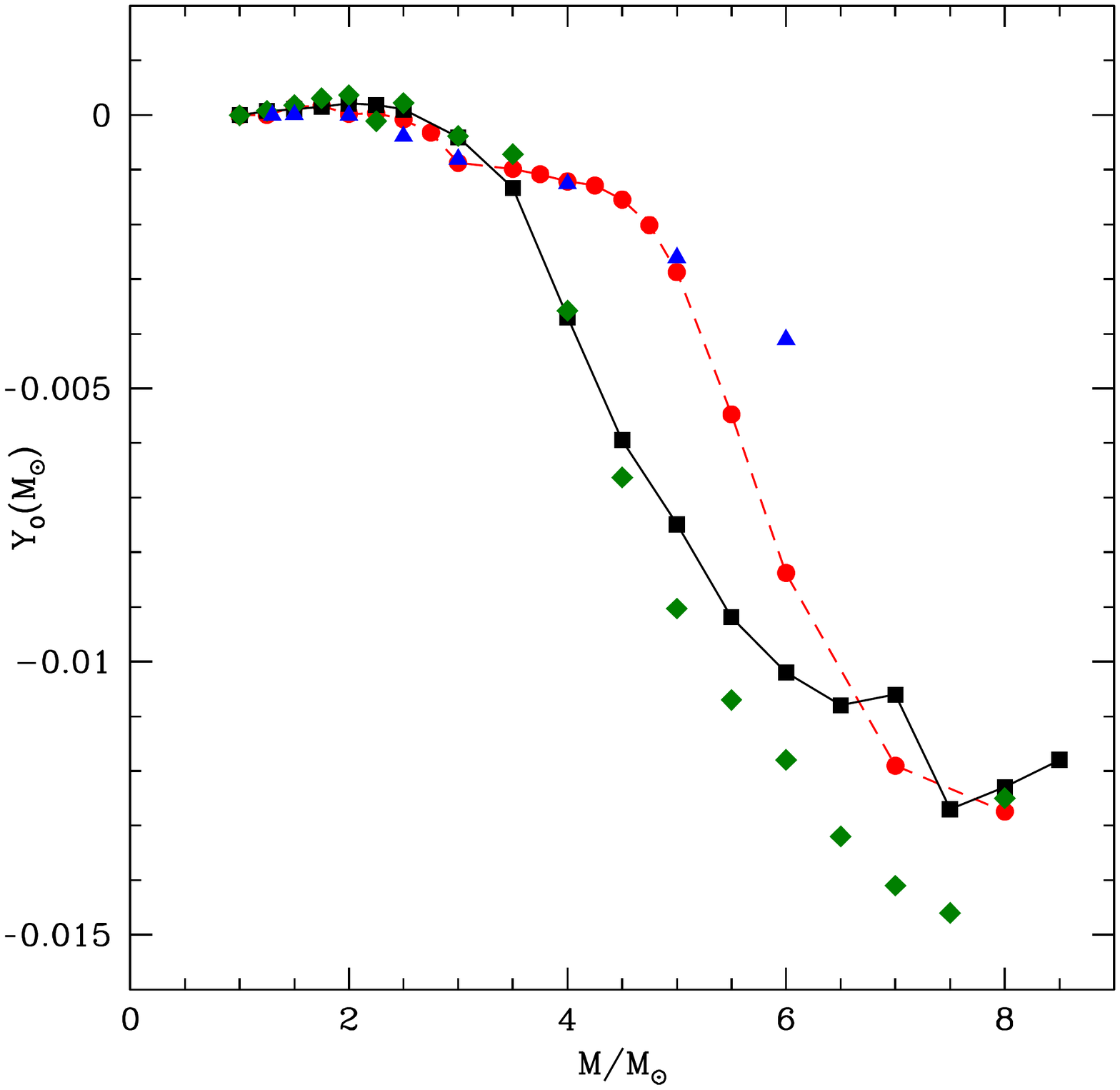}}
\end{minipage}
\begin{minipage}{0.48\textwidth}
\resizebox{1.\hsize}{!}{\includegraphics{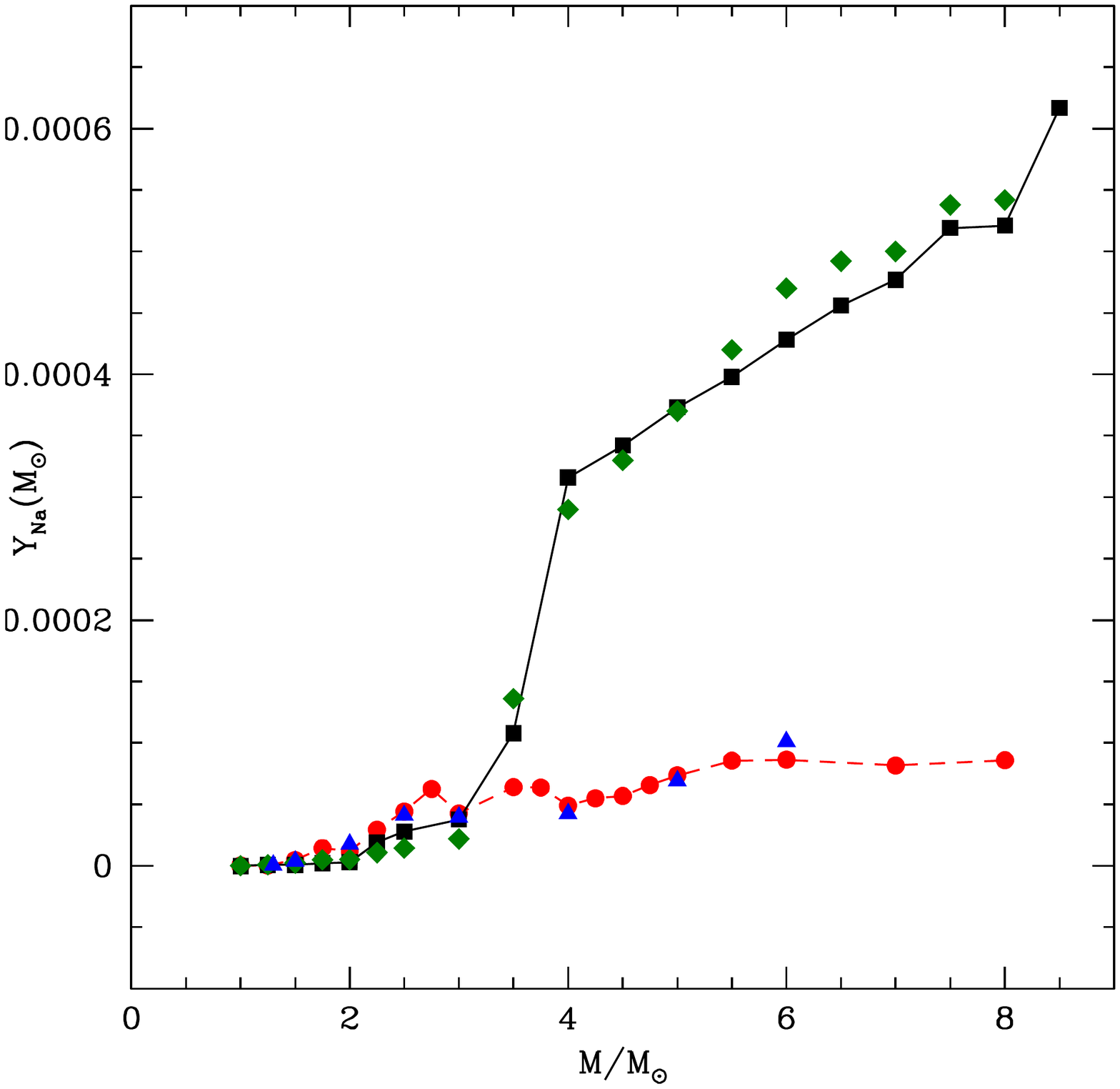}}
\end{minipage}
\vskip-50pt
\caption{The yields of the CNO elements and of sodium for the AGB models
presented here (shown as black squares), compared with the results from 
\citet{karakas16} (red points), \citet{cristallo15} (blue triangles) and \citet{marcella16} 
(green diamonds). The 4 panels report the yields of $^{12}C$ (top, left panel), $^{14}N$ 
(top, right), $^{16}O$ (bottom, left) and sodium (bottom, right).
}
\label{fyield}
\end{figure*}

The yields of the main CNO species and of sodium are shown in Fig.~\ref{fyield}.
The carbon yields can be understood based on the different behaviour of low-mass stars
and massive AGB stars, outlined in the previous sessions. We find carbon production for 
stars of mass $M_{init} \leq 3~M_{\odot}$, with a maximum yield of $\sim 0.01~M_{\odot}$
for $2.5-3~M_{\odot}$ stars, which experience the largest enrichment of carbon at
the surface (see Fig.~\ref{fchim}). $Y_{\rm C}$ increases with stellar mass,
in agreement with the discussion in section \ref{chimagb}. 
For stars experiencing HBB the carbon yields are negative. 
In this mass domain $Y_{\rm C}$ decreases from $Y_{\rm C} = -0.005~M_{\odot}$ 
(for the $3.5~M_{\odot}$ star) to $Y_{\rm C} = -0.015~M_{\odot}$ ($3.5~M_{\odot}$).

The oxygen yields of low-mass stars are almost zero because little
oxygen is produced in the He-intershell (see Fig.~\ref{fchim}). For stars of
mass $M_{init} \geq 3.5~M_{\odot}$ with HBB, the yields of oxygen are
negative and range from $-0.002~M_{\odot}$ to  $-0.012~M_{\odot}$, for
masses between 3.5 and 8.5$\Msun$, respectively.

The yields of nitrogen, $Y_{\rm N}$, are positive for all the stars.
Low-mass stars produce nitrogen via the FDU and the corresponding yields are below
$0.005~M_{\odot}$ (see top, right panel of Fig.~\ref{fyield}). The N yields of stars
experiencing HBB are higher, owing to nitrogen synthesis via CNO cycling. For these
stars we find $0.01~M_{\odot} < Y_{\rm N} < 0.03~M_{\odot}$. 

The behaviour of sodium, shown in the right, bottom panel of Fig.~\ref{fyield},
is qualitatively similar to nitrogen. In the low mass domain we have 
$Y_{\rm Na} < 5\times 10^{-5}~M_{\odot}$, whereas for stars with HBB 
we find a gradual increase with stellar mass, from
$Y_{\rm Na} \sim 3\times 10^{-4}~M_{\odot}$ to $Y_{\rm Na} \sim 6\times 10^{-4}~M_{\odot}$.
This is consistent with the results shown in Fig.~\ref{fchim}.

In the comparison with the models published in \citet{karakas16} we find that the
helium yields, shown in Fig.~\ref{fyieldhe}, are extremely similar (as
discussed in Section~\ref{chimagb}). The same holds for the results taken from
the FRUITY database, published by \citet{cristallo15}.

The carbon yields of stars of mass $M \geq 3~M_{\odot}$ by \citet{karakas16} are higher 
than those presented here. This can be clearly seen in the top, left panel of 
Fig.~\ref{fyield}. The largest discrepancy is found for masses 
$3~M_{\odot} < M < 5~M_{\odot}$, for which the present yields are negative whereas the
MONASH yields are positive. Similar differences, though of minor extent, are found when
the present yields are compared with results from the FRUITY database. In the
massive AGB domain all the sets of carbon yields are negative, but the ATON yields are
lower, because the HBB found in these models is stronger and the surface chemistry is 
affected almost entirely by HBB. Conversely, in the MONASH and FRUITY cases, TDU events 
increase the surface carbon mass abundance. 

The oxygen yields of low-mass stars by \citet{karakas16} and \citet{cristallo15}
are similar to ours, whereas in the higher mass domain they are higher than those 
presented here (see Fig.~\ref{fyield}). This is a consequence of the different efficiency 
of HBB, which in turn, is determined by convection modelling. The two lines indicating the 
ATON and MONASH yields tend to converge towards the most massive models. This is an indication
that for large core masses an efficient HBB is found, independently of convection
modelling. It is difficult to make a similar comparison to the
FRUITY models because the most massive initial masses
considered are of $6~M_{\odot}$, lower than the most massive models
considered in the ATON or Monash codes.

The results shown in Fig.~\ref{fyield} indicate that 
a larger production of nitrogen is found by \citet{karakas16}.
For stars of mass around $6~M_{\odot}$ the MONASH yields are almost a factor of $2$ 
higher than ATON. As discussed in section \ref{pne}, this is due to
TDU events which mix primary carbon from the He-shell to the envelope,
which allows the production of primary nitrogen. In the 
ATON case the nitrogen produced is essentially secondary.
The nitrogen FRUITY yields are even smaller than ATON, because
HBB is not particularly efficient even in their most massive
models.

By looking at Fig.~\ref{fyield} we notice that the largest difference between ATON,  
MONASH and FRUITY models is found in the sodium yields. In the present models we find some 
sodium production in the higher mass domain, whereas in the models by \citet{karakas14a} 
and \citet{cristallo15} the
excess of sodium with respect to the initial quantity in the ejecta is negligible. 
This is consistent with the results shown in Fig.~\ref{f5hbb}.

Fig.~\ref{fyieldhe} and \ref{fyield} also show the comparison between the present yields 
and those found in \citet{marcella16}, which are based on the solar
composition by \citet{gs98}. The differences are small and far lower
than those introduced by the use of a different
description of the convective instability.

\section{Dust formation}
\label{agbdust}
We applied the formulae described in Section~\ref{dustin} to calculate the 
dust formed in the wind of the models presented here, as well as in the models
of solar metallicity by \citet{karakas14a}. 

The dust species formed in the wind are primarily determined by the
surface $C/O$ ratio.  In the
wind of oxygen rich stars the majority of the dust produced is in the form of
silicates and alumina dust, whereas in carbon stars the main species are carbon and
SiC. We will focus our attention on the size of the dust grains formed, on the degree
of condensation of the key elements to form dust, and on the total mass of dust produced.

In the previous sections we outlined that low-mass stars with mass in the range
$1.5~M_{\odot} \leq M_{init} \leq 3~M_{\odot}$ become C-rich, whereas more massive
stars evolve as oxygen-rich objects. This holds for both the ATON and the MONASH models.
Because most of the mass loss in low-mass stars takes place during the C-rich phase
(see Fig.~\ref{fchim}), we find that most of the dust produced by these stars is under 
the form of carbonaceous solid particles: solid carbon and SiC.

The evolution of the grain size of the various particles formed in the wind of AGB stars
is thoroughly documented in the literature \citep{fg06, nanni13a, nanni13b, nanni14, 
ventura14} and we do not repeat the details here.
We will discuss how dust production works in oxygen-rich and carbon stars, and eventually
describe the overall dust formation phenomenon in AGB stars of solar metallicity.

\begin{figure*}
\begin{minipage}{0.48\textwidth}
\resizebox{1.\hsize}{!}{\includegraphics{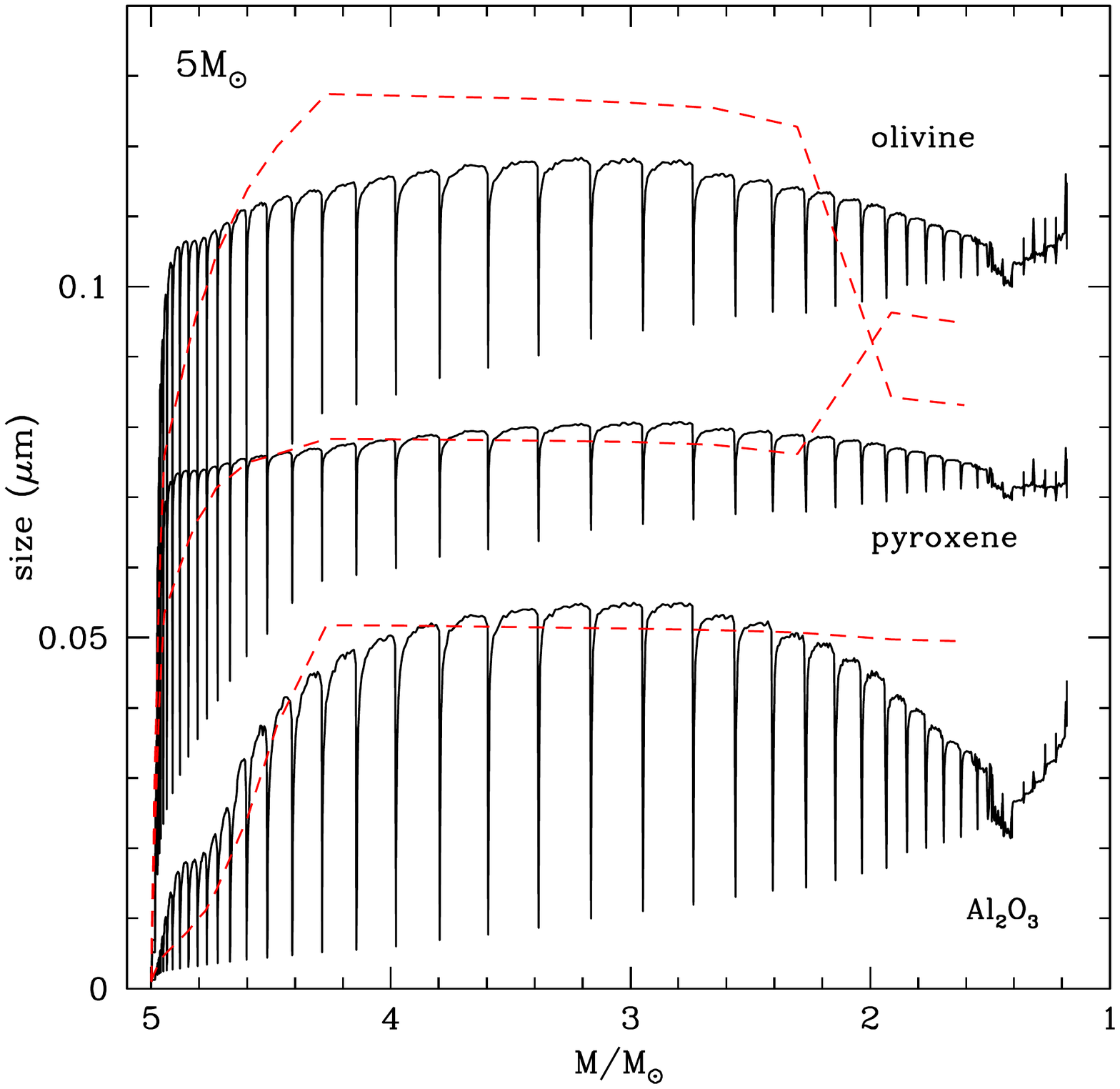}}
\end{minipage}
\begin{minipage}{0.48\textwidth}
\resizebox{1.\hsize}{!}{\includegraphics{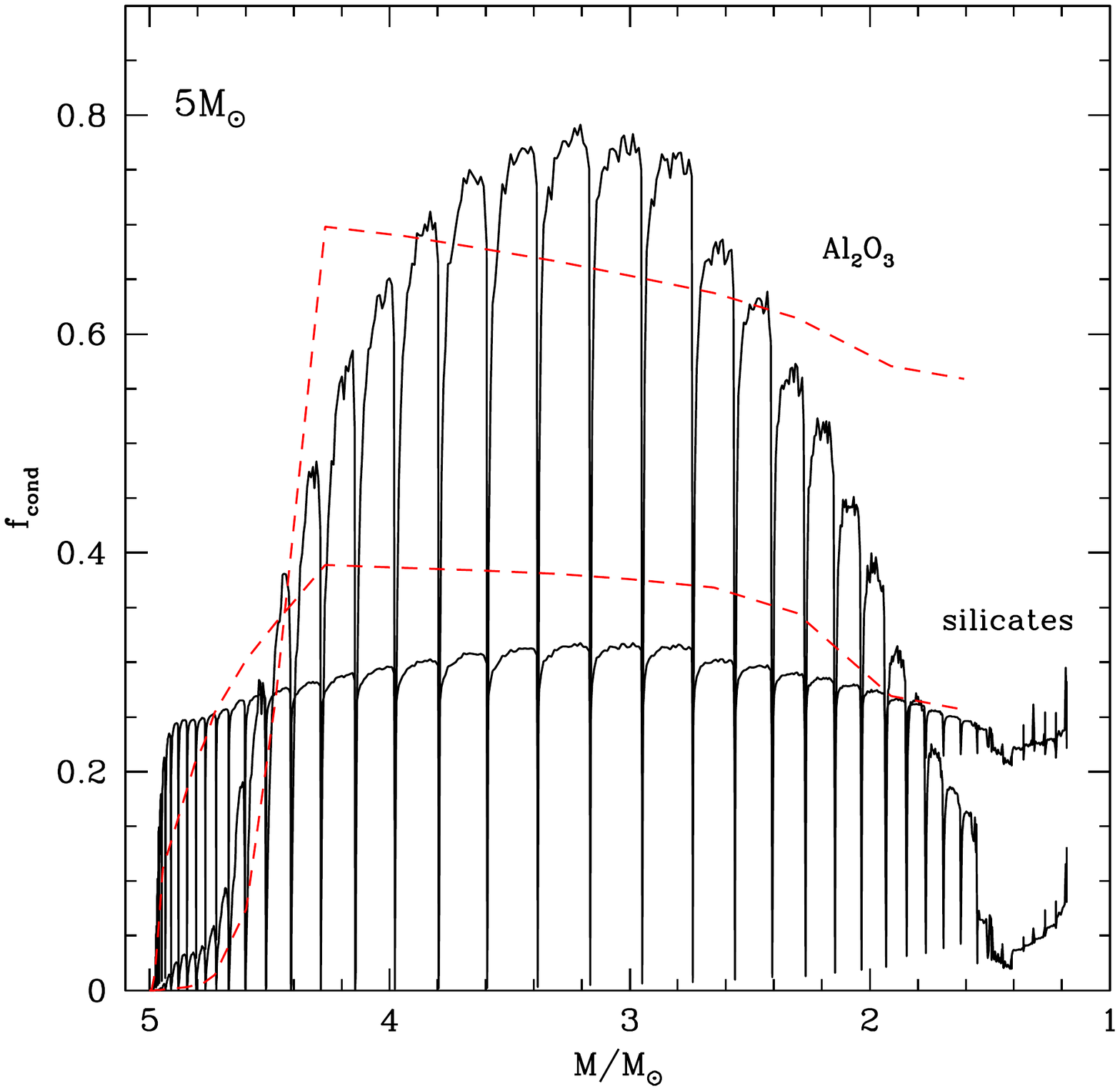}}
\end{minipage}
\vskip-80pt
\begin{minipage}{0.48\textwidth}
\resizebox{1.\hsize}{!}{\includegraphics{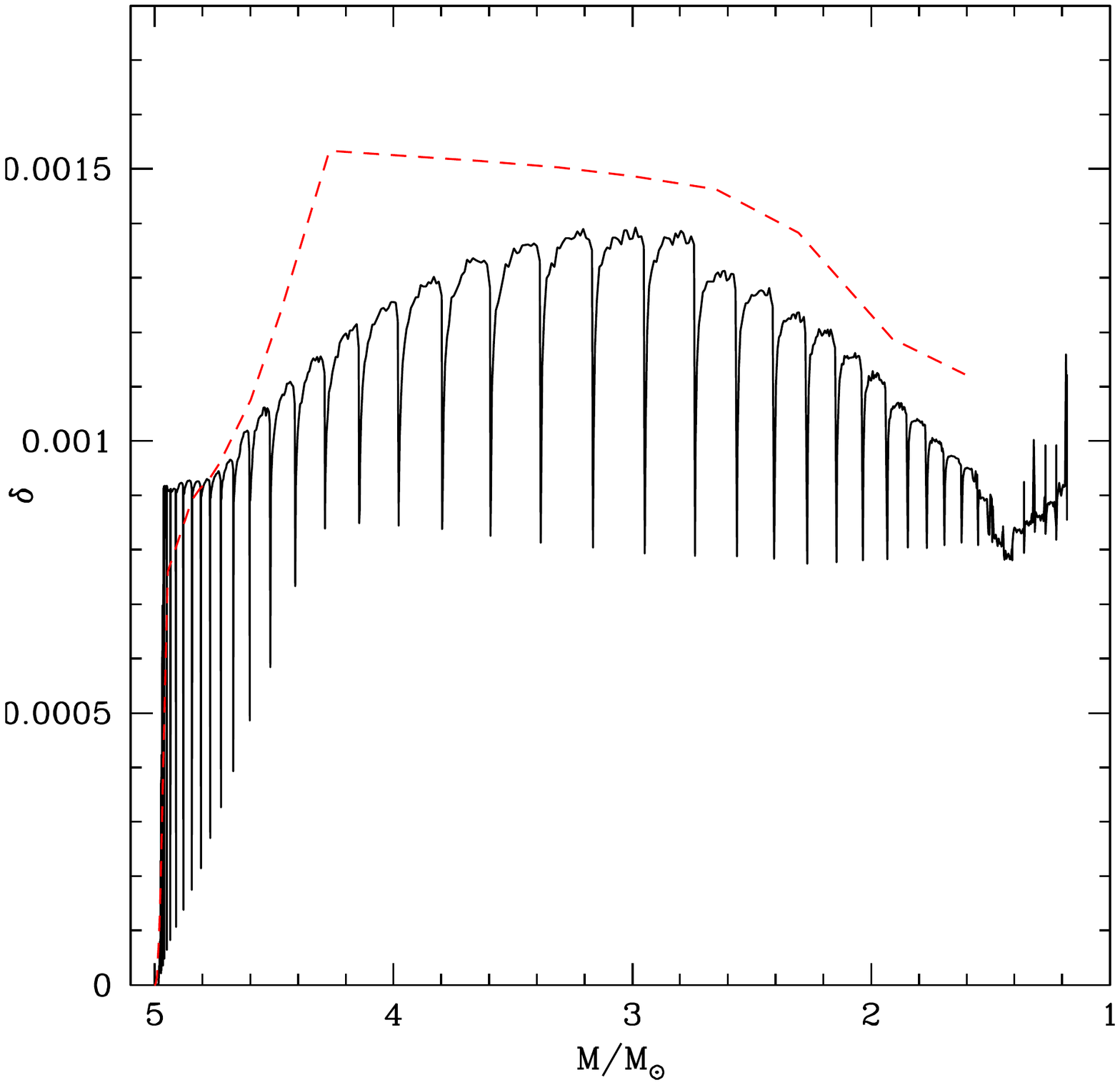}}
\end{minipage}
\begin{minipage}{0.48\textwidth}
\resizebox{1.\hsize}{!}{\includegraphics{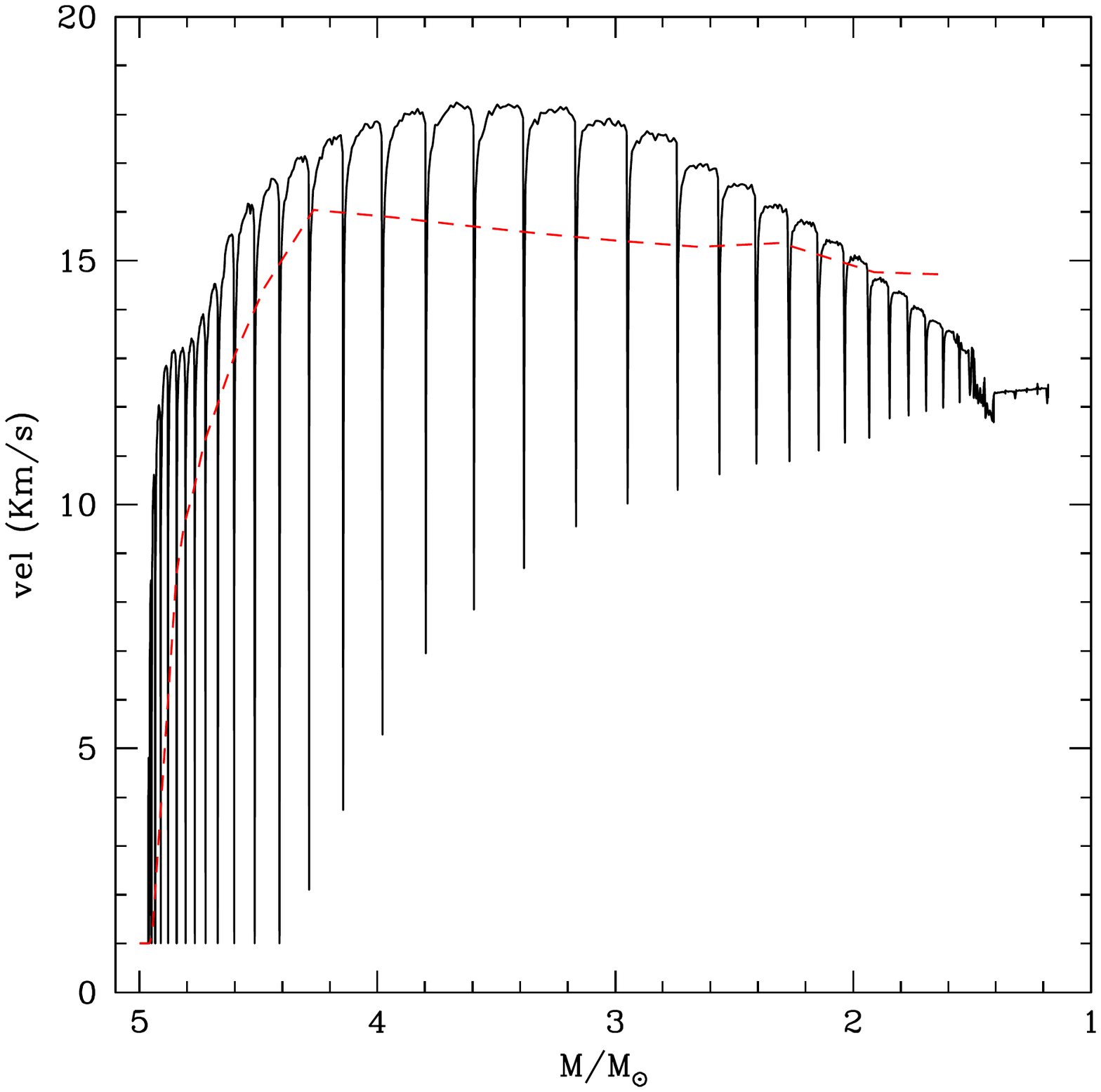}}
\end{minipage}
\vskip-50pt
\caption{Left: the variation of the grain size of olivine, pyroxene and
alumina dust formed in the wind of a $5~M_{\odot}$ during the AGB phase. The results
discussed in this work are indicated with a black, solid line and are compared with
the results obtained on the basis of the AGB models by
\citet{karakas16} are indicated with
a red, dashed line. Right: The variation of the fraction of silicon condensed into 
silicates and of aluminium condensed into $Al_2O_3$ for the same models shown in the
left panel.}
\label{fd5}
\end{figure*}

\subsection{Dust formation under HBB conditions}
The top, left panel of Fig.~\ref{fd5} shows the evolution of the grain size of two out of the 
three silicate species considered (olivine and pyroxene; quartz is not shown, for clarity
reasons) and of alumina dust ($Al_2O_3$) in the $5~M_{\odot}$ model presented in 
Fig.~\ref{f5fis}.  The top, right panel of the same figure shows the condensation factor of 
silicon and  aluminium\footnote{We refer to the condensation factor of a given element 
as the fraction 
of the element in the gaseous state that condensed into dust. For these specific cases we
refer to the fraction of gaseous silicon condensed into olivine, pyroxene and quartz and 
to the fraction of gaseous aluminium condensed into $Al_2O_3$.}. 
These results confirm earlier findings \citep{ventura14}: a) most of the dust produced 
is in the form of silicates, the dominant species being olivine, followed by pyroxene; 
b) during the HBB phase around $\sim 40\%$ of silicon is condensed onto dust, whereas the 
condensation factor of aluminium is higher, around $\sim 80\%$ \citep{flavia14}.
In the same figure we see that the typical dust to gas ratio ranges from $10^{-3}$ to
$1.5 \times 10^{-3}$, whereas the velocities of the wind fall in the range
$15-20$ km/s.

The comparison with the models by \citet{karakas16} highlights 
that these results are not greatly affected by the input physics used, particularly
the treatment of convection and mass loss. While in the previous sections we found
significant differences for the chemistry of these stars and the stellar yields,
the results in terms of dust production are fairly similar. 
The explanation is in the dust formation process, in particular the relationship
between the growth of dust particles and the dynamics of the wind. ATON models evolve
at larger luminosities (see Fig.~\ref{f5fis}), which based on Eq.2, enhances the effects 
of radiation pressure on dust particles. This provokes a fast acceleration of the wind
(note in the bottom, left panel of Fig.~\ref{fd5} that the MONASH velocities are smaller),
which in turn favours the decrease in the gas density (see Eq.~4), hence in the number
of molecules available for condensation into dust. The dust formation mechanisms for
silicates is self-regulated and this is the reason for the similarity in the results 
shown in Fig.~\ref{fd5}; the findings concerning the production of silicates in the winds 
of massive AGB stars given here are rather general and independent of AGB modelling.

\begin{figure*}
\begin{minipage}{0.48\textwidth}
\resizebox{1.\hsize}{!}{\includegraphics{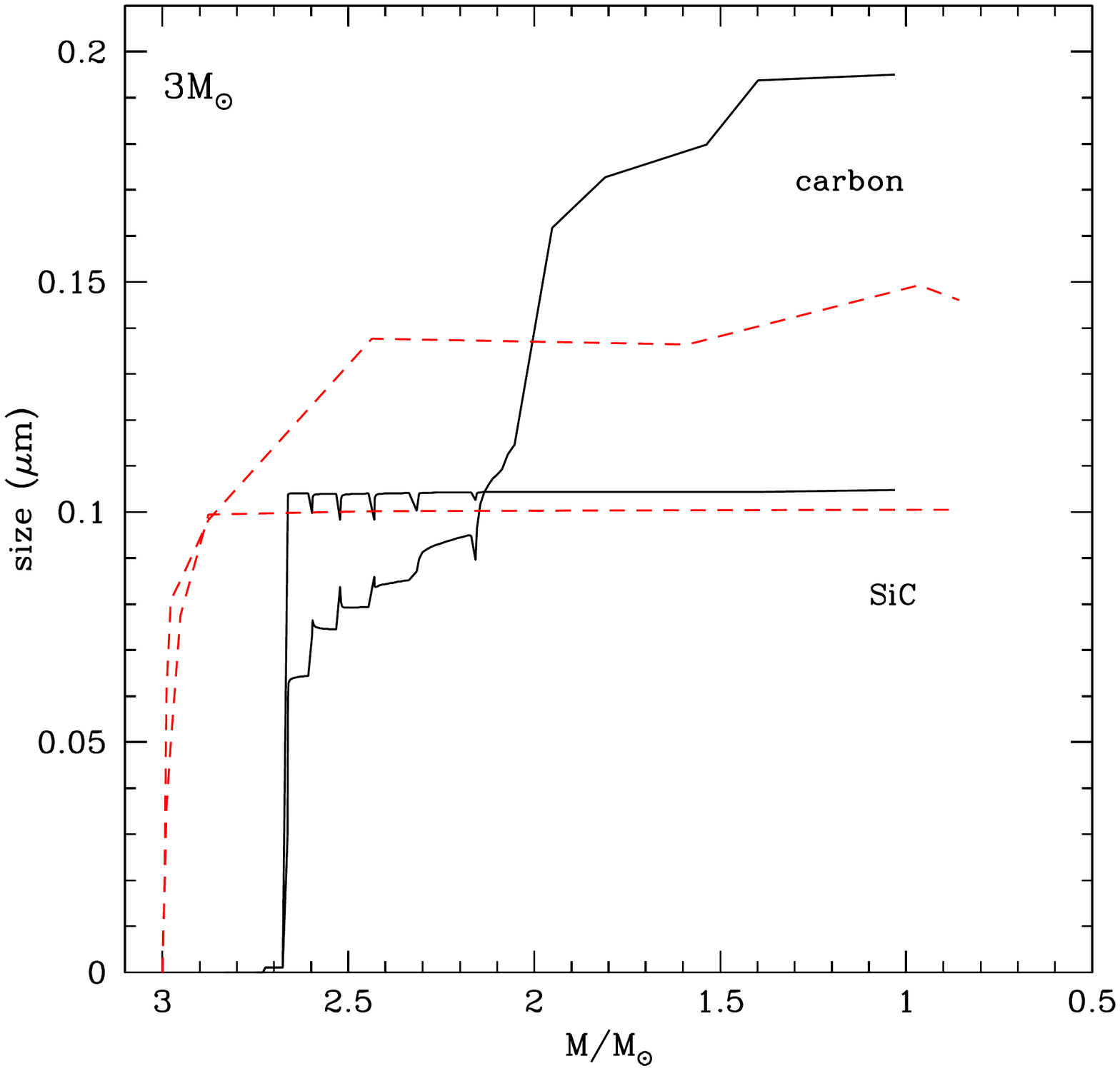}}
\end{minipage}
\begin{minipage}{0.48\textwidth}
\resizebox{1.\hsize}{!}{\includegraphics{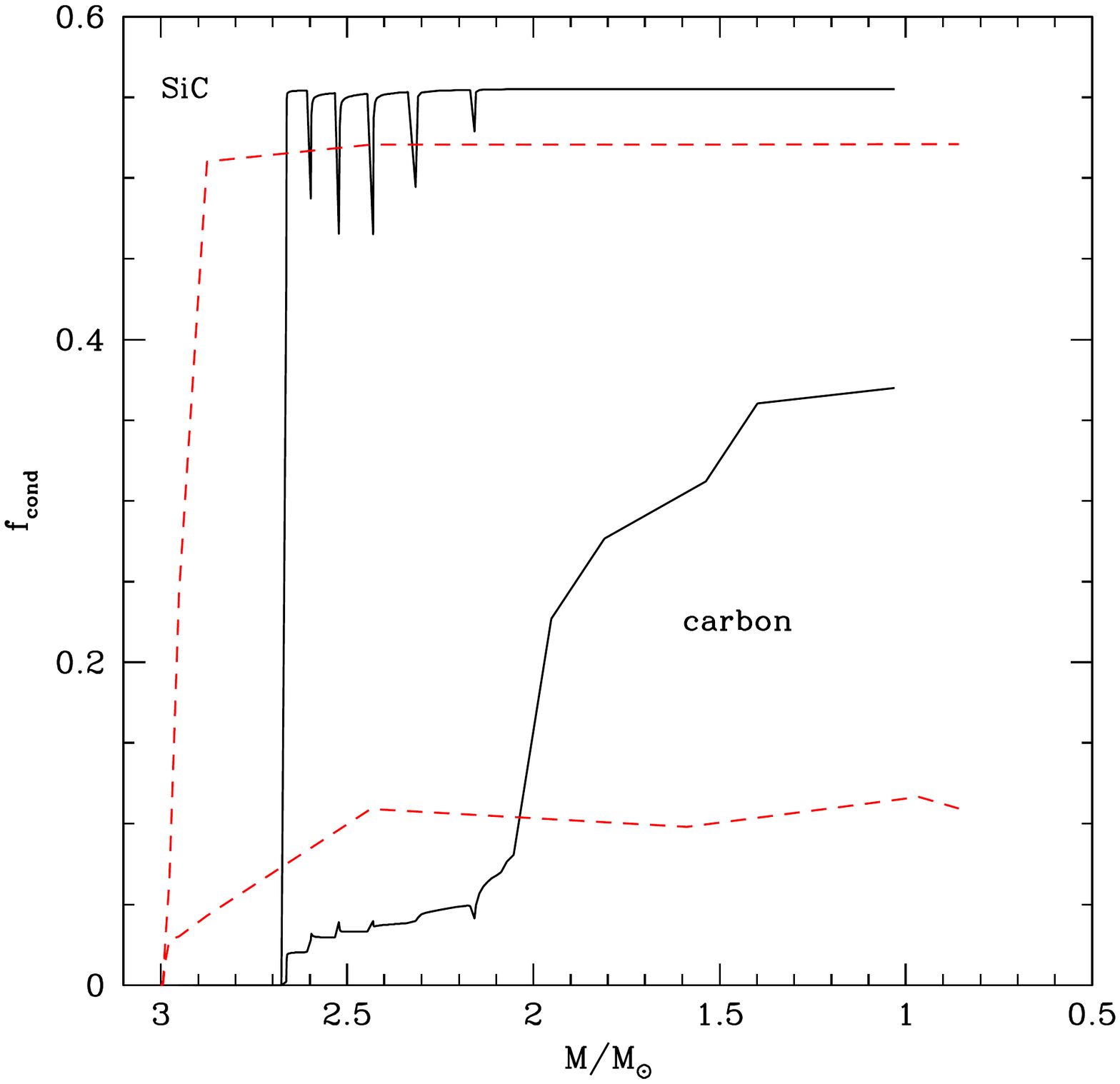}}
\end{minipage}
\vskip-80pt
\begin{minipage}{0.48\textwidth}
\resizebox{1.\hsize}{!}{\includegraphics{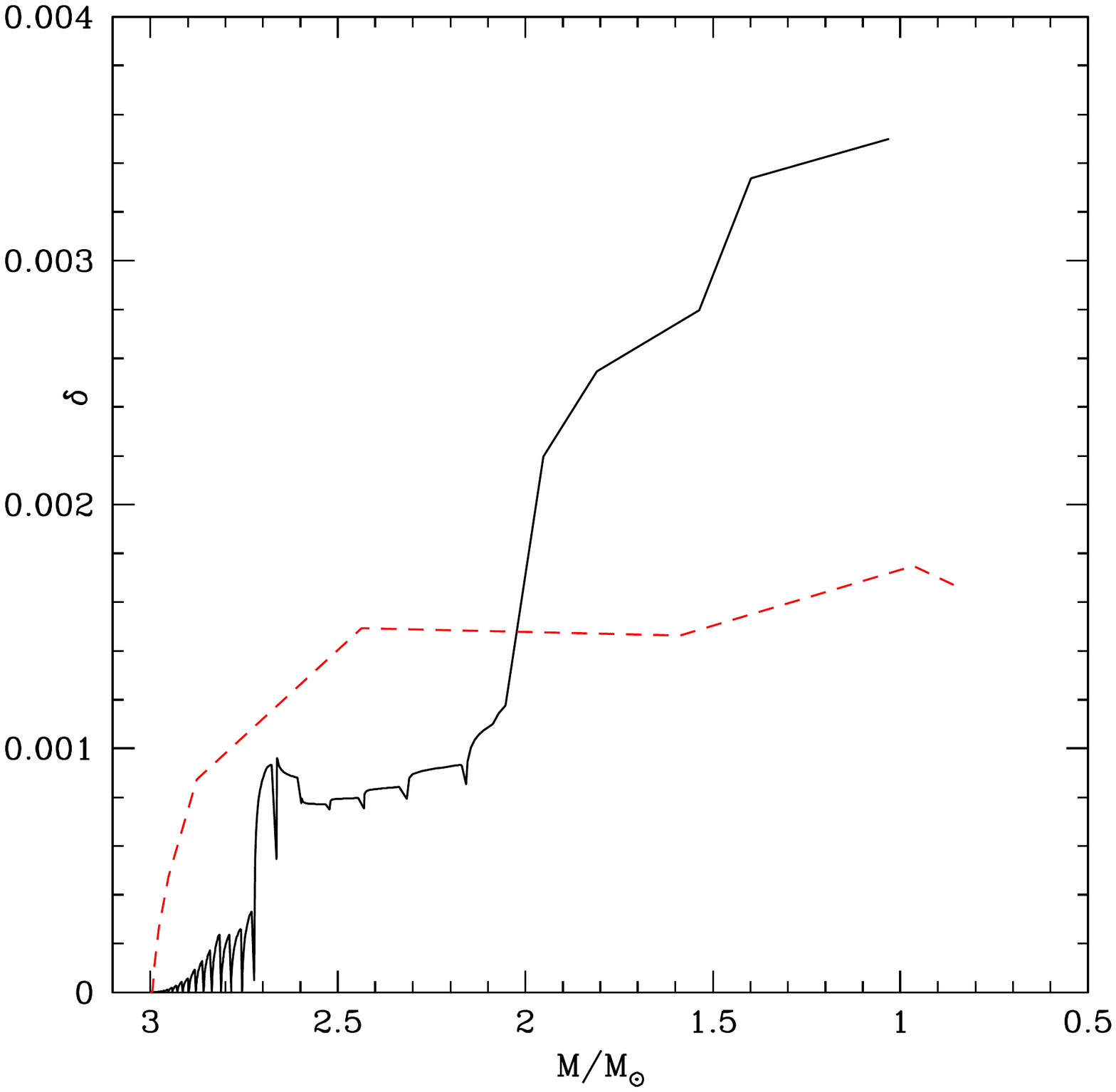}}
\end{minipage}
\begin{minipage}{0.48\textwidth}
\resizebox{1.\hsize}{!}{\includegraphics{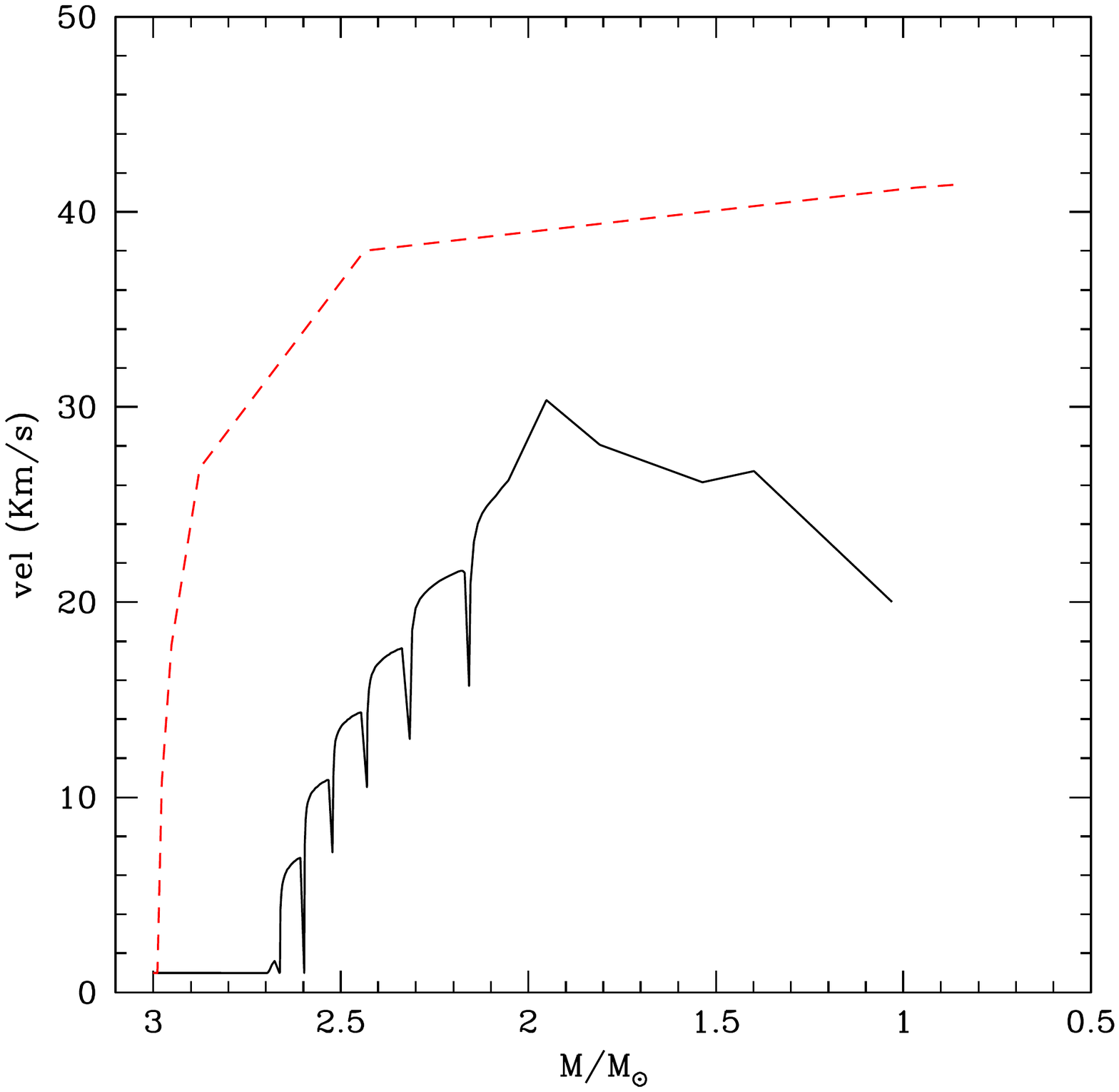}}
\end{minipage}
\vskip-50pt
\caption{Left: the variation of the grain size of solid carbon and SiC
formed in the wind of a $3~M_{\odot}$ during the AGB phase. The results
discussed in this work are indicated with a black, solid line and are compared with
the results obtained on the basis of the AGB models by \citet{karakas16}, indicated with
a red, dashed line. Right: The variation of the fraction of gaseous carbon condensed into 
solid carbon and of silicon condensed into SiC for the same models shown in the
left panel.}
\label{fd3}
\end{figure*}

\subsection{Dust formation in carbon stars}
\label{dustc}
Fig.~\ref{fd3} shows the results for the dust produced by a $3~M_{\odot}$ star. In this 
case, for the reasons given above, we focus our attention on SiC and solid carbon grains.

SiC is produced efficiently in the wind of
carbon stars, owing to the thermodynamic stability of this solid compound \citep{fg06}. 
The typical size of SiC particles is $\sim 0.1 \mu$m.
Indeed we find a saturation condition, such that the residual silicon not 
bound into the very stable SiS 
molecules \citep{fg06}, i.e. around $55 \%$, condenses 
into dust. This corresponds to grain sizes of the order of $0.1 \mu$m. The saturation 
process is the reason why the grain size of SiC grains and the corresponding condensation
fraction of silicon remain constant for the whole AGB phase, after the
achievement of C/O $\ge 1$. Saturation occurs shortly after the beginning of the 
carbon star phase and is fairly independent of the stellar parameters and the amount of
carbon accumulated. This is the reason why the ATON and MONASH results are extremely 
similar in this regards, as shown in Fig.~\ref{fd3}.

The dust production process in the winds of carbon stars can be divided into two
phases. At the beginning of the C-star phase, when the
excess of carbon with respect to oxygen is smaller than the amount of silicon in the
envelope, the dominant dust species is SiC. The progressive increase in the surface carbon
eventually makes carbon production dominant with respect to SiC. While SiC particles of 
the size given above form in an internal region of the circumstellar envelope, 
solid carbon particles of bigger size are produced in a more external zone. 
This can be seen in the left panel of Fig.~\ref{fd3}, where the size of carbon 
particles grow bigger and bigger, until reaching dimensions slightly below $0.2 \mu$m.
In the very final phases the fraction of carbon condensed into dust approaches 
$40 \%$ and the dust-to-gas ratio increases up to $\sim 0.0035$. We reiterate here that
these late evolutionary phases, though extremely important for the dust pollution by these
stars (as shown in Fig.~\ref{fd3}) are extremely short relative to the whole TP-AGB phase,
and more importantly, to the duration of the C-star phase. 

When comparing Fig.~\ref{fd3} with the results based on MONASH models, 
we find significant differences for the production of solid carbon. 
In the MONASH case the fraction of carbon condensed into dust barely
exceeds $10 \%$, the maximum size of the carbon grains formed is $0.15 \mu m$ and
the dust-to-gas ratio is below $0.002$. The reason for this difference can be deduced 
based on the results shown in the top, right panel of Fig.~\ref{f3msun} and discussed  
in Section~\ref{lmagb}. In the ATON case, owing to the presence of a narrow and very 
high overadiabaticity peak, the effective temperatures become extremely cool, which
favours very large mass-loss rates, which in turn favours dust production. This does
not affect the formation of SiC, as there is no more gaseous silicon available, but
strongly enhances the formation of solid carbon. In the MONASH case the effective
temperature are hotter, thus no enhanced formation of solid carbon is found.
Note that this difference holds despite the fact that the surface carbon is higher in the MONASH
model. This is because the growth of carbon grains is much more
sensitive to the rate of mass-loss than the surface carbon abundance
\citep{ventura16a}.

\begin{figure*}
\begin{minipage}{0.48\textwidth}
\resizebox{1.\hsize}{!}{\includegraphics{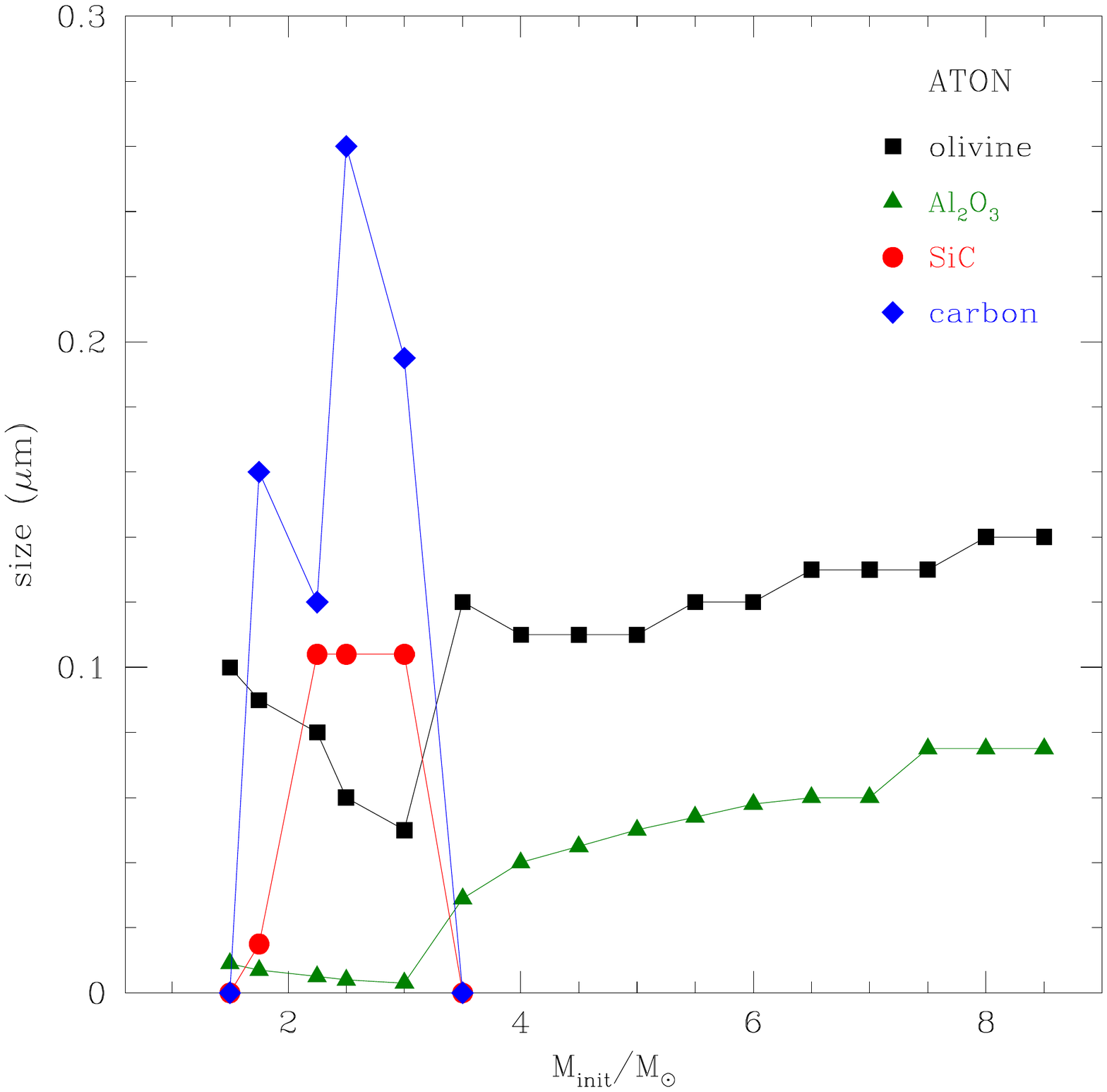}}
\end{minipage}
\begin{minipage}{0.48\textwidth}
\resizebox{1.\hsize}{!}{\includegraphics{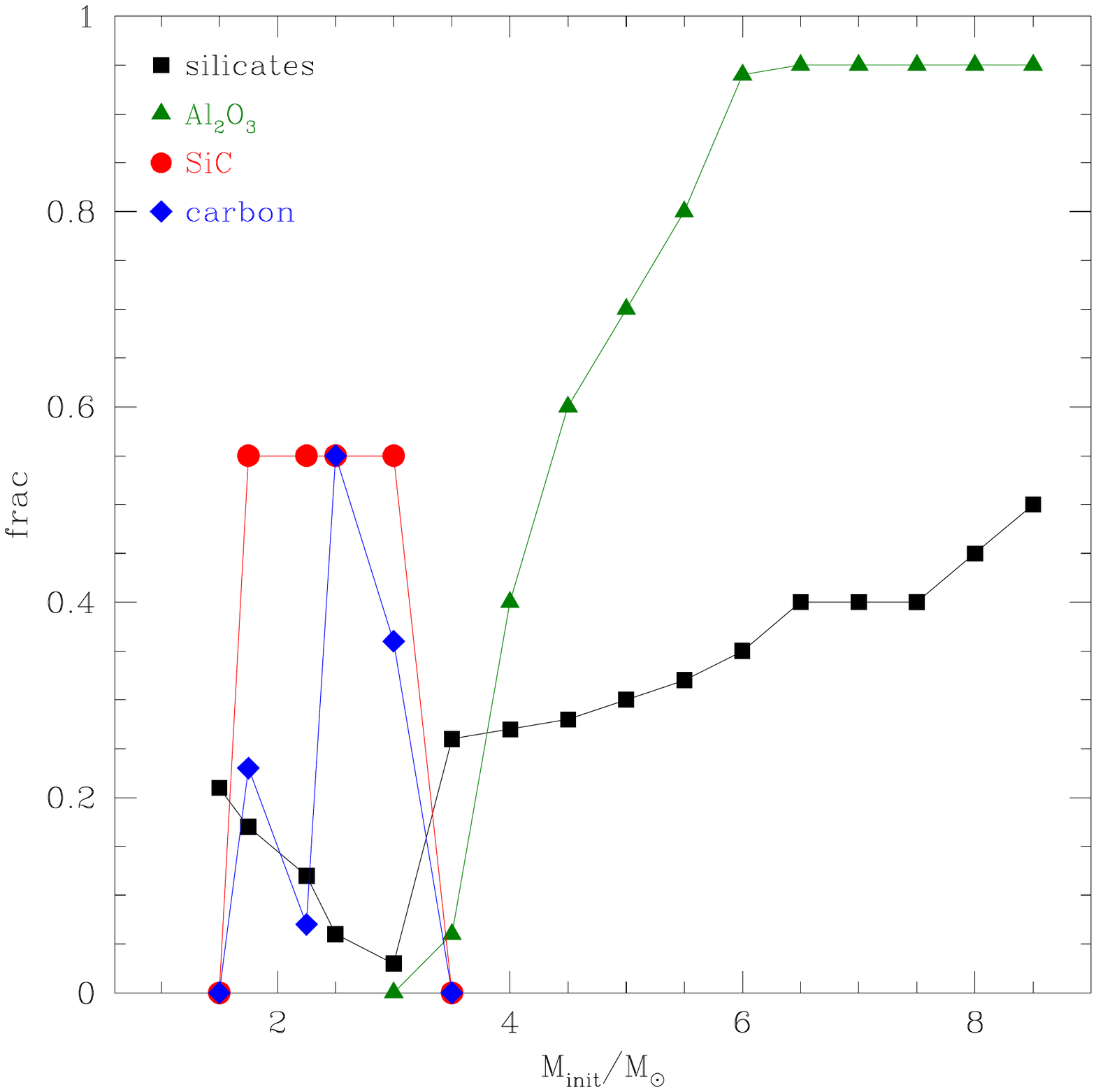}}
\end{minipage}
\vskip-60pt
\caption{Left: The typical size of dust particles of different species formed in the
wind of AGB stars of different mass. The meaning of the various symbols is the following:
olivine - black squares; alumina dust - green triangles; SiC - red points; solid carbon -
blue diamonds. Right: The fraction of the key elements condensed into solid particles
during the AGB phase. The fraction of silicon condensed into silicates and SiC is
indicated, respectively, with black squares and red points; green triangles and blue
diamonds indicate the fraction of aluminium involved in $Al_2O_3$ and the fraction of
gaseous carbon condensed into carbon grains.
}
\label{fsize}
\end{figure*}

\subsection{The properties of dust particles in solar metallicity AGB stars}
Fig.~\ref{fsize} shows the typical grain size of the various dust particles formed
in the wind of stars of different mass during the AGB phase. We also show the fraction
of the key species condensed into dust (silicon for silicates and SiC, aluminium for
$Al_2O_3$ and carbon).

Low-mass stars with mass $1.75~M_{\odot} \leq M_{init} \leq 3~M_{\odot}$ produce mainly solid 
carbon and SiC. Owing to the saturation effect discussed above, the size of the SiC grains
formed and the silicon condensation fraction are $\sim 0.1 \mu$m and
$55 \%$ respetively, independent of mass. The size of carbon grains span the range 
$0.1\mu m < a_{\rm C} < 0.28 \mu m$, while the carbon condensation factor is within
$10\% < f_{\rm C} < 55\%$. These values depend on the amount of carbon accumulated into the 
envelope and the rate of mass loss, which change with stellar mass.
The carbon grains with the largest size form in the winds of the stars which experience
a large number of TDU events and accumulate the largest amount of carbon; as shown in
Fig.~\ref{fchim}, this occurs for stars of initial mass $2.5-3~M_{\odot}$, which are 
expected to exhibit extremely large infrared excesses towards the final AGB phases.
Note that stars in this mass range have been invoked to reproduce the most obscured sources
in the LMC \citep{flavia15a, flavia15b, ventura16a}.

Fig.~\ref{fsize} shows that low-mass stars also produce some silicates. These
are produced in the evolutionary phases before the achievement of the C-star
stage, when the star is still oxygen-rich. For masses around the lower limit to
become carbon stars, namely $\sim 1.5~M_{\odot}$, the quantity of silicates
produced is higher than SiC and carbon, because the C-star stage is reached
only at the very end of the AGB phase, when a significant fraction of the
(oxygen-rich) envelope is already lost.

Interestingly, these stars could be the progenitors of the so-called
mixed-chemistry PNe, where both C- and O-rich IR dust features have been
detected simultaneously \citep{guten08,perea09,liz11,liz14,liz15}. These objects
show polycyclic aromatic hydrocarbon (PAH) features (C-rich) as well as
(amorphous/crystalline) silicates features (O-rich) and their stellar origin is
not understood \citep[see e.g., recent discussions in][and references therein]{garcia14,garcia16}.  
One of the explanations to this mixed-chemistry phenomenon, especially for the
case of the PN BD$+$30 36 39 \citep{liz15}, is to invoke a fatal thermal pulse
at the very end of the AGB phase \citep{waters98,perea09,liz15}. The
evolutionary models presented here, show that this mixed-chemistry could
naturally arise as part of the evolution of at least some (i.e., those converted
to C-rich) low-mass stars. Indeed, \citet{rojas17} have very recently reported
precise C/O ratios (homogeneously derived) for a sample of 23 mixed-chemistry
PNe, which combined with the He abundances and N/O ratios, suggest for the first
time that some mixed-chemistry PNe may be the descendants of very low-mass (M
$<$ 1.5 M$_{\odot}$) stars.

In the wind of stars of mass $M \geq 3.5~M_{\odot}$ the only  dust species
formed are silicates and alumina dust. As shown in the left panel of
Fig.~\ref{fsize}, the typical size of the grains formed are $0.12-0.14 \mu$m and
$0.03-0.08 \mu$m, respectively, for olivine and alumina dust. The fraction of
silicon condensed into silicates ranges from  $\sim 25\%$ to $\sim 50\%$,
whereas for alumina dust the fraction of aluminium condensed into dust ranges
from $\sim 10\%$ to $\sim 95\%$. The large  percentages of gasesous aluminium
condensed into dust stems from the large stability of $Al_2O_3$, which forms at
temperatures $T \sim 1400$K in regions of the circumstellar envelope very close
to the surface of the star \citep{flavia14}. Both the amounts of silicates and
of alumina dust formed increase with the mass of the star, because stars of
higher mass also experience higher mass-loss rates, which leads to denser winds
with a higher number of gaseous molecules available to form dust. This in
agreement with previous studies focused on lower metallicity AGB stars
\citep{paperI, paperII,  ventura14}.

\begin{figure}
\resizebox{1.\hsize}{!}{\includegraphics{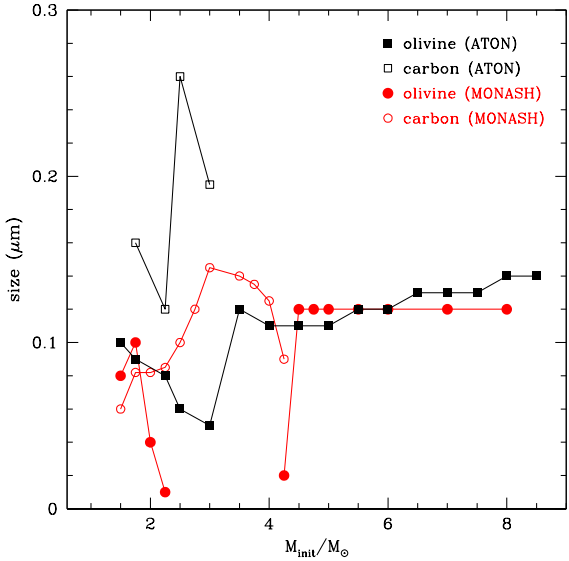}}
\vskip-60pt
\caption{The size of solid carbon (open squares) and olivine (full squares) particles
formed in the wind of the AGB models discussed here. Open, red circles and full, red
points indicate, respectively, the dimension of solid carbon and olivine grains
formed when the MONASH results for the description of the AGB phase are used.
}
\label{fconf}
\end{figure}

Fig.~\ref{fconf} shows the comparison between the grain sizes of the dust particles
found when using the present models and the AGB models by \citet{karakas16}.
For clarity we only show the most relevant silicate species, olivine, and carbon.
Following the discussion above and the results shown in Fig.~\ref{fd3}, we know that the 
size of SiC particles are extremely similar in the two cases.

The dimension of the olivine grains is fairly similar between the ATON and MONASH models.
This holds both in the massive AGB domain and for low-mass stars, with $M < 2~M_{\odot}$.
This is consistent with the results shown in Fig.~\ref{fd3} and with our previous study 
on the oxygen-rich stars in the LMC with the largest infrared emission \citep{ventura15a}.
The only difference holds in the range of mass $2~M_{\odot} < M < 4~M_{\odot}$, where we
find some silicate production in the present models, whereas a
negligible amount of silicates form in the MONASH case. This is because in the latter case the achievement of 
the C-star stage occurs when only a tiny fraction of the envelope was lost, thus all the
dust formed and ejected into the interstellar medium is under the form of carbonaceous
dust.

\begin{table*}
\setlength{\tabcolsep}{6pt}
\caption{Dust masses (in solar masses) produced by solar metallicity AGB stars.}                                       
\begin{tabular}{c c c c c c c c c}        
\hline\hline                        
M &     $M_{\rm ol}$ &  $M_{\rm py}$ &  $M_{\rm qy}$ &$M_{Al_2O_3}$
  &$M_{\rm SiC}$ & $M_{\rm C}$  & $M_{\rm ir}$ & $M_{\rm dust}$ \\
\hline        
1.50 &  1.05E-04 &  3.67E-05 &  1.13E-05 &  1.28E-07 &  0.00E+00 &  0.00E+00 &  3.73E-04 &  5.26E-04  \\
1.75 &  8.00E-05 &  2.82E-05 &  9.21E-06 &  7.28E-08 &  2.22E-04 &  5.48E-04 &  1.19E-04 &  1.01E-03  \\
2.00 &  2.59E-05 &  9.42E-06 &  3.94E-06 &  2.64E-08 &  5.00E-04 &  0.00E+00 &  1.27E-03 &  1.81E-03  \\
2.25 &  2.33E-05 &  1.12E-05 &  4.38E-06 &  1.25E-08 &  7.55E-04 &  2.97E-04 &  2.50E-04 &  1.34E-03  \\
2.50 &  6.70E-06 &  2.65E-06 &  1.37E-06 &  6.19E-09 &  6.68E-04 &  4.30E-03 &  1.22E-04 &  5.10E-03  \\
3.00 &  2.77E-06 &  1.21E-06 &  5.21E-07 &  3.94E-09 &  2.95E-03 &  1.05E-02 &  8.52E-05 &  1.35E-02  \\
3.50 &  1.43E-03 &  4.41E-04 &  5.63E-05 &  1.90E-05 &  0.00E+00 &  0.00E+00 &  2.54E-04 &  2.19E-03  \\
4.00 &  1.99E-03 &  6.12E-04 &  5.48E-05 &  1.13E-04 &  0.00E+00 &  0.00E+00 &  1.45E-04 &  2.92E-03  \\
4.50 &  2.41E-03 &  7.35E-04 &  4.81E-05 &  2.36E-04 &  0.00E+00 &  0.00E+00 &  9.83E-05 &  3.52E-03  \\
5.00 &  2.96E-03 &  9.09E-04 &  5.18E-05 &  4.59E-04 &  0.00E+00 &  0.00E+00 &  9.75E-05 &  4.48E-03  \\
5.50 &  3.33E-03 &  9.98E-04 &  4.61E-05 &  5.67E-04 &  0.00E+00 &  0.00E+00 &  6.78E-05 &  5.01E-03  \\
6.00 &  4.09E-03 &  1.19E-03 &  4.11E-05 &  8.34E-04 &  0.00E+00 &  0.00E+00 &  5.92E-05 &  6.21E-03  \\
6.50 &  4.83E-03 &  1.41E-03 &  4.17E-05 &  1.16E-03 &  0.00E+00 &  0.00E+00 &  6.29E-05 &  7.50E-03  \\
7.00 &  5.47E-03 &  1.55E-03 &  4.08E-05 &  1.26E-03 &  0.00E+00 &  0.00E+00 &  6.58E-05 &  8.39E-03  \\
7.50 &  6.55E-03 &  1.85E-03 &  3.73E-05 &  1.52E-03 &  0.00E+00 &  0.00E+00 &  7.70E-05 &  1.00E-02  \\
8.00 &  8.13E-03 &  2.16E-03 &  3.10E-05 &  1.72E-03 &  0.00E+00 &  0.00E+00 &  1.06E-04 &  1.21E-02  \\
8.50 &  1.01E-02 &  2.58E-03 &  2.40E-05 &  1.89E-03 &  0.00E+00 &  0.00E+00 &  1.24E-04 &  1.48E-02  \\
\hline     
\end{tabular}
\label{tabdust}
\end{table*}

The ATON and MONASH results are similar for carbon stars of mass
$M \leq 2~M_{\odot}$, whereas they differ for $M \sim 2-2.5~M_{\odot}$.
In the present models we find a much larger formation of solid carbon particles, which
reach sizes in the range $0.2 \mu m < a_C < 0.28 \mu m$.  Conversely, when using the
MONASH models, we find carbon grain dimensions below $0.15 \mu m$. This is the only
relevant difference found among the two sets of models, which has been extensively 
discussed in Section~\ref{dustc}.

\begin{figure}
\resizebox{1.\hsize}{!}{\includegraphics{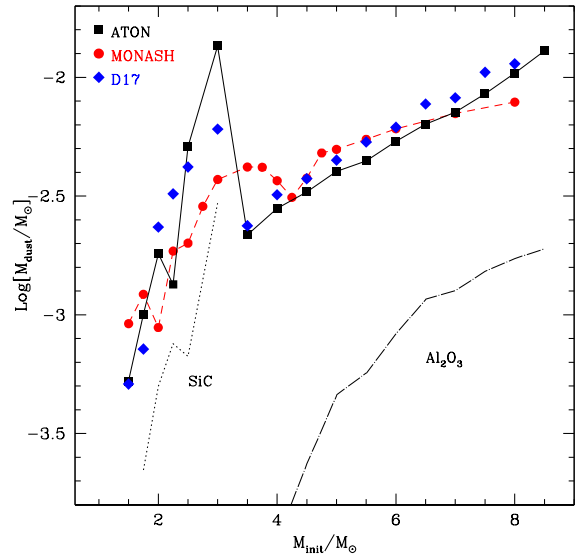}}
\vskip-60pt
\caption{The total dust mass produced by solar metallicity, AGB models is shown as
a function of the initial mass of the star and indicated with black, full squares.
The results based on the MONASH models for the AGB evolution are indicated
with full, red points, whereas the results from D17 are shown as blued diamonds.
}
\label{fmdust}
\end{figure}

\subsection{The overall dust mass budget by AGB stars}
We conclude this analysis with the discussion on the dust mass produced by AGB stars of
solar metallicity.  In Fig.~\ref{fmdust} we show the total dust mass
produced by stars  of different mass during the AGB
phase. The values of the dust mass of the individual species formed are reported in
Table~\ref{tabdust}; the total dust mass produced is indicated in the last column of the 
Table. As discussed previously, most of the dust produced by low-mass stars is 
solid carbon, whereas for massive AGB stars the majority of the dust formed are
silicates. The mass of SiC and alumina dust produced is significantly smaller than
the mass of carbon and silicates, and are shown separately in Fig.~\ref{fmdust}.

Stars with mass in the range $3.5~M_{\odot} < M < 8.5~M_{\odot}$ 
produce dust masses in the range $2\times 10^{-3}~M_{\odot} < M < 10^{-2}~M_{\odot}$.
This dust is mainly silicates: the contribution of alumina dust is below
$\sim 20\%$ (see Fig.~\ref{fmdust}). Low-mass stars in the range 
$1.75~M_{\odot} \leq M \leq 3~M_{\odot}$ produce carbonaceous particles, in quantities 
above $10^{-3}~M_{\odot}$. We note the peak of $\sim 10^{-2}~M_{\odot}$ in the dust mass 
produced for stellar masses $2.5 - 3~M_{\odot}$, due to the
abundant production of solid carbon in the wind of this stars. The contribution of SiC 
to the total dust produced ranges from $\sim 15\%$ to $\sim 30\%$.

These results are in substantial agreement with the results published by D17, indicating
that the details of the solar mixture adopted has a minor effect on the dust mass
expected. 

The comparison with the results based on the AGB models by \citet{karakas16} shows that
the results in the massive AGB domain are extremely similar.
In the low-mass domain the amount of carbon dust expected when using the MONASH
models is a factor $\sim 3$ smaller, owing to the differences in convection modelling.
Note that this dissimilarity is found only in a limited range of mass, namely for 
$2.5 - 3~M_{\odot}$ stars.

\begin{figure*}
\begin{minipage}{0.48\textwidth}
\resizebox{1.\hsize}{!}{\includegraphics{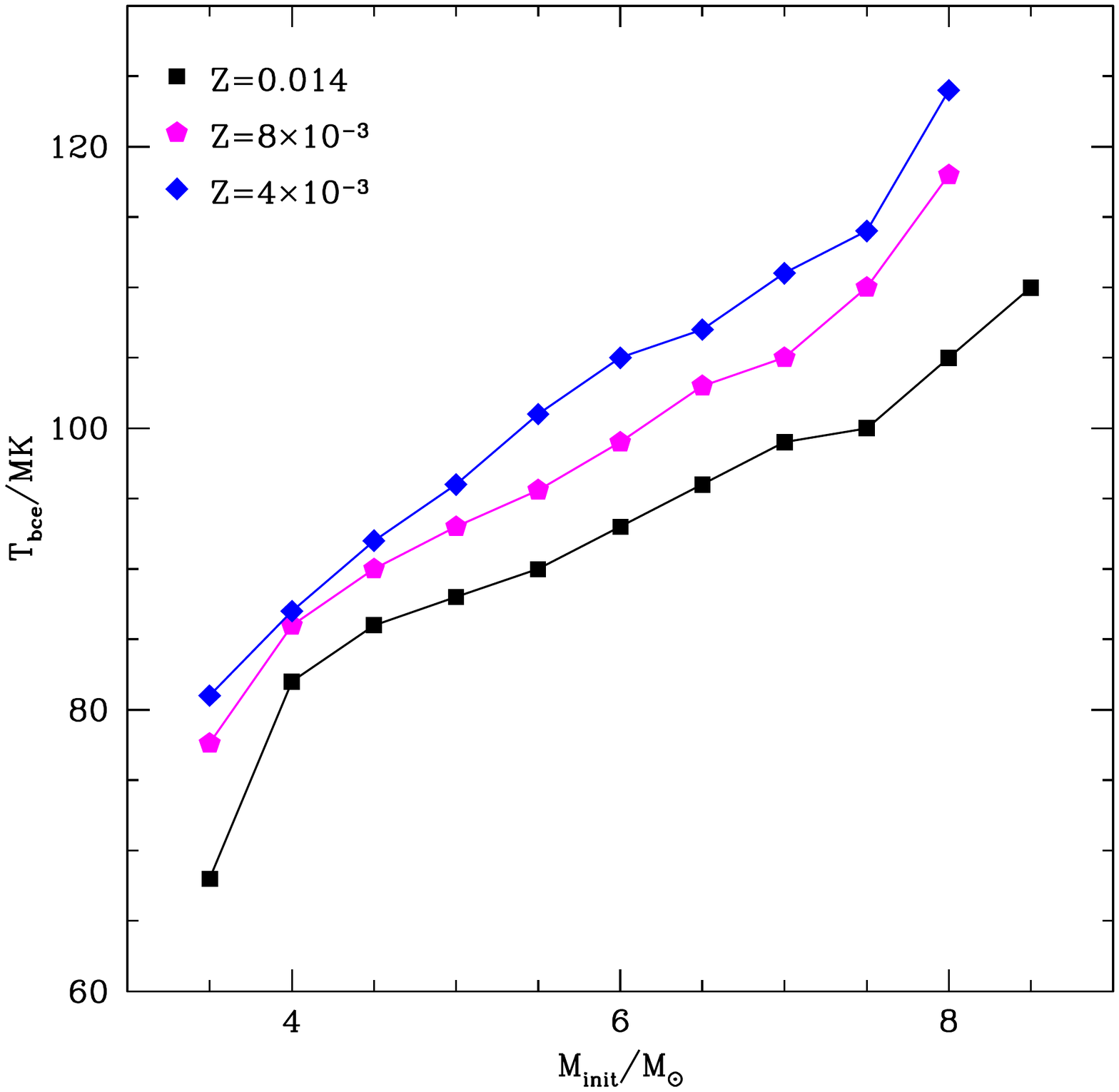}}
\end{minipage}
\begin{minipage}{0.48\textwidth}
\resizebox{1.\hsize}{!}{\includegraphics{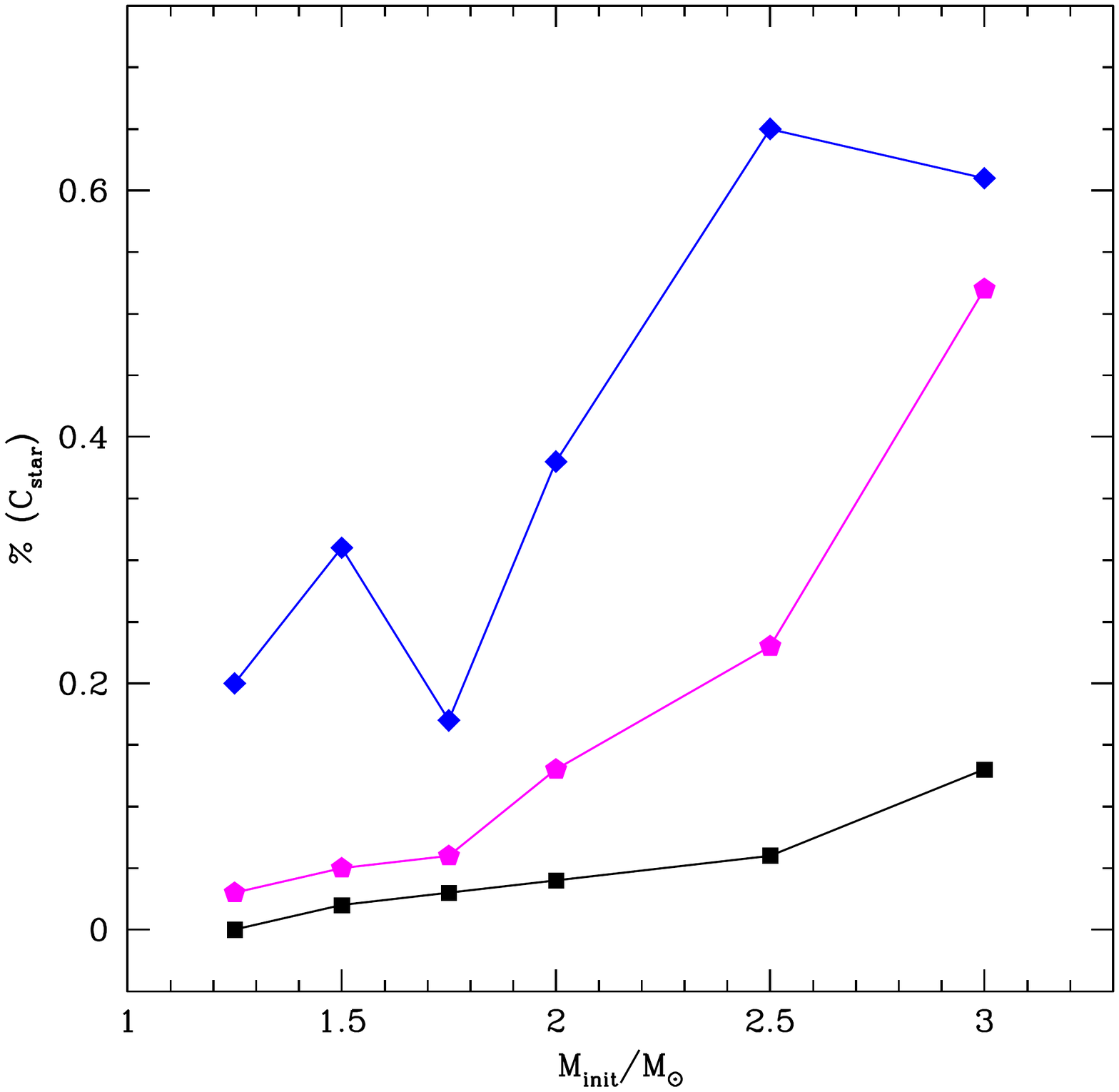}}
\end{minipage}
\vskip-60pt
\caption{Left: The temperature at the base of the convective envelope experienced 
by the solar metallicity models presented in this work (black squares), compared
with their counterparts of metallicity $Z=4\times 10^{-3}$ (blue diamonds) and
$Z=8\times 10^{-3}$ (magenta pentagons). Right: The time fraction of the C-star phase
in relation to the total duration of the AGB phase of low-mass, AGB models of
different metallicity. The meaning of the symbols is the same as in the left panel.}
\label{fmetfis}
\end{figure*}

\section{The role of metallicity on the evolutionary properties of AGB stars}
In our previous studies we used AGB models of sub-solar metallicity 
($Z=4,8\times 10^{-3}$) to interpret the evolved stars in the MC with the largest 
infrared excess \citep{ventura15a, ventura16a}. 
The comparison of theoretical results with the observational evidence is more tricky 
in the present case, because the poor knowledge of the distances of Galactic sources
prevents a straightforward interpretation of the currently available observations of
solar metallicity, AGB stars. 

Gaia and future space missions are likely alter this framework, with highly accurate 
determination of the parallaxes of several AGB sources and the availability of high-quality
data. To be prepared for the interpretation of these results we discuss the differences 
between the present models and the cases discussed in \citet{ventura15a, ventura16a}.

On the physical side, if we focus on stars with mass
above $\sim 3.5~M_{\odot}$, we find that the HBB experienced at the
bottom of the convective mantle is weaker in the present models than
in \citet{ventura15a, ventura16a}. This result is in agreement with previous studies, 
focused on the sensitivity of the strength of HBB to the metallicity \citep{ventura13}.  
Fig.~\ref{fmetfis} shows the temperature at the base of the envelope of stars 
of various initial mass and metallicity. The differences between the models discussed here
and their $Z=4\times 10^{-3}$ counterparts reach
$\delta T \sim 15$ MK in the $M > 6~M_{\odot}$ domain. Given the steep sensitivity of the nuclear
proton capture cross sections for temperatures in the range $70-100$ MK, this reflects into
a much more advanced p-capture nucleosynthesis in lower metallicity stars.   

Turning to the low-mass regime, all the stars with initial mass $1~M_{\odot} < M < 3~M_{\odot}$ 
become 
carbon stars during the AGB evolution. In the present models the C-star condition is
reached in a more advanced phase compared to the models used in \citet{ventura15a} and
\citet{ventura16a}, owing to the initial higher oxygen content, which delays the 
$C/O>1$ condition.
As shown in the right panel of Fig.~\ref{fmetfis}, the time fraction of the phase during which
these stars are C-rich is significantly shorter in the solar case, ranging from
$\sim 2\%$ to $\sim 15\%$, compared to the lower metallicity chemistries. We expect to
detect a smaller fraction of carbon stars in the solar metallicity environment than in the
MC.

\begin{figure}
\resizebox{1.\hsize}{!}{\includegraphics{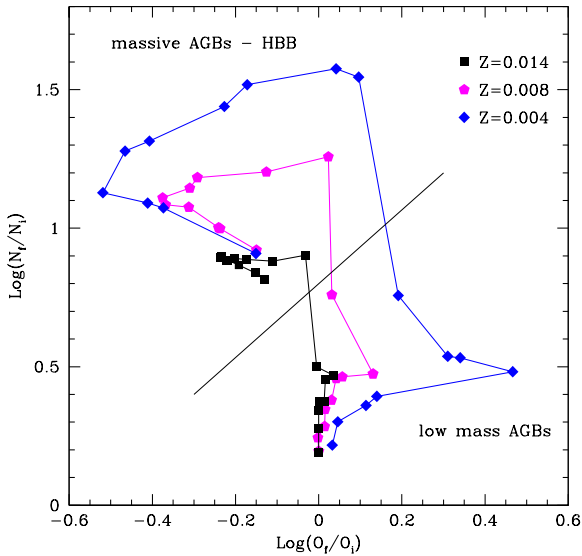}}
\vskip-60pt
\caption{The final surface chemical composition of the same models presented in 
Fig.~\ref{fmetfis}, in terms of the variation of the oxygen (reported on the abscissa)
and nitrogen (y-axis) mass fraction, compared to the quantities initially present in 
the star. The three lines connect models of the same metallicity, with the mass 
increasing counter-clock wise. The diagonal line marks the distinction
between low-mass AGB stars, whose chemical composition is determined by TDU, and
massive AGB stars, affected by HBB.
}
\label{fmetchem}
\end{figure}

The differences outlined above affect the evolution of the surface chemical composition of 
AGB stars. To show this, we report in Fig.~\ref{fmetchem} the final surface abundances of 
oxygen and nitrogen, in terms of the ratio with respect to the initial values. 

In the solar case, indicated by black squares, the surface oxygen of low-mass 
stars (reported in the lower, right region of the diagram) remains practically unchanged 
during the AGB life, because the initial oxygen is too large to be meaningfully affected 
by TDU. Conversely, for sub-solar metallicities, the amount of oxygen transported to 
the surface regions via repeated TDU events is sufficient to determine a significant 
variation, with a maximum increase of the order of a factor $\sim 3$, for the 
$Z=4\times 10^{-3}$, $2~M_{\odot}$ model. 

In the three lines shown in Fig.~\ref{fmetchem} we note the transition to the massive 
domain, in the sudden increase in the surface nitrogen, a clear signature of the effects 
of HBB, as also the decrease in the surface oxygen abundance. The latter effect
is more evident in lower metallicity models, because of the more advanced 
nucleosynthesis experienced, according to the discussion above and the results shown in 
the left panel of Fig.~\ref{fmetfis}.

\begin{figure}
\resizebox{1.\hsize}{!}{\includegraphics{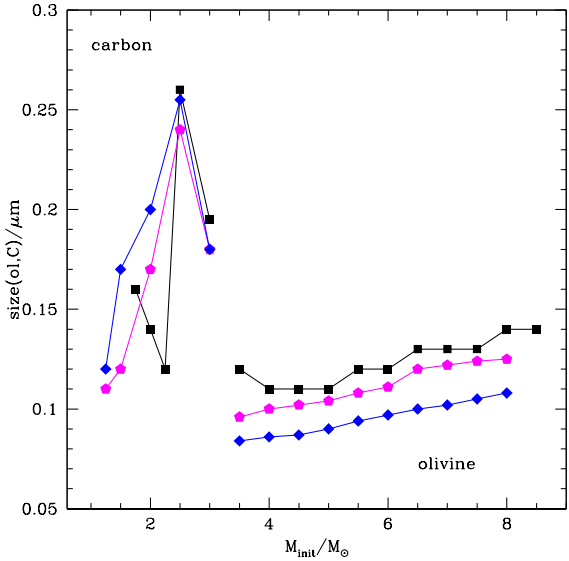}}
\vskip-60pt
\caption{The typical size of the dust particles formed in the wind of AGB stars of
different mass and metallicity. The colour coding is the same as in Fig.~\ref{fmetfis}.
For low mass stars we give the size of solid carbon grains, whereas for massive AGB
stars we report the size of olivine particles.
}
\label{fmetdsize}
\end{figure}

The differences in the physical behaviour and in the variation of the surface chemical
composition of AGB stars of different metallicity reflect into the dust production.
Fig.~\ref{fmetdsize} shows the typical size of solid carbon (for
stars of mass $M \leq 3~M_{\odot}$) and olivine grains ($M > 3~M_{\odot}$).
We focus on these two compounds because they are the dust species formed in the
largest quantities in C-rich and O-rich gas, respectively, and, more important,
because these are the solid particles providing the most important contribution to
the degree of obscuration of the star \citep{fg06}.

The size of the olivine grains formed is bigger in the solar metallicity models
compared to their lower-$Z$ counterparts, owing to the larger amount of silicon available. 
This trend, though limited to smaller metallicities, was discussed in \citet{ventura14}.
For what attains the carbonaceous particles formed, the largest
dimensions reached, of the order of $\sim 0.27 \mu$m, are substantially independent of
metallicity: this is because the carbon dredged-up during the TDU events is of
primary origin and is almost independent of the metallicity of the stars.
On the other hand, in the solar case only AGB stars of mass $2.5-3~M_{\odot}$
produce significant quantities of carbon dust; this is due to the later achievement of
the C-star stage, as discussed earlier in this section (see the right panel of
Fig.~\ref{fmetfis}).

\section{Conclusions}
We present updated models of stars with masses between $1~M_{\odot} - 8.5~M_{\odot}$,
evolved through the AGB phase. The chemical composition of the models, with metallicity
$Z=0.014$, reflects the chemistry of the Sun, thus it is suitable to
interpret the observations of Galactic AGB stars in the solar neighborhood and beyond.
To evaluate how the various results presented are
affected by the description of the input physics adopted, we compare the present results
with those obtained with the MONASH evolution code, which are based on
different input physics.

On the physical side, the behaviour of the models is mainly determined by the mass of the
star, with an abrupt transition occurring around $\sim 3.5~M_{\odot}$. Lower mass stars
experience TDU, which produces a gradual enrichment in carbon of their
surface layers.  Stars with initial mass $1.5~M_{\odot} \leq M \leq 3~M_{\odot}$ reach
the C-star stage with $1.2 < $C/O ratios $ < 3$ at the tip of the
AGB. While the  TP-AGB phase of these stars lasts between 1 Myr and 5
Myr,  the duration of the C-star  phase is below $\sim 15\%$ of the
overall TP-AGB evolution. The ejecta are enriched in 
carbon and show a modest increase in the nitrogen content. 

Stars of mass above $3.5~M_{\odot}$ experience 
HBB at the base of the convective envelope. Their evolution times, decreasing with the
mass of the star, are in the range $3\times 10^4 yr < \tau_{TP-AGB} < 3\times 10^5 yr$.
Their surface chemistry shows the signature of HBB nucleosynthesis, with
a significant depletion of carbon and a considerable production of 
nitrogen; oxygen and heavier species are almost unchanged at the present metallicity.
The final C/O $\sim 0.05 $ and N/O $\sim 1.8$ ratios are almost independent of the initial 
mass. The pollution from these stars is made up of C-poor gas, significantly enriched in nitrogen. 

Dust production by AGB stars is also very sensitive to their initial stellar mass.
Stars of mass $1.5~M_{\odot} \leq M \leq 3~M_{\odot}$ form mainly solid carbon particles,
of size $0.1-0.2 \mu$m; the solid carbon mass produced during the whole stellar life ranges
from $10^{-3}M_{\odot}$ to $2\times 10^{-2}M_{\odot}$, according to the initial mass
of the star. The second most abundant dust species formed in these stars is silicon carbide: 
SiC grains reach typical size of $\sim 0.1 \mu$m, whereas the total mass of SiC dust produced
is in the range $3\times 10^{-4}-3\times 10^{-3}~M_{\odot}$.

Massive AGB stars experiencing HBB form silicates and alumina dust. The amount of silicates
produced increases with the initial mass of the star, ranging from 
$2\times 10^{-3}M_{\odot}$ to $10^{-2}M_{\odot}$; the typical size of silicate particles
are $0.1-0.15 \mu$m. Owing to the small content of aluminium compared to silicon, the mass 
of alumina dust produced is $5-10$ times smaller than the mass of silicates, whereas the
dimension of alumina dust grains is about half of that of silicates.

The comparison between the present results with those obtained with the MONASH code outlines
strong similarities in the results of low-mass stars, in terms of the evolution of 
luminosity and core mass, the variation of the surface chemical composition and the 
duration of the whole AGB phase. Use of either ATON or MONASH models would lead to
similar conclusions in the interpretation of observations of Galactic low-mass AGB 
stars with solar or nearly solar metallicity.

The description of the AGB evolution of stars of mass above $3.5~M_{\odot}$ is more
uncertain, because the results obtained depend critically on the convective model adopted,
which affects directly the overall energy release, the growth rate of the core mass
and the variation of the surface chemistry. In the near future, observational results
from ongoing space missions will likely allow a better understanding of the main
properties of these stars.

The predictions concerning silicates, alumina dust and SiC produced by AGB stars of solar 
metallicity are also similar, allowing a model-independent interpretation of IR 
observations of of both O- and C-rich, dust-enshrouded stars,
and the determination of the global dust budget from AGB stars, particularly in galaxies with 
recent star formation. The amount of carbon dust produced is still affected by
significant uncertainties, primarily associated to the description of overadiabatic
convection on the outermost regions of C-rich stars.

\section*{Acknowledgments}
DAGH was funded by
the Ram\'on y Cajal fellowship  number RYC$-$2013$-$14182. DAGH and FD
acknowledge support provided by the Spanish Ministry of Economy and
Competitiveness (MINECO) under grant AYA$-$2014$-$58082-P. 
AIK acknowledge support provided by the Australian Research Council
(DP170100521 and FT110100475).

\end{document}